\documentclass[twocolumn]{aastex702}

\usepackage{booktabs}
\usepackage{amssymb}
\newcommand{\gaia}{\textit{Gaia}}
\newcommand{\g}{\ensuremath{G}}
\newcommand{\gbp}{\ensuremath{G_{\mathrm{BP}}}}
\newcommand{\grp}{\ensuremath{G_{\mathrm{RP}}}}
\newcommand{\bprp}{\ensuremath{G_{\mathrm{BP}}\!-\!G_{\mathrm{RP}}}}

\newcommand{\teff}{\ensuremath{T_{\mathrm{eff}}}}

\newcommand{\dcmd}{variability-encoded CMD}
\newcommand{\vecmd}{variability-encoded CMD}
\newcommand{\dscuti}{$\delta$ Scuti}
\newcommand{\mesa}{\texttt{MESA}}
\newcommand{\gyre}{\texttt{GYRE}}

\begin{document}


\title{Gaia in Technicolor. I. A Simple Variability-Encoded Color-Magnitude Diagram }

\correspondingauthor{Brad N. Barlow}

\author[0000-0002-8558-4353]{Brad N. Barlow}
\altaffiliation{These authors contributed equally to this work.}
\affiliation{Department of Physics \& Astronomy, University of North Carolina at Chapel Hill, Chapel Hill, NC 27599, USA}
\email[show]{bbarlow@unc.edu}

\author[0009-0003-4021-0934]{Bennett C. Kirby}
\altaffiliation{These authors contributed equally to this work.}
\affiliation{Department of Physics \& Astronomy, University of North Carolina at Chapel Hill, Chapel Hill, NC 27599, USA}
\email{kirbyben@unc.edu}

\begin{abstract}
The Hertzsprung--Russell diagram is one of the most important conceptual and analytical tools in stellar astrophysics. In its modern, empirical form as the \gaia\ color--magnitude diagram (CMD), each star is represented with a point determined from its time-averaged brightness and color. However, many stars are intrinsically variable, and the fixed-point representation ignores the variety of ways in which a star can vary. As a system pulsates, eclipses, is irradiated, erupts, or varies in other ways, it \textit{moves} through the CMD along a trajectory whose size, slope, and shape are determined by the physics driving the brightness and color variability. Here we exploit the near-simultaneous, three-band, multi-epoch photometry (MEP) released for $\sim$11.7~million sources in \gaia\ Data Release~3 (DR3) to construct a \dcmd\ (VECMD). In this updated take on the traditional CMD, each star is represented not by a point but by a linear segment approximating the path it traces in absolute magnitude and color as it varies. Because these trajectories are built from the raw epoch data, they are agnostic to period and do not require phase-folding. We show that radial-mode pulsators trace looped, Lissajous-like trajectories that can encode the phase lag between radius and temperature. The Hertzsprung bump is even visible in the CMD loops of some Cepheids. Variable binary systems can trace tight color-correlated lines and/or hooked paths. Accreting and dust-obscured systems trace long, high-amplitude tracks aligned with the outburst or reddening vector. Distinct regions of the VECMD are dominated by distinct trajectory morphologies, immediately drawing the eye to the variable sky. We argue that these trajectories are period-independent fingerprints well suited to machine learning classification, which we will pursue on the much larger epoch-photometry sample expected in \gaia\ DR4. Finally, we present a free, downloadable app called \texttt{IVECMD} for finding, classifying, and analyzing variable stars in the VECMD.
\end{abstract}


\section{Introduction}
\label{sec:intro}

More than a century after the work of \citet{Hertzsprung1911} and \citet{Russell1914}, the H-R diagram remains the central framework of stellar astrophysics. Its empirical form, the color--magnitude diagram (CMD), encodes the physics of stellar structure and evolution, the initial mass function, the ages of stellar populations, and more. Ever since the observational CMD was first developed, its resolution and completeness have been limited by the difficulty of obtaining precise, homogeneous photometry, and, more importantly, accurate distances through astrometry. The \textit{Hipparcos} mission delivered the first precise all-sky parallaxes and, with them, the first genuinely calibrated empirical CMD \citep{Perryman1997}. Since then, the many wide-field photometric surveys that followed have transformed stellar astronomy into a panchromatic, time-domain enterprise. Ground-based imaging surveys like the Sloan Digital Sky Survey \citep{York2000}, OGLE \citep{Udalski2015}, Catalina Sky Survey \citep{Drake2009}, ASAS-SN \citep{Shappee2014}, Pan-STARRS \citep{Chambers2016}, and the Zwicky Transient Facility \citep{Bellm2019}, together with high-precision space-based light curves from \textit{Kepler} \citep{Borucki2010} and TESS \citep{Ricker2015}, cataloged millions of variable stars and mapped the sky's variability in unprecedented detail. Regarding the empirical CMD, the culmination of these efforts has been the European Space Agency's \gaia\ mission \citep{Gaia2016}, which has delivered astrometry and three-band photometry for nearly two billion sources. The \gaia\ CMD, built from parallaxes and mean \g, \gbp, and \grp\ photometry, is the most precise and complete empirical CMD ever created \citep{GaiaHRD2018,Gaia2023}. It has helped researchers resolve and explore fine-tuned features like the Jao gap in the lower main sequence \citep{Jao2018}, bifurcation of the white dwarf cooling track \citep{GaiaHRD2018, Kilic2018}, the dual halo main sequences produced by the Gaia-Enceladus/Sausage merger \citep{Haywood2018,Helmi2018}, and more.

As powerful as it certainly is, the standard \gaia\ CMD encodes each of its $\sim10^{9}$ stars as a single, static point, located at the star's time-averaged magnitude $\langle \g \rangle$ and time-averaged color $\langle \bprp \rangle$. This representation is exact only for a star whose brightness is constant over the lifetime of the \gaia\ mission. However, essentially all stars vary in brightness, to some degree. \gaia\ DR3 alone classified more than ten million variable sources into two dozen distinct types \citep{Rimoldini2023,Eyer2023}. The true variable fraction is even higher once low-amplitude and non-periodic variables are included. When a star is reduced to its mean point in the CMD, the details of its variability are averaged away. Two stars that occupy the same mean CMD location could have vastly different physical natures.

Recognizing the value of \gaia\ multi-epoch photometry (MEP), several groups have used the data to identify and classify variables. The \gaia\ DR3 variability pipeline trained supervised classifiers on around thirty hand-engineered features (time-series statistics, periodicity indicators, and summary color information) calibrated against a cross-match of more than a hundred literature catalogs \citep{Rimoldini2023}. More recent efforts have applied unsupervised and representation-learning methods to multi-modal \gaia\ data. Some machine-learning studies targeting specific stellar populations, such as the hot subluminous variables analyzed by \citet{Ranaivomanana2025}, have shown that clustering in feature space can separate physically distinct classes. These approaches are highly effective, but they share a common feature. Each star is compressed into a vector of scalar attributes or into a learned latent code, in which the geometry of how the star moves through the CMD is implicit at best. The exact path a star traces in the CMD, perhaps the most physically transparent fingerprint, is never explicitly represented.

In this paper, we promote that fingerprint to an object of study in its own right. Using the near-simultaneous three-band epoch photometry in \gaia\ DR3, we replace each star's point in the CMD with a representation of the approximate \textit{trajectory} it traces as it varies, producing what we refer to as the ``\dcmd '' (VECMD). Because trajectories are built directly from the epoch measurements without using their time stamps, they are agnostic to period and phase and can be constructed without a periodogram, without phase-folding, and even for aperiodic variables. The shape of the trajectory is determined by the physics of the variability, and we show that it cleanly separates the major variability classes across the entire H-R diagram. 
We note that \citet{Gaia2019MECMD} already presented a ``time-dependent CMD'' in their Figure 11 that included individual \gaia\ MEP measurements for selected variable stars. Each star's MEP points were shown in a connect-the-dots fashion, and it was clear in their examples that CMD motions can be detected using \gaia\ DR3 MEP. However, their connect-the-dots presentation style, along with outliers, the limited number of stars represented, and noise, makes it difficult to see the overarching patterns that are present across the entire time-resolved CMD without simplifying the presentation style of each star's MEP. As such, we provide a simple alternative presentation style that, while oversimplifying the details present in each star's MEP, allows one to efficiently explore the entire CMD for variability. We note that \citet{Spano2009} briefly explored a similar idea using variable star morphologies in the CMD, but their work was limited to select stars in the Large Magellanic Cloud observed by OGLE-II/III.

Here we introduce a simple variability-encoded CMD constructed using data from \gaia\ DR3 and review the CMD trajectory morphologies of the principal variable classes. We also present an interactive tool called \texttt{IVECMD} that can be used to explore the diagram and analyze new and known variables. We stress that this first paper is intended as a methods-and-atlas paper. Our goal is to establish the VECMD representation, present the diagram and tool that makes it easily explorable, and show {\em qualitatively} how CMD trajectory shapes encode the physics of the principal variable star classes. We reserve for subsequent papers the detailed statistical and quantitative analyses that the diagram invites, especially once \gaia\ DR4 photometry is available.  In Section \ref{sec:mep} we review the basic properties of \gaia\ DR3 MEP that make CMD trajectories possible. In Section \ref{sec:mecmd} we describe the construction of the VECMD, present the diagram for stars with available MEP from DR3, and introduce a simple app for exploring it. In Section \ref{sec:analysis} we briefly examine the trajectories of pulsating stars, binaries, and aperiodic systems and connect their shapes to the underlying physics. In Section \ref{sec:future} we outline the machine-learning and stellar-modeling directions this representation enables, and in Section \ref{sec:summary} we summarize. 

\begin{figure*}[ht!]
\centering
\includegraphics[width=0.99\textwidth]{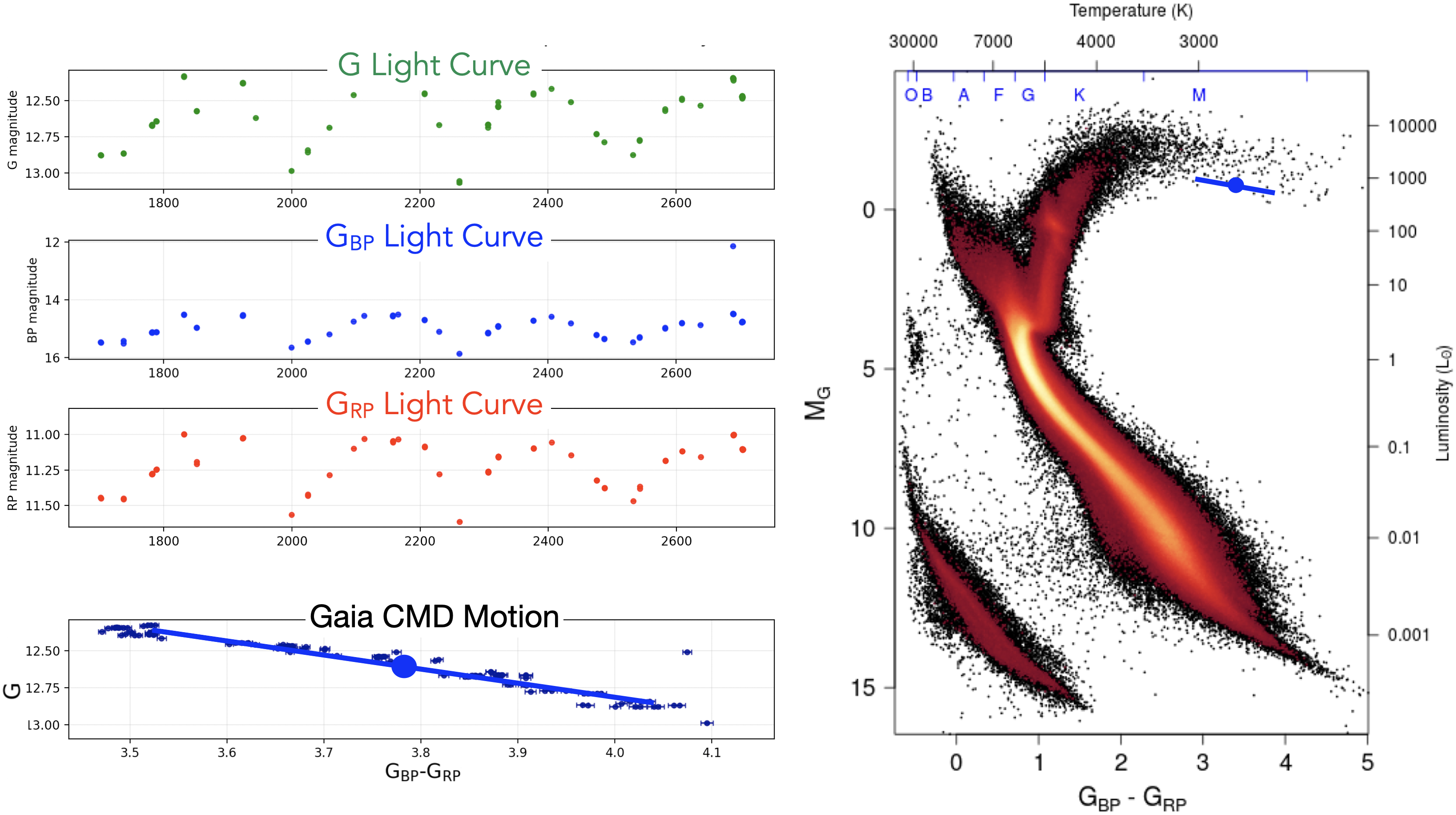}
\caption{Raw ingredients for building a variability-encoded CMD. {\em Left Panel:} The full \g, \gbp, and \grp\ light curves from \gaia\ DR3 for an example long-period variable (LPV). By combining together the \g\ magnitudes with their contemporaneous \bprp\ colors, one can plot the back-and-forth motion of the LPV through the CMD (bottom-left panel) as it pulsates. The LPV shown becomes bluer as it brightens and redder as it fades. The blue line segment is a simple linear fit to its CMD points. {\em Right Panel:} The traditional \gaia\ CMD from DR3, taken from \citet{Gaia2023}, with modifications. The example long-period variable is highlighted with a blue point, and its approximate movement through the CMD as it pulsates is shown as a blue line segment (the same as in the left panel). }
\label{fig:mep_example}
\end{figure*}

\section{Gaia DR3 Multi-Epoch Photometry}
\label{sec:mep}

The European Space Agency's \gaia\ spacecraft measured the precise positions, distances, motions, and properties of nearly two billion stars to create the most detailed three-dimensional map of the Milky Way ever made \citep{Gaia2016}. From its launch in 2013 until science operations ceased in early 2025, \gaia\ scanned the sky continuously, and at each field-of-view transit, a source was measured in three photometric bands, including the broad, white-light \g\ band (330-1050 nm), the blue \gbp\ band (330-680 nm), and the red band \grp\ (640-1050 nm). At each epoch, data in all three photometric bands were obtained within seconds of one another, as the source crossed the focal plane \citep{Riello2021}. In DR3, the \gaia\ Collaboration released the full multi-color MEP for 11.7 million objects \citep{Eyer2023}. These sources were not a random subset of the full \gaia\ catalog, but only those systems flagged as candidate variables by the DR3 variability processing pipeline, supplemented by calibration and special-purpose samples like the \gaia\ Andromeda Photometric Survey. The DR3 MEP spans the 34-month DR3 observing period from July 2014 to May 2017, giving a baseline of $\sim$ 2.8 years. Each source has approximately 40-50 measurements per band \citep{Eyer2023} over this time period, with the \g\ band typically being the best sampled and most precise.

Because the number of transits at a given position was set by the scanning law, the photometric sampling for any given star is highly non-uniform. The average time between consecutive epochs is tens of days, but this spacing is highly irregular. Normally, such sparse and irregular sampling would be a liability to periodogram analyses of time-resolved photometry. However, two features of \gaia\ MEP make the situation more promising than one would expect. First, the per-epoch photometric precision is extraordinary. For the bright, well-measured sources that dominate the MEP sample, individual-epoch uncertainties go down to the milli-magnitude level, such that even low-amplitude variability ($\sim$ppt) can be resolved within a single epoch. Second, the quasi-random cadence resulting from the scanning law helps break the regular aliasing patterns that plague evenly sampled ground-based data \citep{Eyer1999,VaderPlas2018}. Combined with the long baseline, this allows periodogram analyses to recover true periods for a substantial fraction of periodic sources despite the modest number of points and non-uniform cadence (see Section \ref{sec:analysis}). The bright and faint limits of the usable MEP are respectively set by saturation and calibration near $\g \approx$ 3, and by photon noise near $\g$ $\approx$ 20-21, so that essentially the entire stellar CMD is accessible, from the most luminous supergiants to the faintest resolved white dwarfs and cool dwarfs.

Figure \ref{fig:mep_example} illustrates the raw \gaia\ MEP for an example long-period variable (LPV) star, Gaia DR3 2902864392736607232. The right panel shows the source's average position in the \gaia\ CMD as a blue point. The top three panels on the left show the full \g, \grp, and \gbp\ epoch DR3 light curves, from top to bottom. Although the three color measurements obtained at a given epoch are not recorded at the same instant, the time offset between them is negligible compared to the majority of stellar variability timescales of interest. At each target epoch, the mean time between \gbp\ and \grp\ measurements is just a few seconds. This provides a nearly instantaneous measurement of a target's \bprp\ color at each epoch. The time between a target's \bprp\ color and corresponding \g\ measurement is a few tens of seconds, with the maximum offset being well under a minute. Each matched {\bf \g $+$\bprp} epoch therefore yields a genuine, contemporaneous $(\bprp,\;\g)$ pair. Thus, for each star with available \gaia\ DR3 MEP, we can plot its position in the CMD at several dozen epochs, revealing any potential changes due to intrinsic variability. 
In the example case of Gaia DR3 2902864392736607232, its \bprp\ color changes by more than 1 mag, while its overall brightness changes by more than 2 mag (bottom left panel of Figure \ref{fig:mep_example}).  Even with this single object, the message of the paper is already visible: the color and brightness vary in concert with one another, and the star sweeps out an extended, structured path in the CMD rather than sitting still. 

\section{A Gaia Variability-Encoded CMD}
\label{sec:mecmd}
The single-star CMD trajectory shown in Figure \ref{fig:mep_example} is informative on its own, but its real power emerges when this same operation is applied to all stars in the CMD at once. Doing so requires a simplified representation compact enough that millions of trajectories can be overplotted without overwhelming or saturating the diagram. However, such a representation must at least preserve the direction and overall amplitude of each star's CMD motion.

\subsection{Construction of the Diagram}
For each source, we start with the \g, \gbp, and \grp\ epoch light curves and remove obvious outliers band-by-band with a 5-$\sigma$ clipping about each light curve's median flux. We then form matched, contemporaneous (\g, \bprp) epochs as described in Section \ref{sec:mep} and fit a straight line to the points in the color-magnitude plane. Since a small number of discrepant epochs can dominate a least-squares slope, we refine the fit iteratively. After fitting a line to the matched epochs, we remove the single most deviant point and refit. If the slope changes by more than 10\%, we treat that point as an outlier, leave it out, and repeat. Once removing the most deviant remaining point changes the slope by less than 10\%, we stop. 
We note that the above procedures can on occasion remove very deep mid-eclipse points from the slope determination, thereby underestimating the true slope for that object. In future work with DR4 MEP, we will investigate supplementing our $\sigma$-clipping with the \gaia\ \texttt{variability\_flag\_\{g,bp,rp\}\_reject} flags to better identify discrepant epochs.

We then replace each star's point in the CMD with a simple line segment representing its average motion over all \gaia\ epochs. The segment is centered on the star's average CMD position after both sigma-clipping algorithms. Its orientation is the fitted slope, while its length is set to the maximum difference of the color and magnitude epochs respectively after sigma clipping. The drawn length then reflects the approximate amplitude of the net CMD motion. As discussed previously, this linear representation is deliberately a simplification. We show in Section \ref{sec:analysis} that many stars trace complex, non-linear trajectories rather than straight lines. Some radial pulsators even show {\em looped} motions over each pulsation cycle. A single line segment cannot capture these features. However, it is quick to compute for millions of stars, captures the essence of a system's variability, and most importantly, it produces a legible diagram that is quick and easy to explore.

\begin{figure*}[ht!]
\centering
\includegraphics[width=0.44\textwidth]{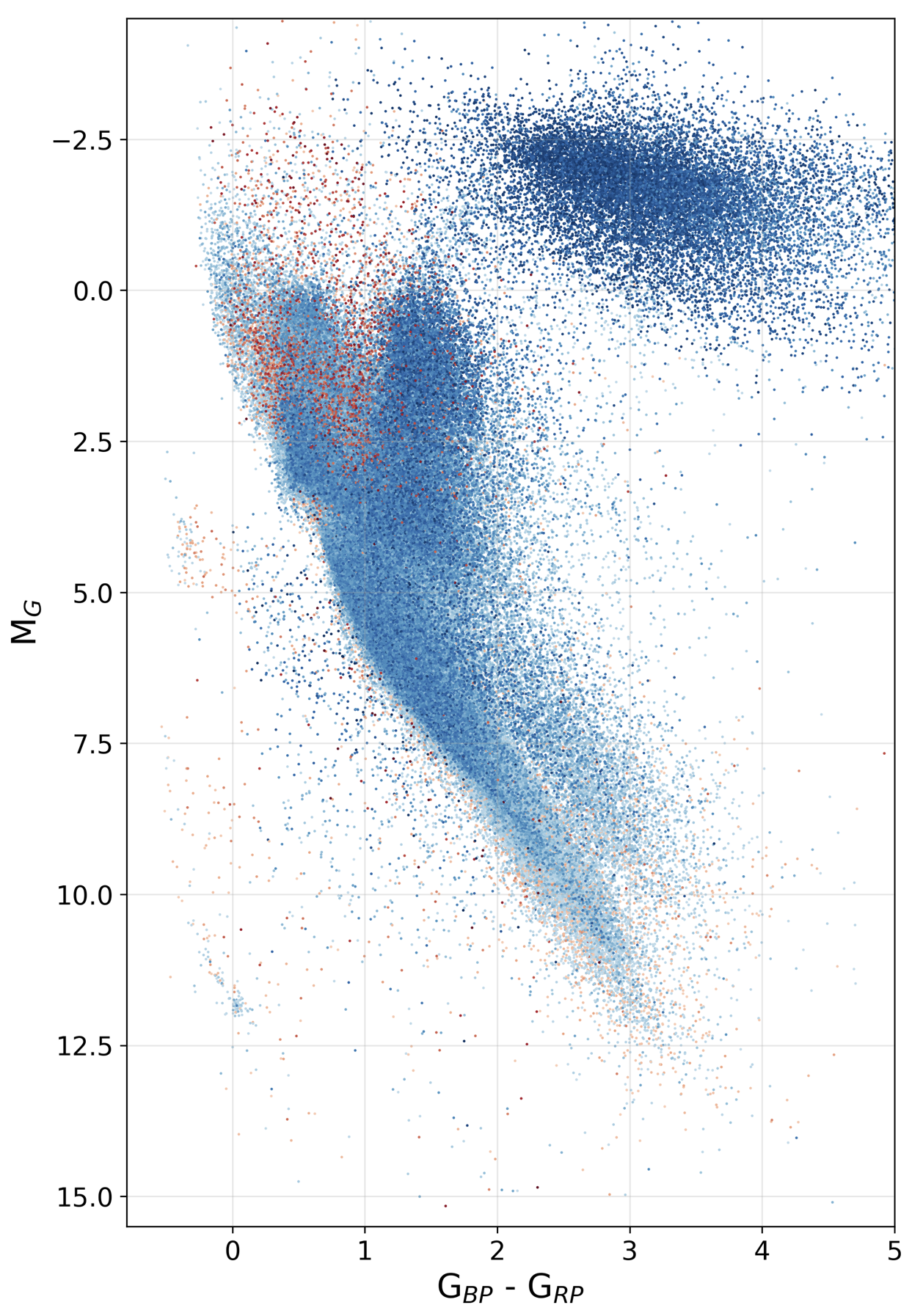} \includegraphics[width=0.53\textwidth]{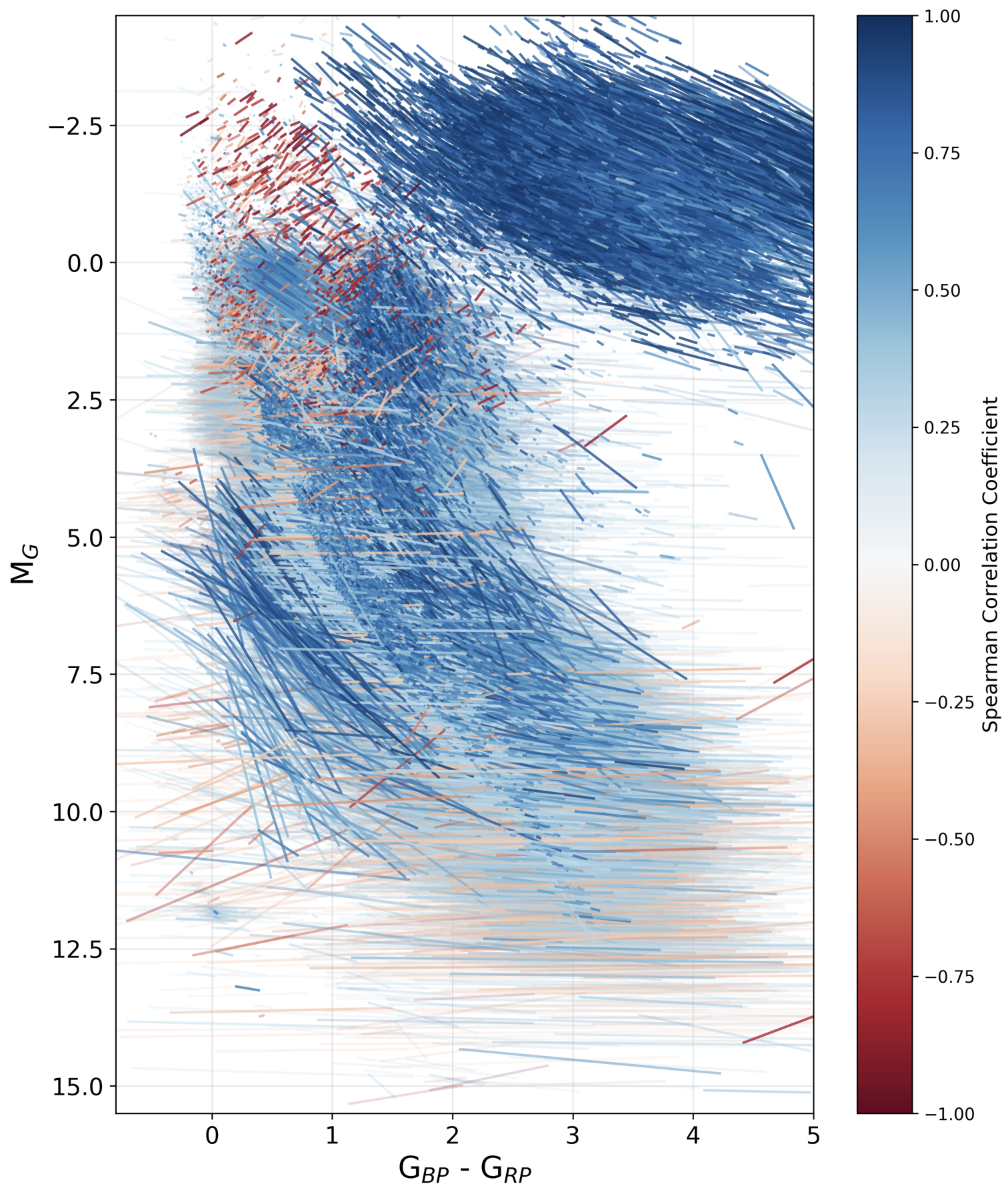}
\caption{A side-by-side comparison of the traditional \gaia\ CMD and our variability-encoded \gaia\ CMD (VECMD). {\em Left Panel:} The traditional \gaia\ CMD is shown for stars with available multi-epoch photometry from DR3. Each star is represented by a static point determined from its average brightness and color. {\em Right Panel:} Our variability-encoded CMD represents the same stars as the left panel not as static points, but instead as line segments representative of the changes in color and brightness they show as they vary. Color indicates whether they are reddest (red) or bluest (blue) when they are brightest, with the hue determined by the Spearman correllation coefficient between \g\ and \bprp. The opacity and z-order of the lines is determined by the p-value of the Spearman coefficient, with more believable correlations producing lines with high opacity closer to the top of the plot. Although line segment representations oversimplify the more complex motions stars exhibit as they move through the CMD, this representation allows one to quickly pick out and even classify different types of variables, including pulsators, binaries, rotators, and aperiodic variables. We provide an interactive tool, \texttt{IVECMD}, for users to explore this diagram in greater detail.}
\label{fig:vecmd}
\end{figure*}

In order to evaluate the sense and quality of each correlation, we compute for every star a variety of correlation coefficients between its \g\ and \bprp\ points. We find the Pearson and Spearman coefficients to be sufficient for this work, continuing the theme of simplicity and efficiency. We can use the results of these analyses to encode additional information into our CMD line segments. First, we use the Spearman correlation coefficient to set the hue of each line segment using a standard red-white-blue diverging color map (e.g., matplotlib's RdBu). We color a segment blue if the star is bluest at maximum light and red if the star is reddest at maximum light. The saturation of the hue encodes the strength of the correlation, so that segments with strong positive or negative correlations appear deep blue or deep red, while those with weak correlations fade toward white. Second, we use opacity and plotting depth to encode the statistical significance of the correlation. We set each line segment's transparency from the logarithm of its Spearman p-value, so stars with more signifcnat correlations are drawn more opaque and above those with weaker correlations. With millions of sources in the frame, this ordering is essential to letting the coherent variables stand out.  

\subsection{The Diagram}

Figure \ref{fig:vecmd} illustrates the VECMD that results from this work. To keep the diagram from being visually overwhelming, we restrict it to sources with reliable distances, excluding those with negative or null parallaxes, parallax precision worse than 10\%, or distances greater than 10 kpc. Additionally, we only plot points if they have at least six epochs of photometry. Of the $\sim$11.7 million sources with DR3 MEP, this leaves approximately 3 million objects shown in Figure 2. The left panel is the familiar \gaia\ CMD, in which each star is shown with a single point sitting at its time-averaged magnitude and color. The right panel shows our \vecmd\ for the same objects, with each star represented using the colored line segment described above. The two plots contain the same exact stars, but they are not the same diagram. In the traditional CMD, the eye is drawn to the main sequence. However, in the \dcmd, the main sequence no longer dominates. Despite it hosting more stars than any other region of the plot, the main sequence is, for the most part, photometrically quieter than other parts of the CMD on the timescale of \gaia\ DR3, and its line segements are mostly short and faint. What stands out more than the main sequence in the VECMD are the regions of stronger variability.

Several groups of variables dominate the \dcmd\ even at a quick glance. In the upper right-hand  (coord. 3.5, -2), the long-period variables (LPVs) and Miras race out some of the longest segments in the entire diagram. Their lines are deeply saturated and predominantly blue, revealing the high-amplitude, strongly color-correlated brightness changes of these cool, pulsating giant stars. Some LPVs in the DR3 sample change their \bprp\ colors by more than 3 mag over their pulsation cycles. Thus, plotting the position of an LPV in the CMD using only a single photometric epoch taken at random could misrepresent the star's true, mean position by a significant amount. Just to the left of the LPVs, the classical Cepheids (1.5, -3) form another prominent blue locus of points, with shorter line segments but steeper slopes. Even if these famous variables were not already known, they could be immediately distinguished from the redder LPVs from their steeper slopes and shorter line segements, characteristics revealing their must be different physics governing their behavior. Descending the classical instability strip, the RR Lyrae stars (0.5, 0.5) appear as a tight cluster of slightly lighter blue lines near the horizontal branch, and the \dscuti\ stars (0.75, 2.5) can be spotted below them as an even tighter locus of blue lines near the upper main sequence. Both sets of line segments have their own characteristic orientations. Elsewhere in the diagram, cataclysmic variables (0.5, 6.0) and other accreting and outbursting systems trace out long, high-correlation tracks. Hot subdwarf stars (-0.25, 4.0), many of which show variability due to binarity, display several light red segements and sit just to the upper left of the CVs segements, between the main sequence and white dwarf cooling sequence. Collectively, these examples show that the \dcmd\ separates variables not only by where they sit in the CMD, but also by how they move. 

Because a static figure cannot convey the full, curved trajectories that underlie each star's line segement, we have built an interactive web application that renders the \dcmd\ and lets users click any segement to fully inspect that star's \gaia\ raw MEP, periodogram, phase-folded light curve, and CMD trajectory. We describe that tool in Appendix \ref{sec:ITCMD} and encourage others to use it to find and classify new variable stars with ease and with fun. We have used the tool ourselves to find several new post-common envelope hot subdwarf binaries, which we discuss in Paper II of this series \citep{Kirby2026}.

\subsection{IVECMD: A Simple Tool to Explore VECMD}
Because a static figure can only hint at the full, detailed CMD trajectories that underly each line segment, we have developed \texttt{IVECMD}, a free, interactive application for exploring our \gaia\ DR3 \vecmd\ and the raw MEP behind it. The app renders the full VECMD and lets users pan and zoom around the diagram. They can click on any line segment to pull up that star's raw \gaia\ MEP, full and sigma-clipped light curves, Lomb Scargle periodograms, and phase-folded light curves. The diagram can be filtered according to \g\ vs. \bprp\ correlation significance for each star, parallax precision, distance, and Galactic latitude. Users can supply custom lists of objects to highlight in the diagram or choose from pre-determined lists of example variables. Currently, these lists contain only radial mode pulsators with the plan to add more lists in the future. The tool is designed to make discovering, classifying, and inspecting variable stars in \gaia\ DR3 MEP fast and intuitive for both researchers and students. We have already used it ourselves to find new compact binaries \citep{Kirby2026}. Many plots in this manuscript were generated directly with \texttt{IVECMD}. Appendix \ref{sec:ITCMD} describes more of its features in detail. 
Future versions of \texttt{IVECMD} will eventually replace the line segments with more complex representations of the CMD trajectory shapes (see Section \ref{sec:analysis}.

\section{A Detailed Look at CMD Trajectories}
\label{sec:analysis}

The line segments shown in the VECMD in Figure \ref{fig:vecmd} simplify each star's CMD motion to just a slope and a length. However, this is not the full story for most variable stars, whose full CMD trajectories carry much richer physical information. In this section, we briefly explore the full CMD trajectories of example pulsators, binaries, and aperiodic variables and connect the geometry of their trajectories to the physical mechanisms driving their variability.

\begin{figure*}[ht!]
\centering
\includegraphics[width=0.515\columnwidth]{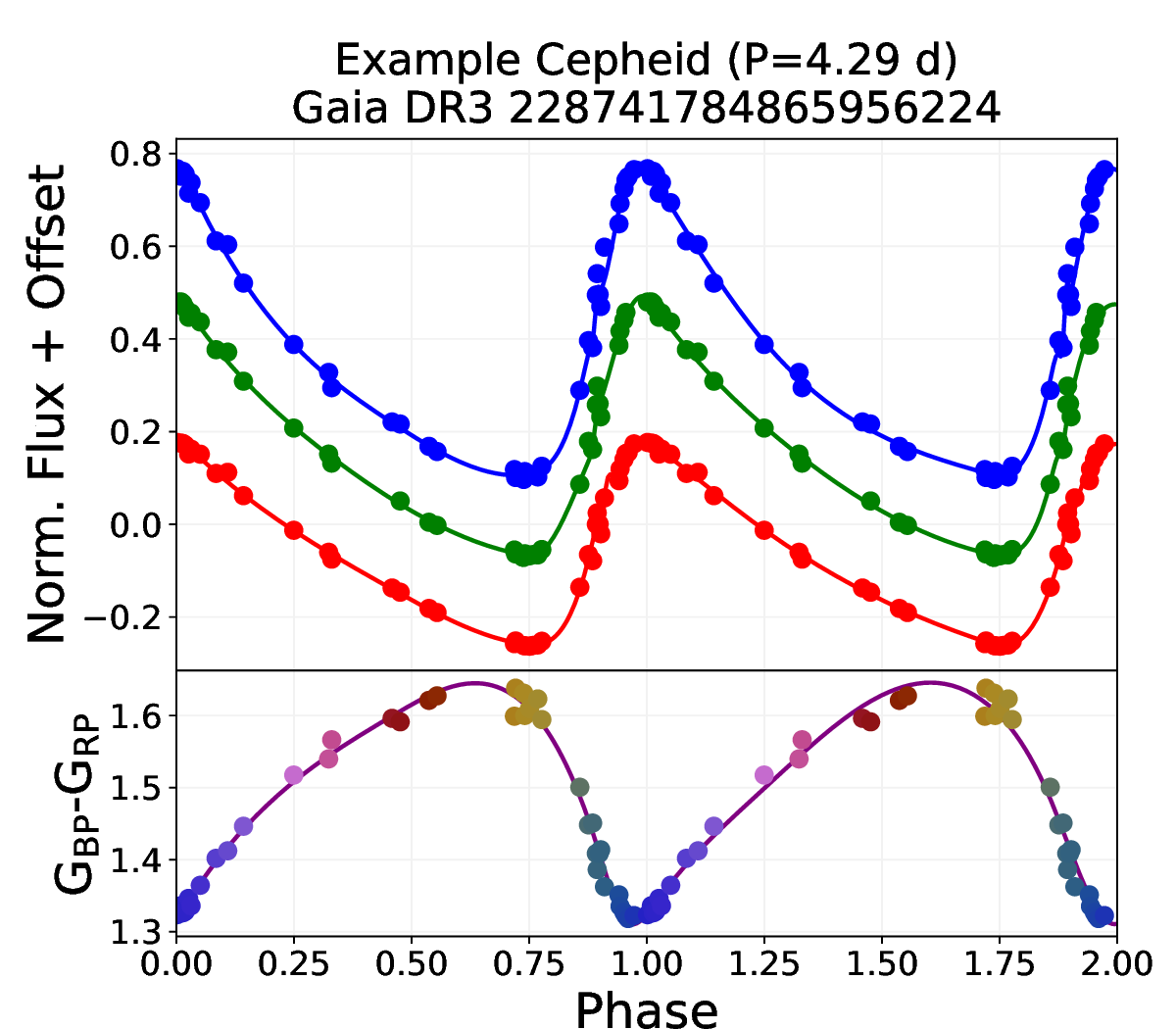} 
\includegraphics[width=0.515\columnwidth]{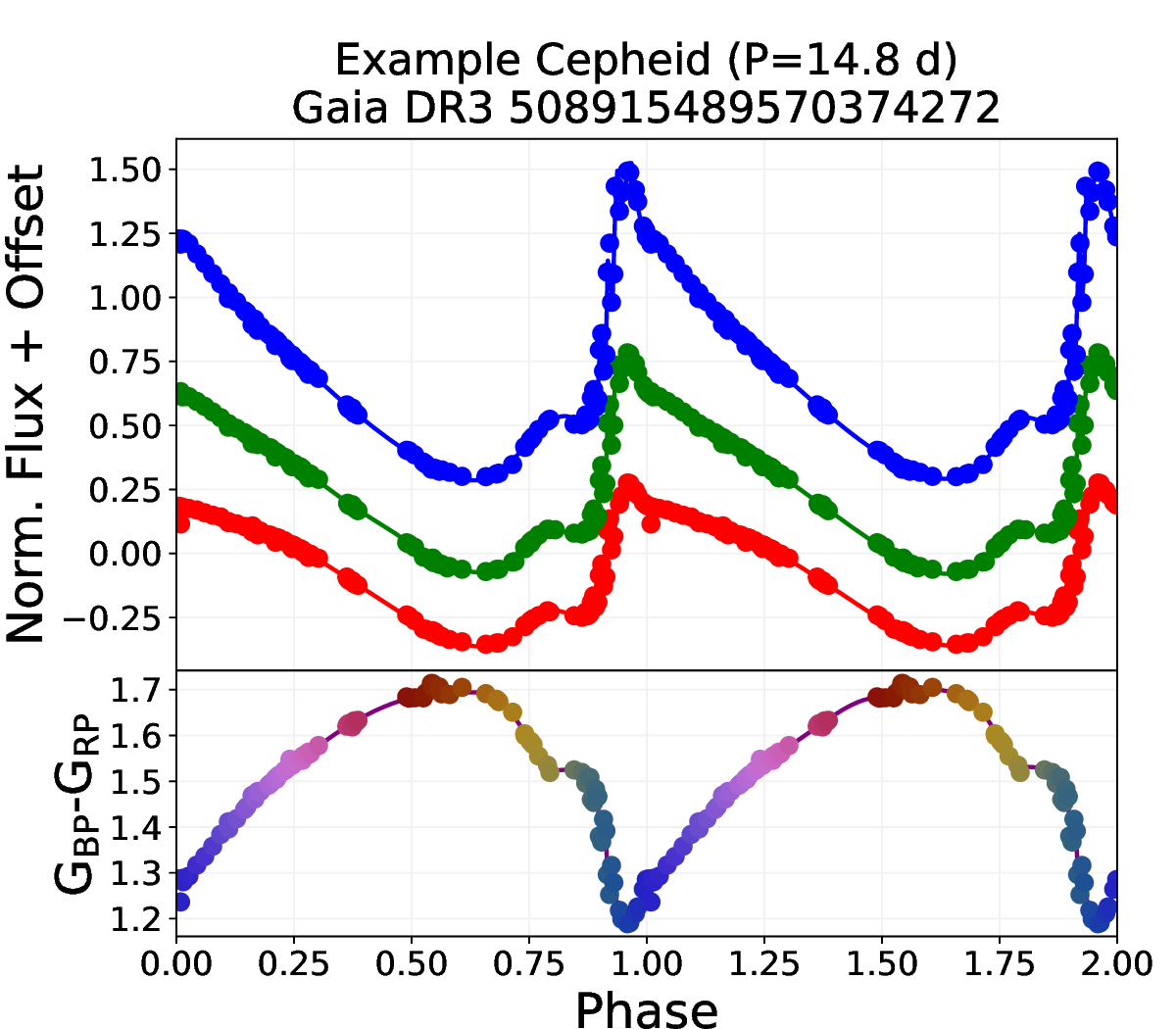} 
\includegraphics[width=0.515\columnwidth]{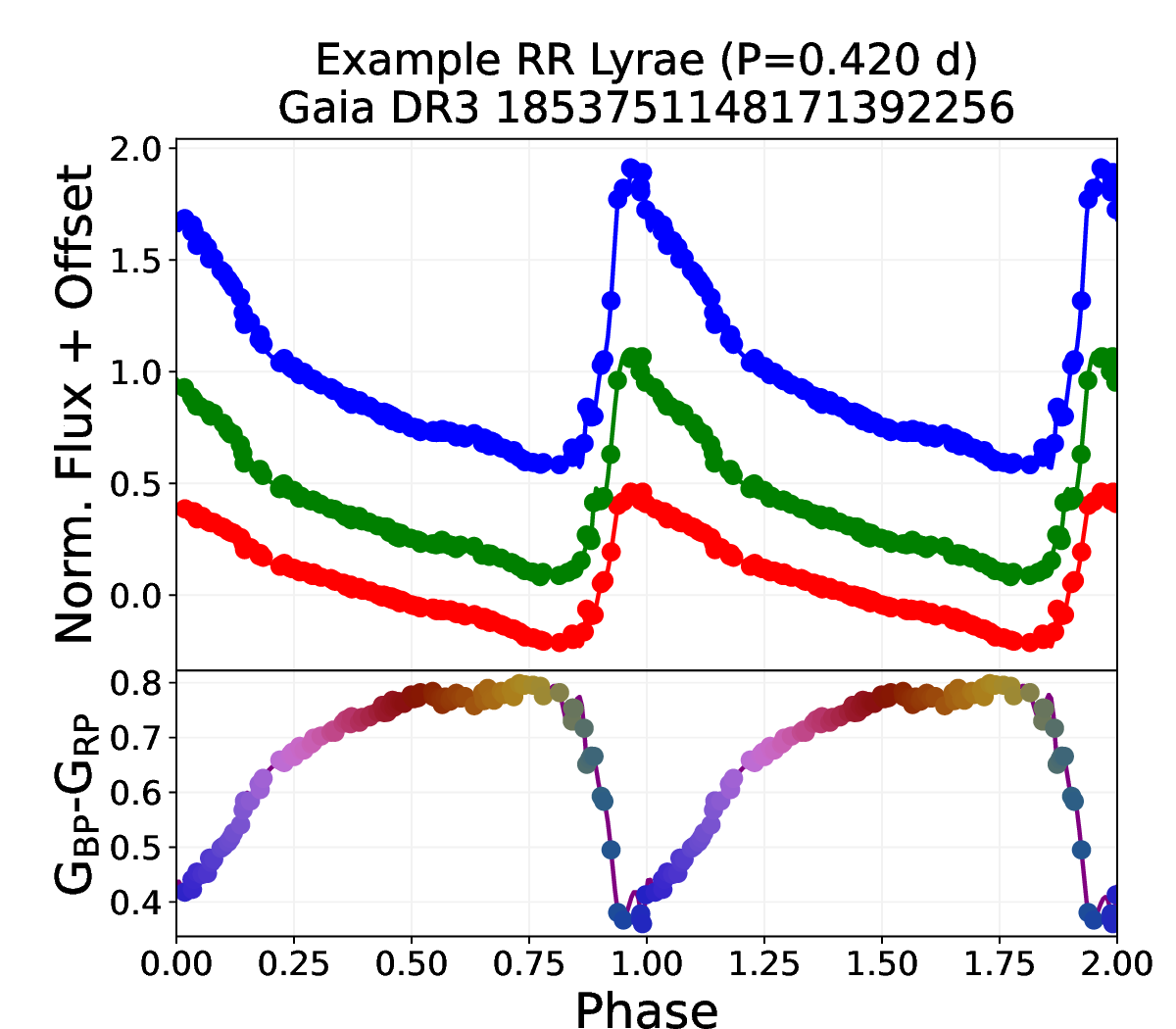} 
\includegraphics[width=0.515\columnwidth]{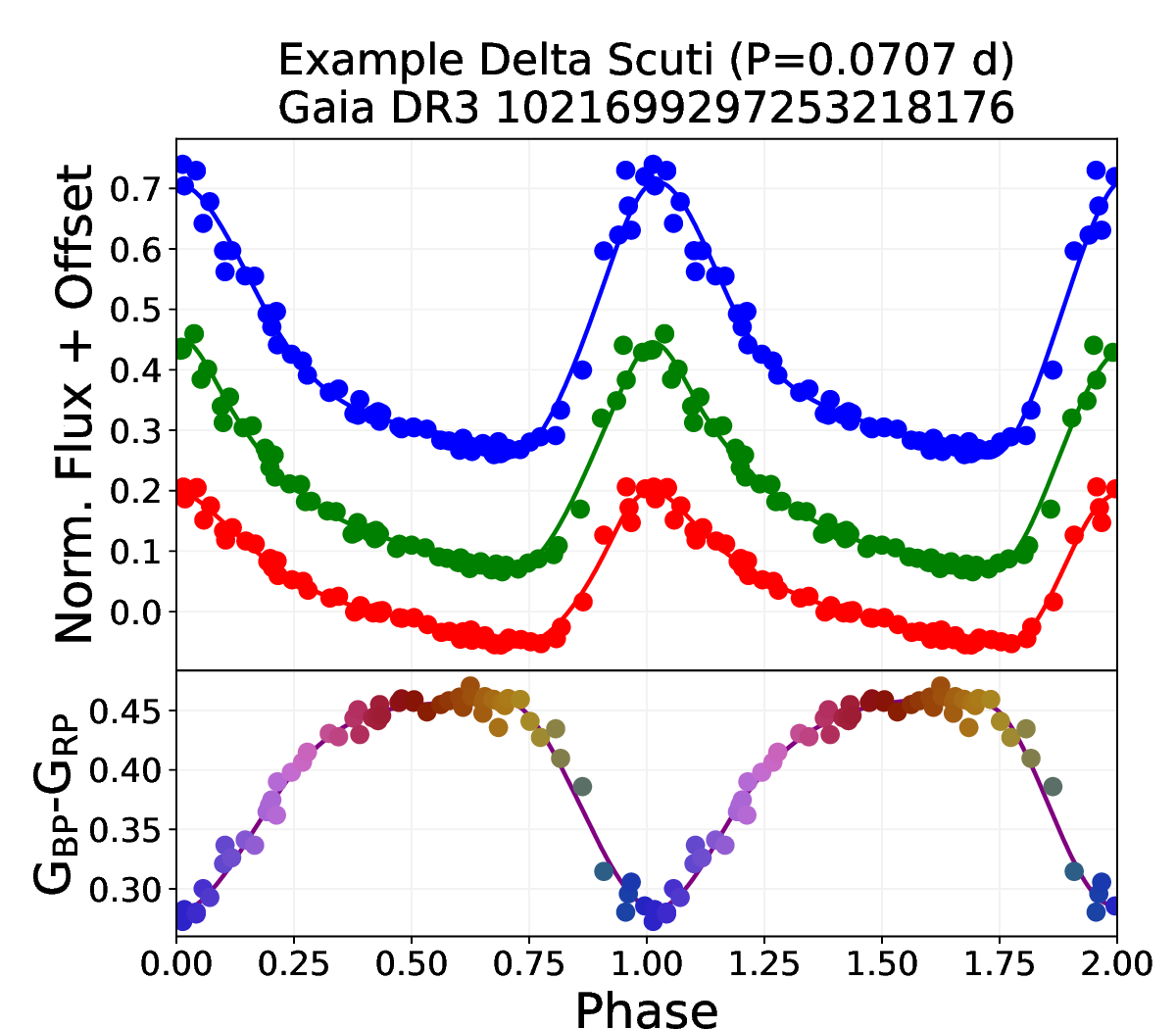} \\

\includegraphics[width=0.515\columnwidth]{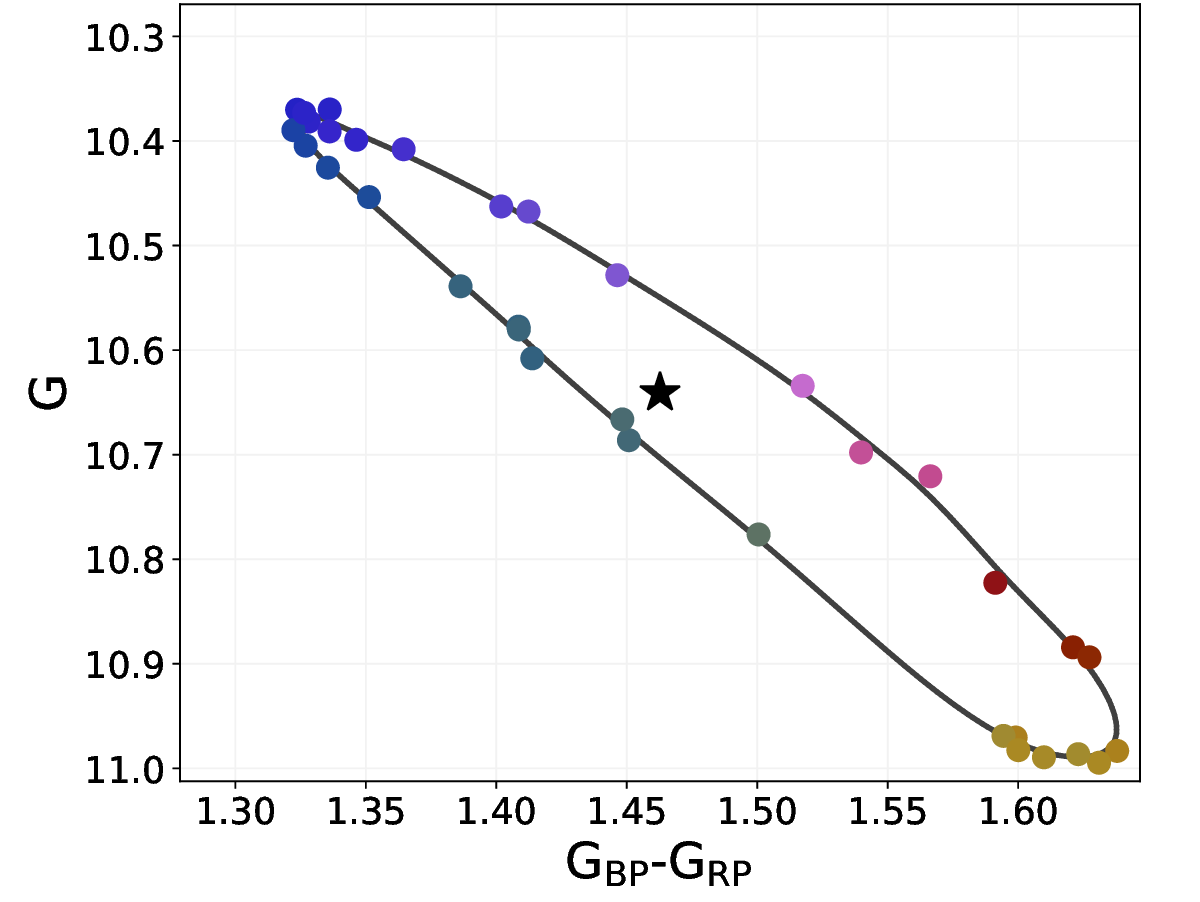} 
\includegraphics[width=0.515\columnwidth]{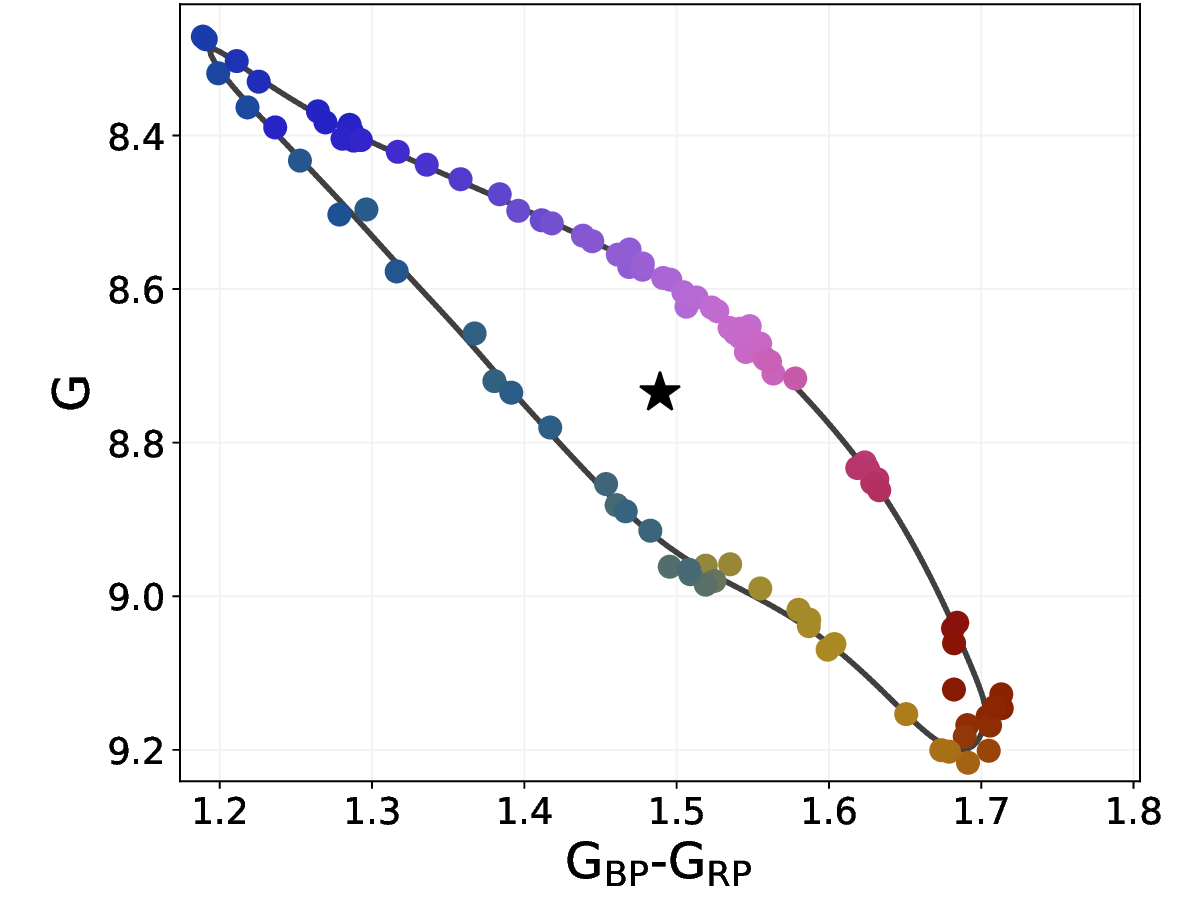}
\includegraphics[width=0.515\columnwidth]{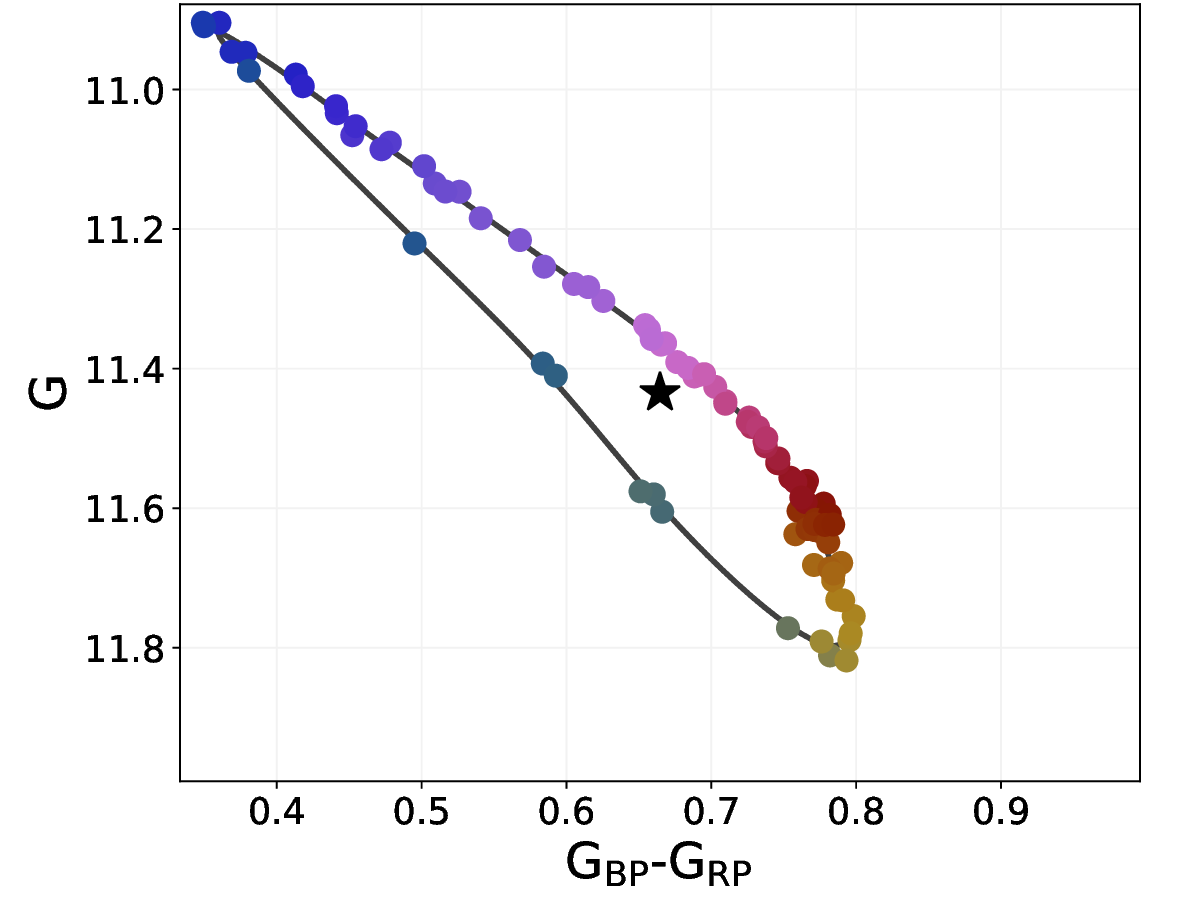}
\includegraphics[width=0.515\columnwidth]{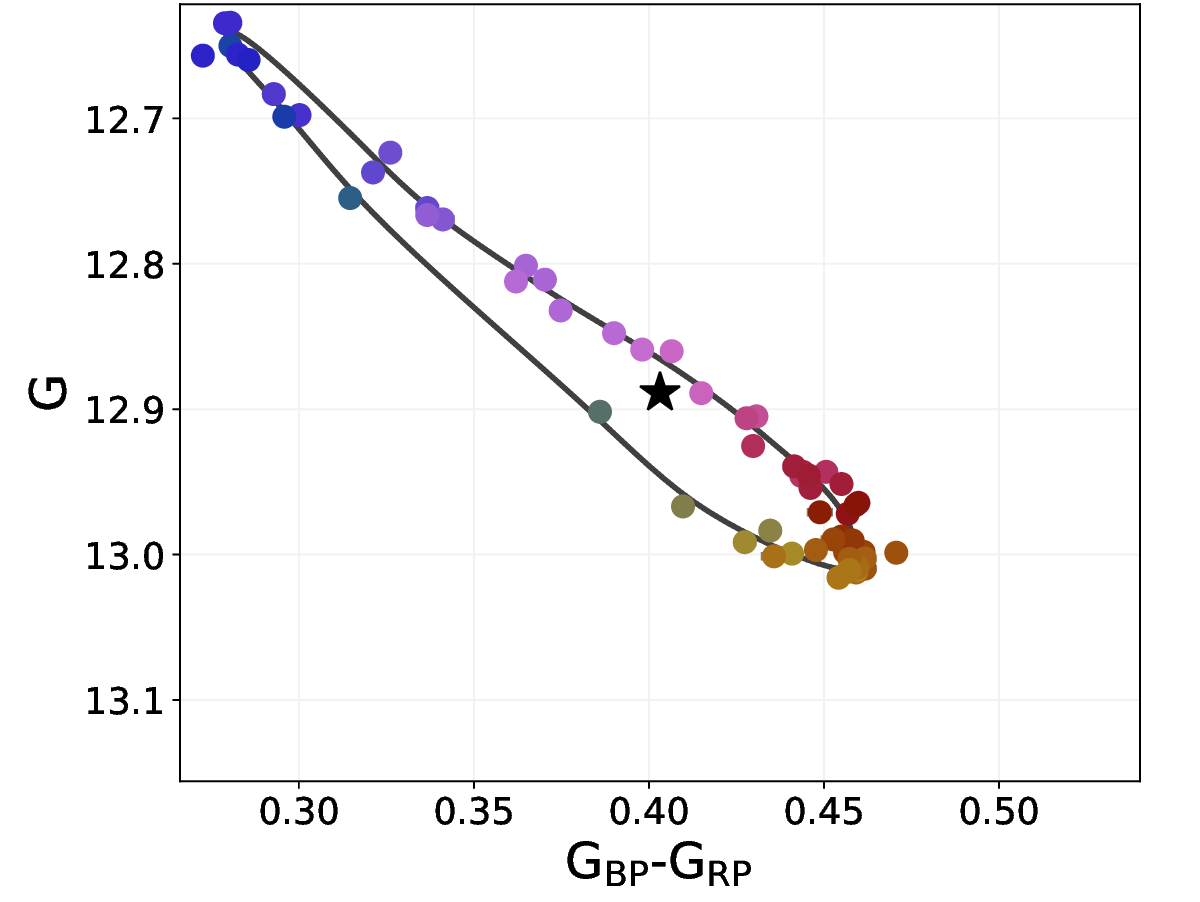}

\caption{\gaia\ photometry and CMD trajectories for four example radial-mode pulsators, including two Cepheids, one RR Lyrae pulsator, and one $\delta$ Scuti pulsator. {\em Top panel:} Phase-folded \gbp\ (blue points), \g\ (green points), and \grp\ (red points) light curves, with simple spline fits to guide the eye. The light curves shown were mean-normalized, and vertical offsets were applied to the \g\ and \gbp\ curves to avoid overlap. The corresponding \bprp\ color curves are shown below with points color-coded according to pulsation phase, for comparison to the CMD points shown below. {\em Bottom panel:} The corresponding CMD points, illustrating how the star moves through a loop in the CMD as it pulsates. Points are color-coded according to pulsation phase, in the same way as the \bprp\ curve  above. For example, a dark blue color (indicating phase = 0.0/1.0) corresponding to the phase of peak flux and bluest color for these objects, while yellow/orange correspond to the phase of reddest color and minimum flux (which occurs shortly before peak flux). The black star symbol represents the average CMD position. The black line shows the best-fitting ElPigraph trajectory fit and is meant to guide the eye through the CMD trajectory shape unique to that type of pulsator.}
\label{fig:pulsators}
\end{figure*}

We would like to reemphasize that the CMD trajectory is constructed {\em without} direct reference to time. The epochs are plotted in the color-magnitude plane, not against a clock. As a result, CMD trajectories can be constructed when the period is unknown or undefined. This is advantageous for observatories like \gaia\, for which the small number of epochs and aliasing pattern may prevent a periodogram from picking out the correct frequency for periodic systems. For periodic stars whose periods {\em can} be determined, we compute Lomb-Scargle periodograms band-by-band over a frequency range spanning from the \gaia\ DR3 baseline up ($\sim$2.8 years) to the shortest known stellar pulsation periods ($\sim$ 1 min). The long \gaia\ baseline results in an effective frequency resolution of $\delta f \approx 1/\Delta T \approx 10^{-3}\, d^{-1}$. The irregular sampling suppresses aliasing well enough that true periods are recovered for a large fraction of periodic sources despite the sparse sampling. In some cases, it is possible to recover periods even down to minutes (see Section \ref{sec:pulsators}). Where a period is found, phase-folding the \gaia\ DR3 light curves provides an independent check on the trajectory's interpretation. Additionally, we can use the ElPiGraph Python package to fit elastic principal curves or closed elastic principal curves (loops) to the \g\ vs. \bprp\ data to better visualize the CMD trajectories that are present \citep{Albergante2020}. Some example fits are shown in the subsections that follow. While we don't do anything {\em quantitative} with these ElPiGraph fits in this manuscript, we note that future machine learning studies will benefit from the parameters of these fitted trajectories (see Section \ref{sec:future}).

\subsection{Pulsating Stars}
\label{sec:pulsators}

Any process that intrinsically changes a star's radius and/or temperature moves it in the luminosity-color plane, in a way governed by the Stefan-Boltzmann law, $L \propto R^2 T^4_{\rm eff}$. The exact path it follows is determined by the relative phase and amplitude of the radius and temperature changes, which is a consequence of the exact physics governing the pulsation. As such, different types of pulsating stars exhibit different shapes in their CMD trajectories.

Radially pulsating stars provide arguably the most physically rich trajectories in the VECMD. As a Cepheid or RR Lyrae star pulsates, for example, its radius and temperature vary simultaneously, but not exactly in phase with one another. The star is generally hottest and bluest not at the moment of smallest radius, but sometime thereafter, and the resulting brightness peak is offset from both. Because the expansion and contraction parts of the pulsation cycle follow different relations between color and magnitude, the star does not retrace its steps in the CMD during both halves of the pulsation cycle. Instead, it follows a closed loop in the CMD, sweeping down one side and back up the other with a Lissajous-like pattern.

Figure \ref{fig:pulsators} shows example CMD loop trajectories and phase-folded light curves for Cepheid, RR Lyrae, and \dscuti\ pulsators. Although the \dcmd\ loop trajectories are time-independent, we chose example pulsators with \gaia\ MEP periodograms clean enough to pick out their periods and phase-folded the light curves. This permits us to represent the pulsation phase in the CMD trajectory plots and demonstrate that the loop direction is {\em clockwise} for most of these fast-rise, slow-decay type pulsators. The corresponding line-segment representations used for these loop trajectories in our global VECMD are shown as gray lines, as are the fitted ElPiGraph trajectories. The exact shape of the loop, the area it encloses, and its tilt are set by the strengths of the radius and temperature variations, and by the phase lag between them. All are functions of the pulsation mode and period. We note that the ability of \gaia\ to detect such patterns was previously noted by \citet{Gaia2019MECMD}, who plotted them for several example systems.

\begin{figure}[ht!]
\centering
\includegraphics[width=0.99\columnwidth]{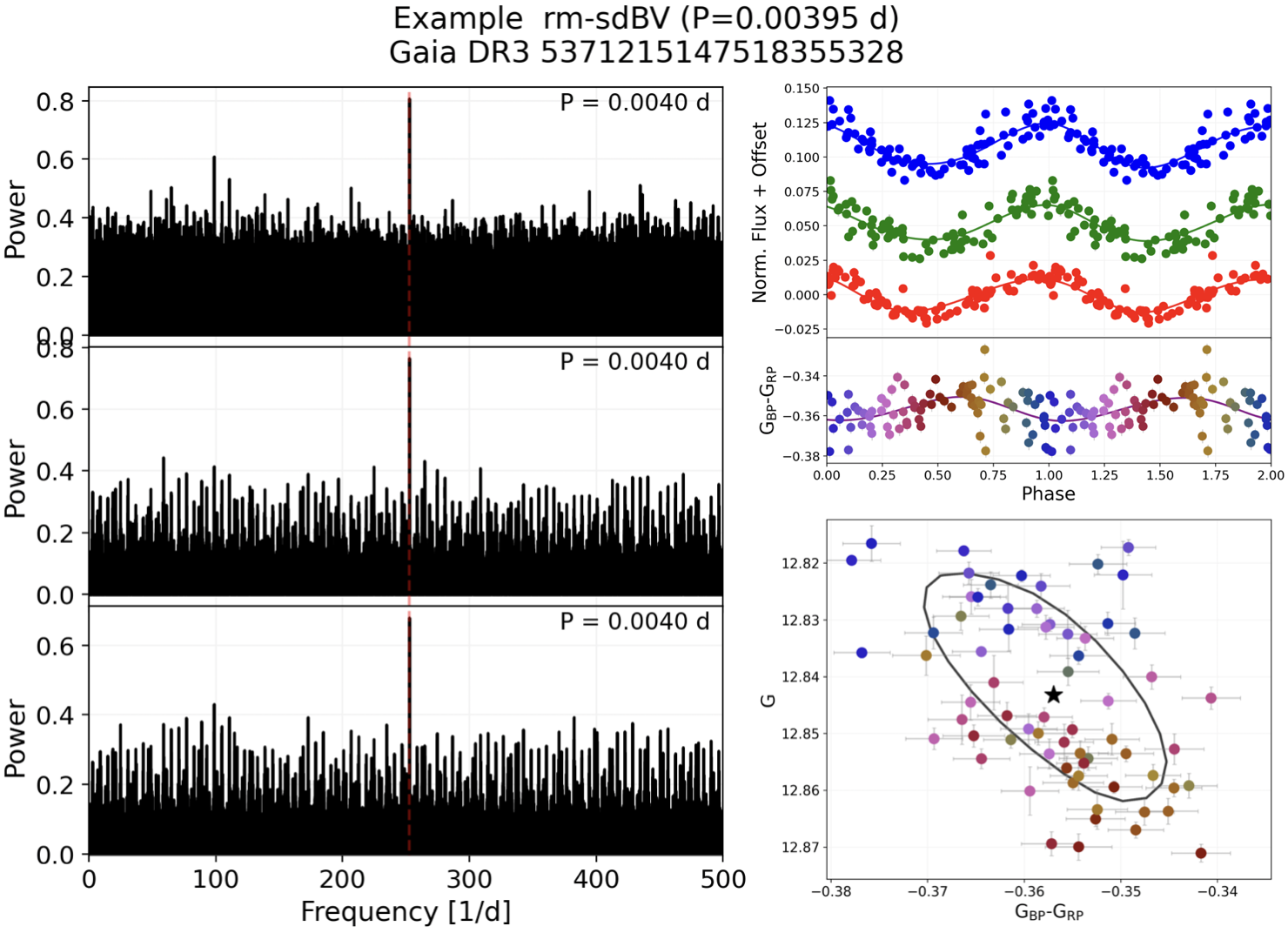}
\caption{\gaia\ photometry and CMD trajectory for an example rapid radial mode sdBV pulsator, BPM 36430. The left panels show the Lomb Scargle periodograms of the individual \gbp, \g\, and \grp\ light curves, from top to bottom. Surprisingly, these periodograms easily pick up on the weak ($\approx$1\%) and extremely short (5.7-minute) pulsation, despite the long average time between consecutive epochs, somewhat random observing cadence, and few data points in the \gaia\ MEP. The upper-right plot shows the phase-folded light and color curves, while the CMD trajectory below them shows BPM 36430 exhibits a looped motion similar to the radial pulsators shown in Figure \ref{fig:pulsators}.}
\label{fig:sdBV}
\end{figure}

Radial mode pulsators exhibit a wide variety of CMD space morphologies between types and subtypes. Classical Cepheids can show quite large CMD loop trajectories with shapes that change drastically with pulsation period. RR Lyrae stars (especially the fundamental-mode RRab subtype) trace loops with the sharp, rapid rise characteristic of their sawtooth light curves. Higher-frequency, lower-amplitude $\delta$ Scuti stars near the upper main sequence trace smaller, smoother loops. Both of their CMD trajectories shapes differ from those of Cepheids. Surprisingly, even extremely rapid pulsations can be detected with \gaia\ MEP, despite its few number of epochs, inconsistent cadence, and long average time between consecutive data points. Figure \ref{fig:sdBV} shows the periodogram, phase-folded light curves, and CMD trajectory for a 5.7-min, radially pulsating hot subdwarf B star (rm-sdBV), previously discovered by \citet{Smith2022}. The amplitude of this pulsation is quite low, at just barely above 10 ppt. Not only can \gaia\ detect the pulsation, it can resolve the star's incredibly subtle looped motion through the CMD. We reiterate that this extremely short-period looped motion would be detectable even if we did not know the pulsation period. This class of rm-sdBV stars likely shows long-term amplitude changes \citep{Bradshaw2026,Hutchens2017}, which should eventually manifest as a spiraling of the looped CMD trajectory.

Figure \ref{fig:pulsators} clearly shows that the CMD loop trajectory shape is a sensitive, time-independent diagnostic of pulsation physics. One striking illustration of this is the famous Hertzsprung progression in classical Cepheids (e.g., \citealt{Marconi2024}). Those with periods from 6-16 days display a secondary bump in their light curves whose phase migrates as the period decreases. This so-called Hertzsprung bump is attributed to a 2:1 resonance between the fundamental mode and the second overtone \citep{Simon1981, Bono2000}. In the CMD trajectory this bump is not inferred from a phase-folded light curve, but it appears directly as a secondary feature in the CMD trajectory shape. Figures \ref{fig:hertzsprung1} and  \ref{fig:hertzsprung2} in Appendix \ref{appendix:hertzsprung} show the phase-folded light curves and corresponding CMD loop trajectories for a sequence of classical Cepheids with increasing periods. The Hertzsprung bump can be seen evolving in lockstep in both the light curves and CMD trajectory plots. The bump migrates around the loop with decreasing period, and the loops becomes progressively cleaner and more regular, approaching fairly simple ovals for the shortest-period Cepheids. Given the famous period-luminosity relationship of Cepheids, Figures \ref{fig:hertzsprung1} and \ref{fig:hertzsprung2} imply that the luminosities and distance of Cepheids could be determined from a time-resolved CMD, even without direct knowledge of the pulsation period.

\begin{figure*}[ht!]
\centering
\includegraphics[width=0.515\columnwidth]{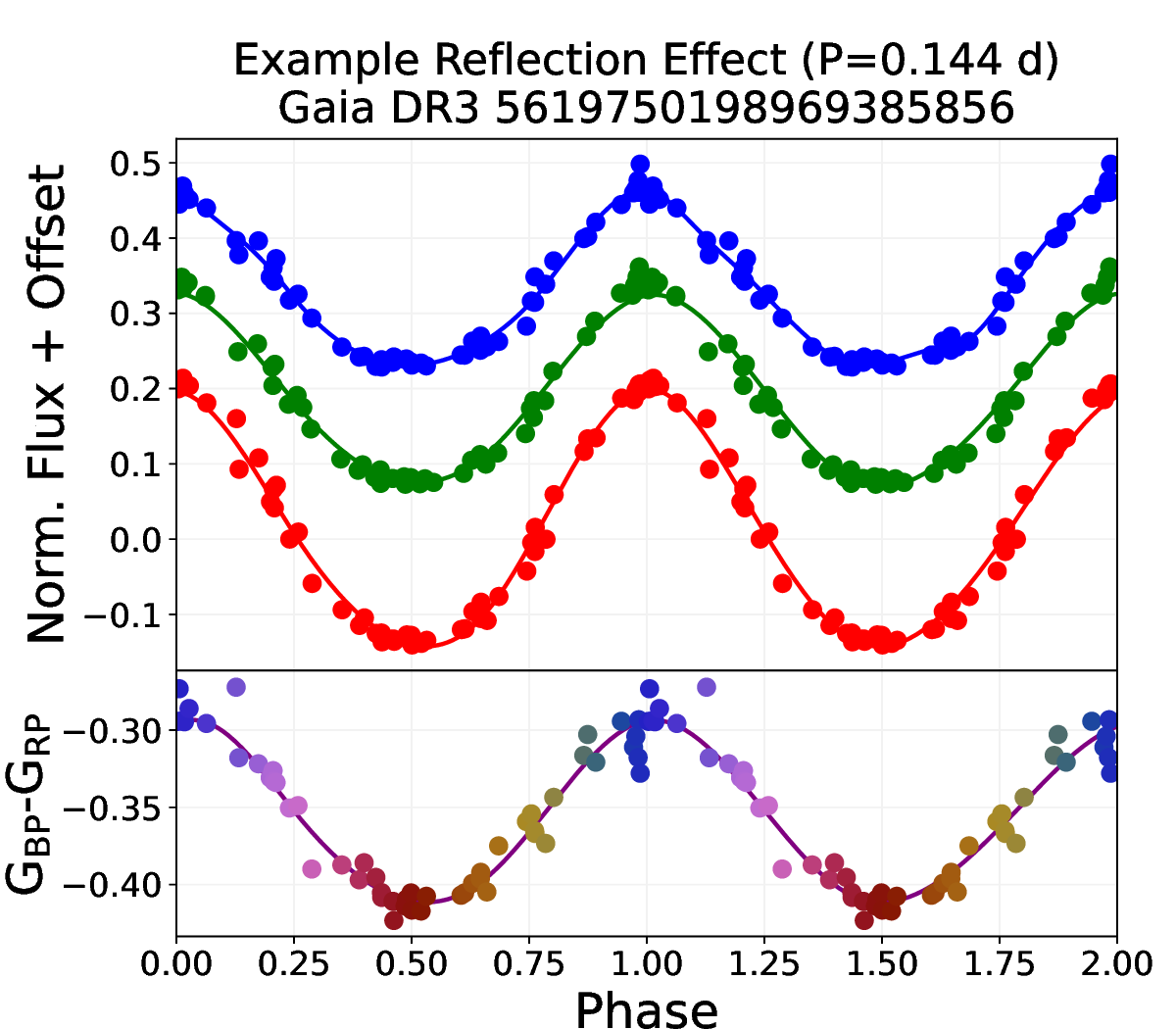}  
\includegraphics[width=0.515\columnwidth]{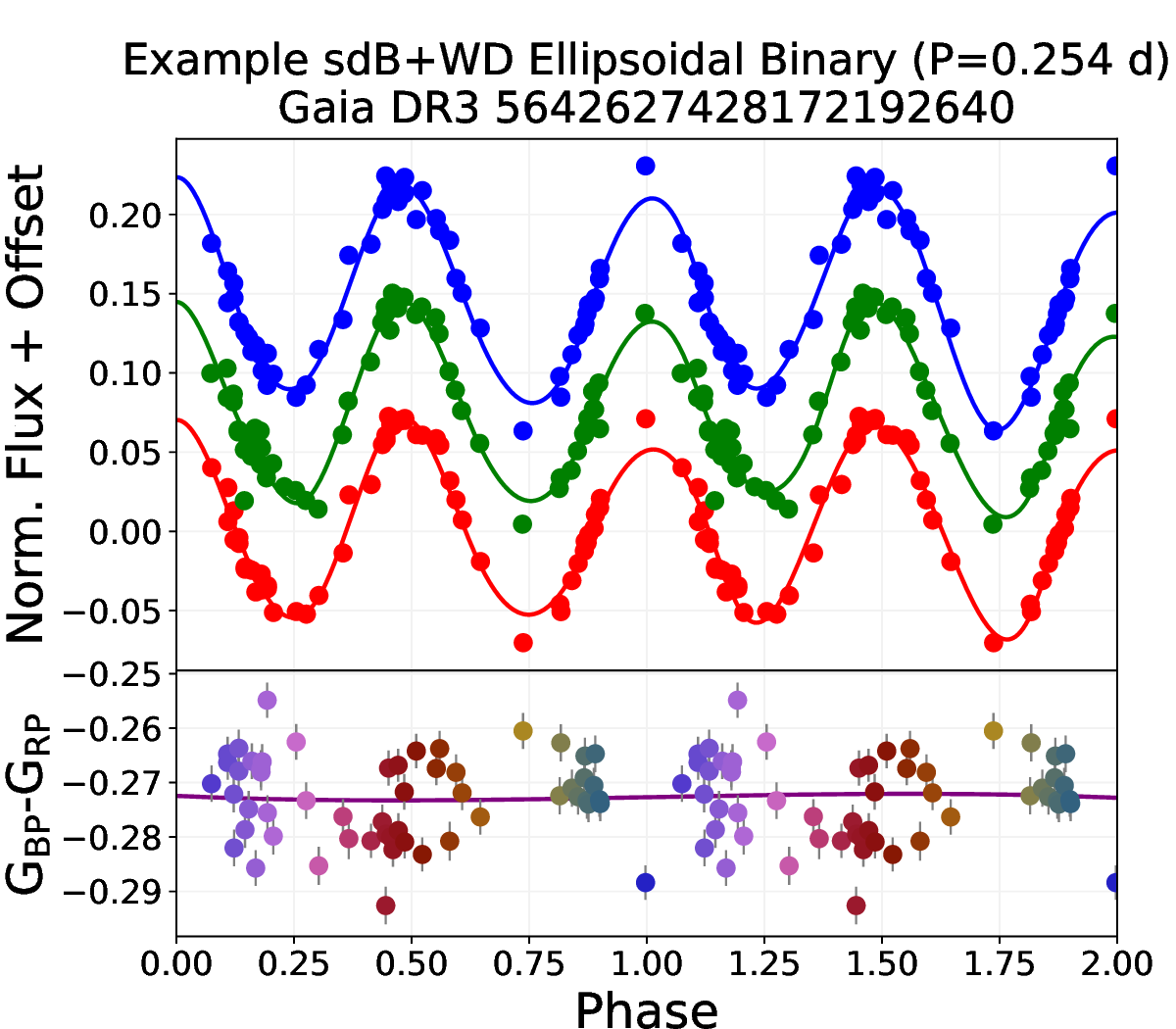}  
\includegraphics[width=0.515\columnwidth]{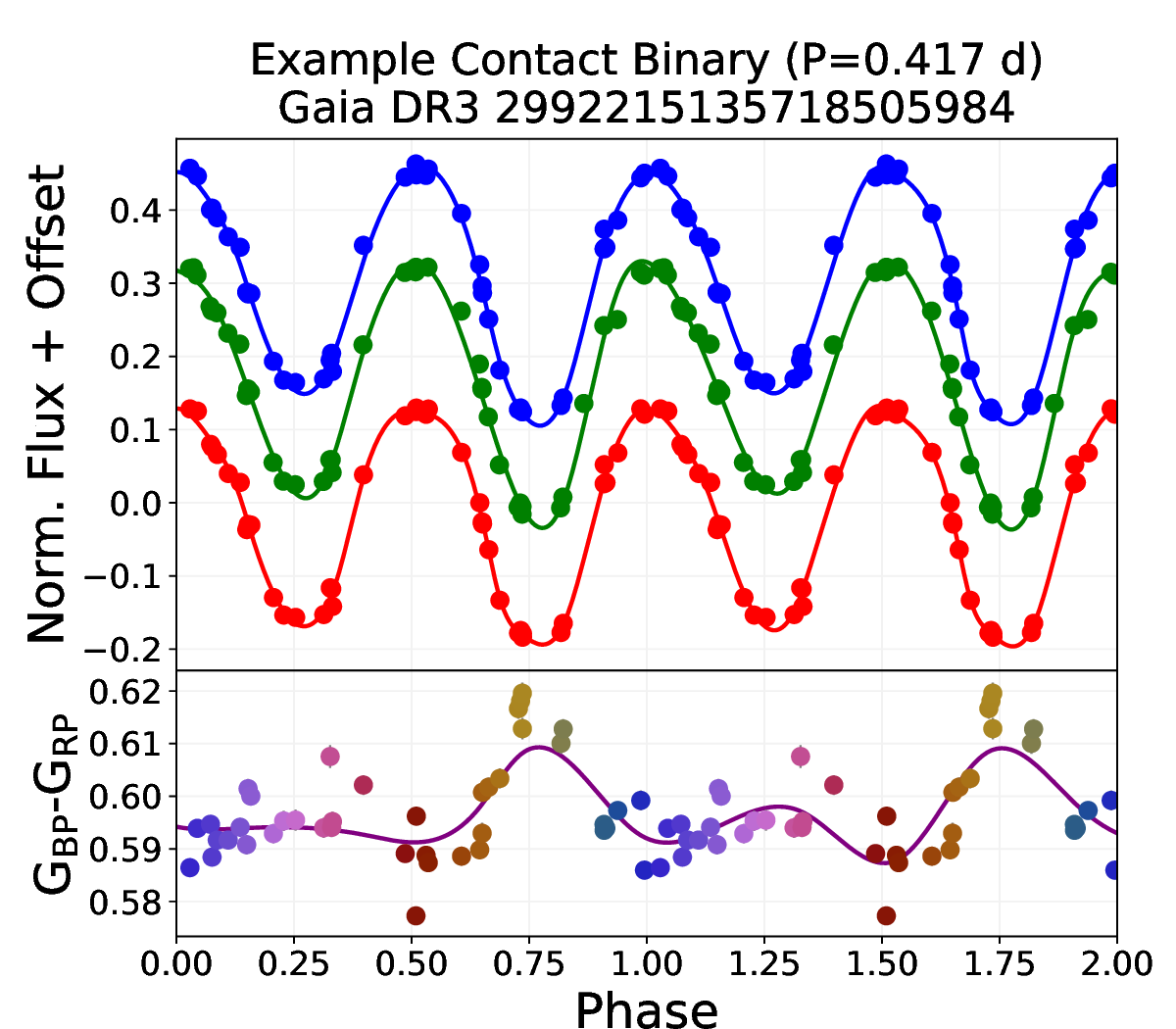} 
\includegraphics[width=0.515\columnwidth]{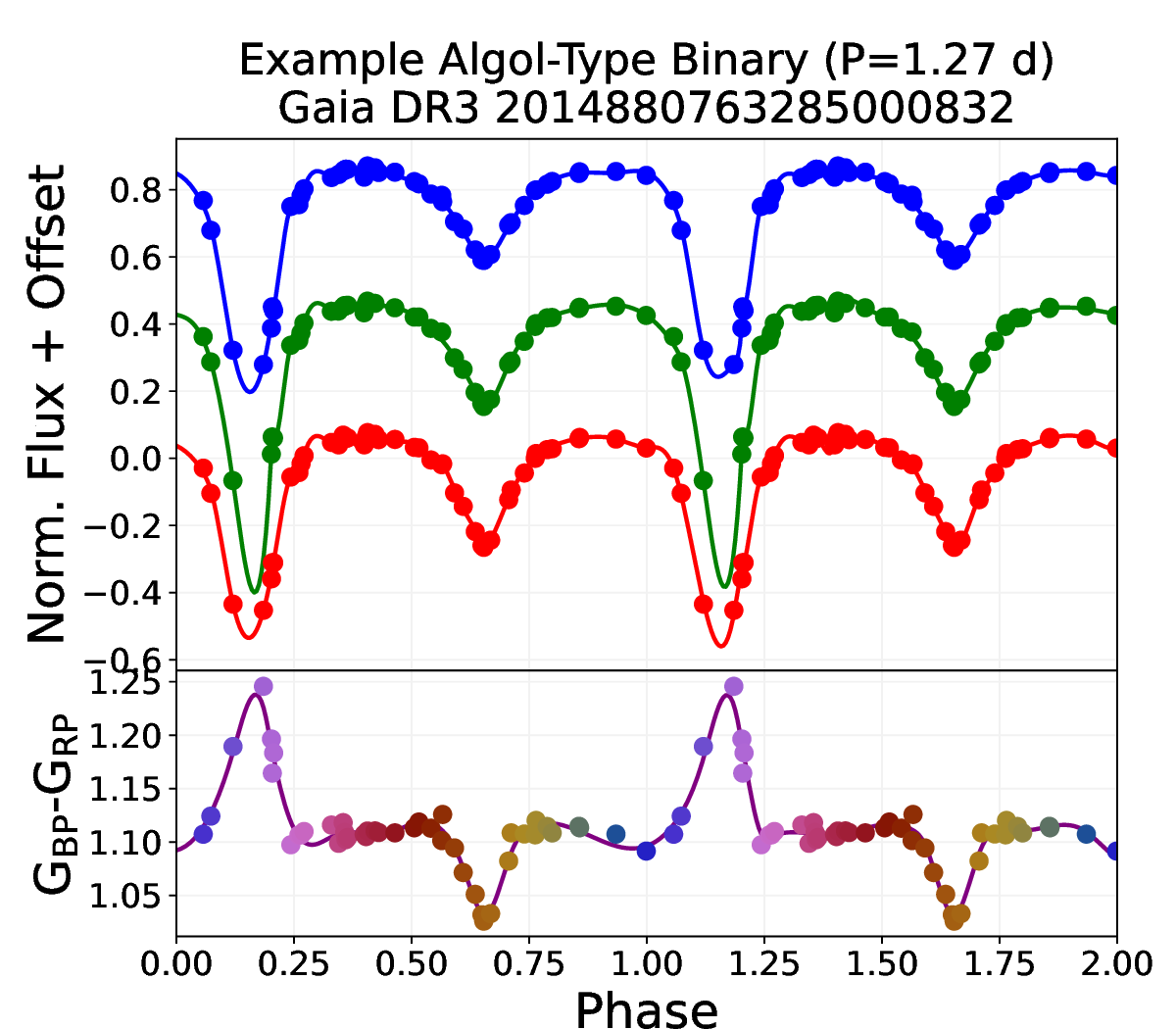}  \\ 

\includegraphics[width=0.515\columnwidth]{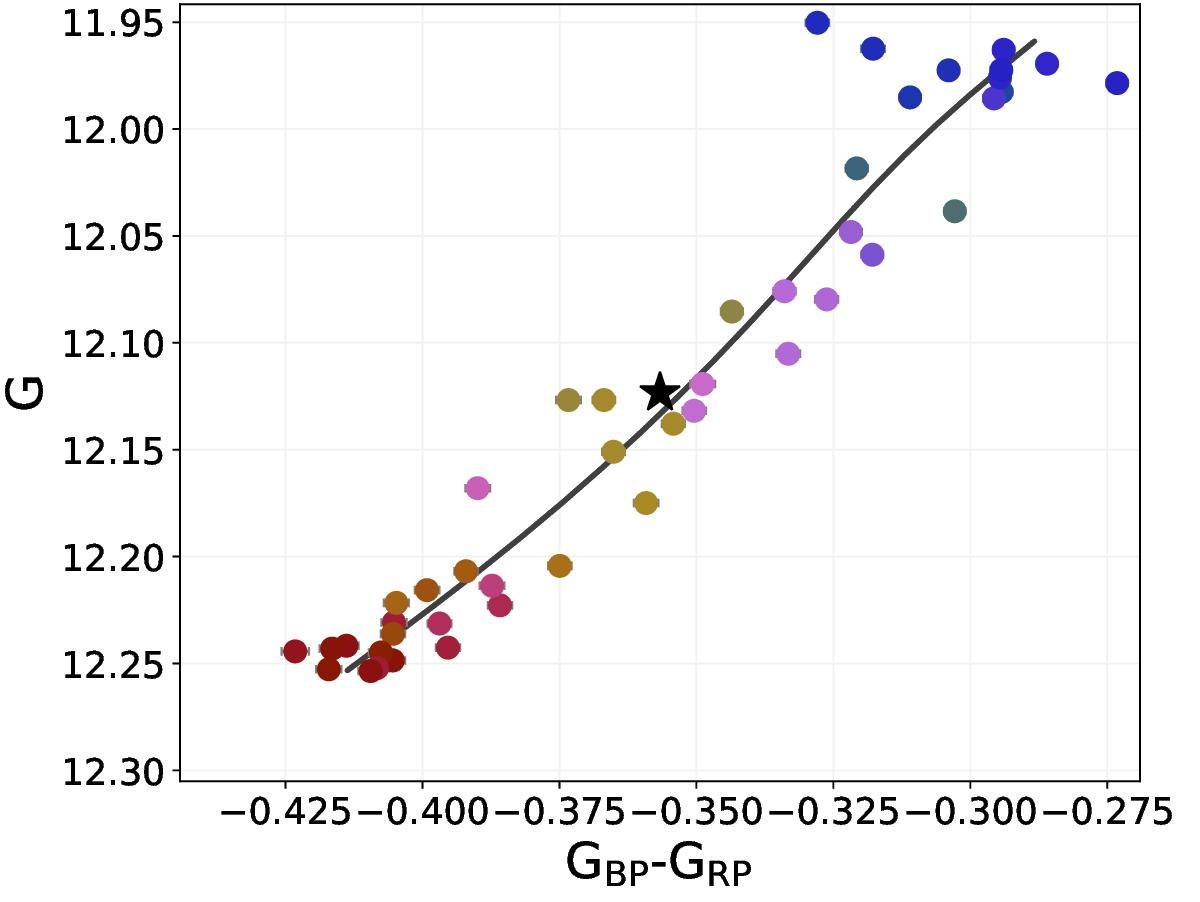}
\includegraphics[width=0.515\columnwidth]{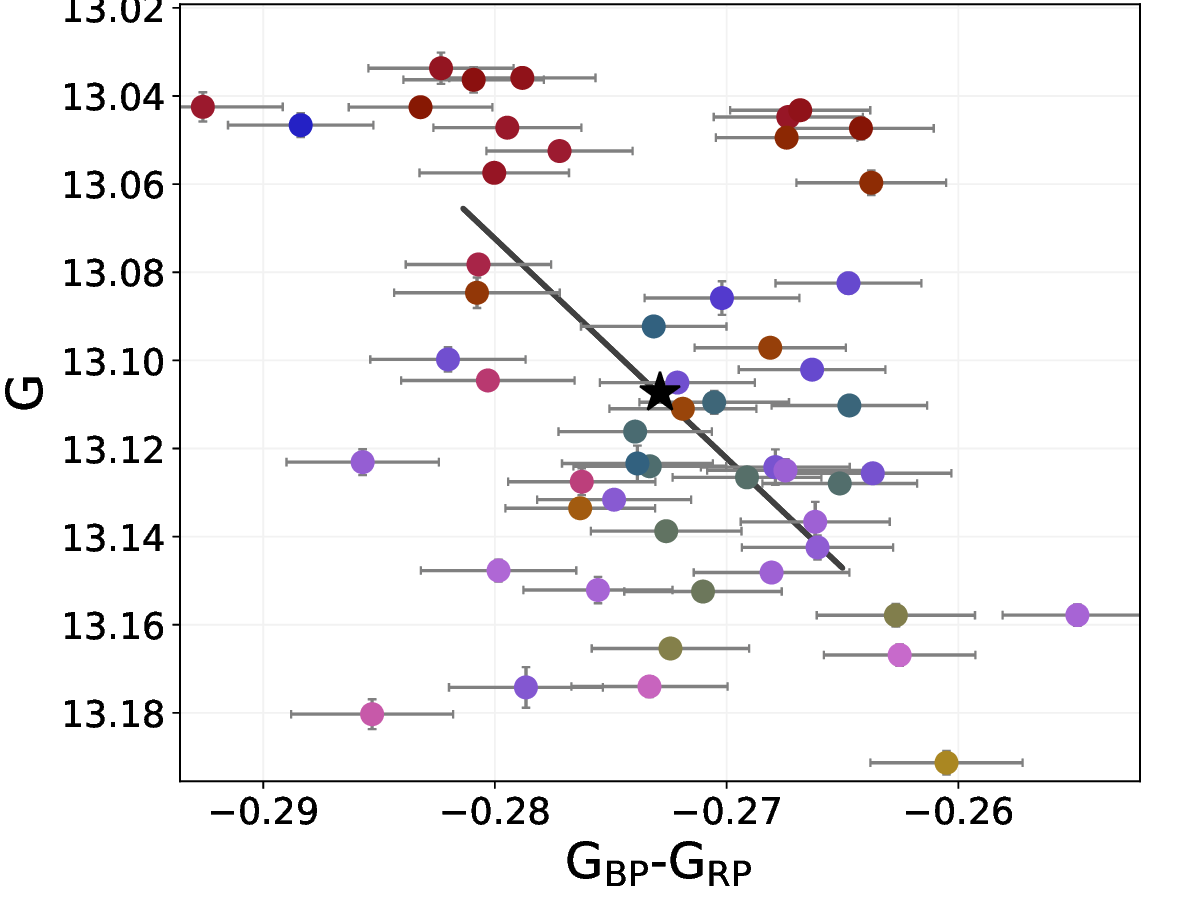} 
\includegraphics[width=0.515\columnwidth]{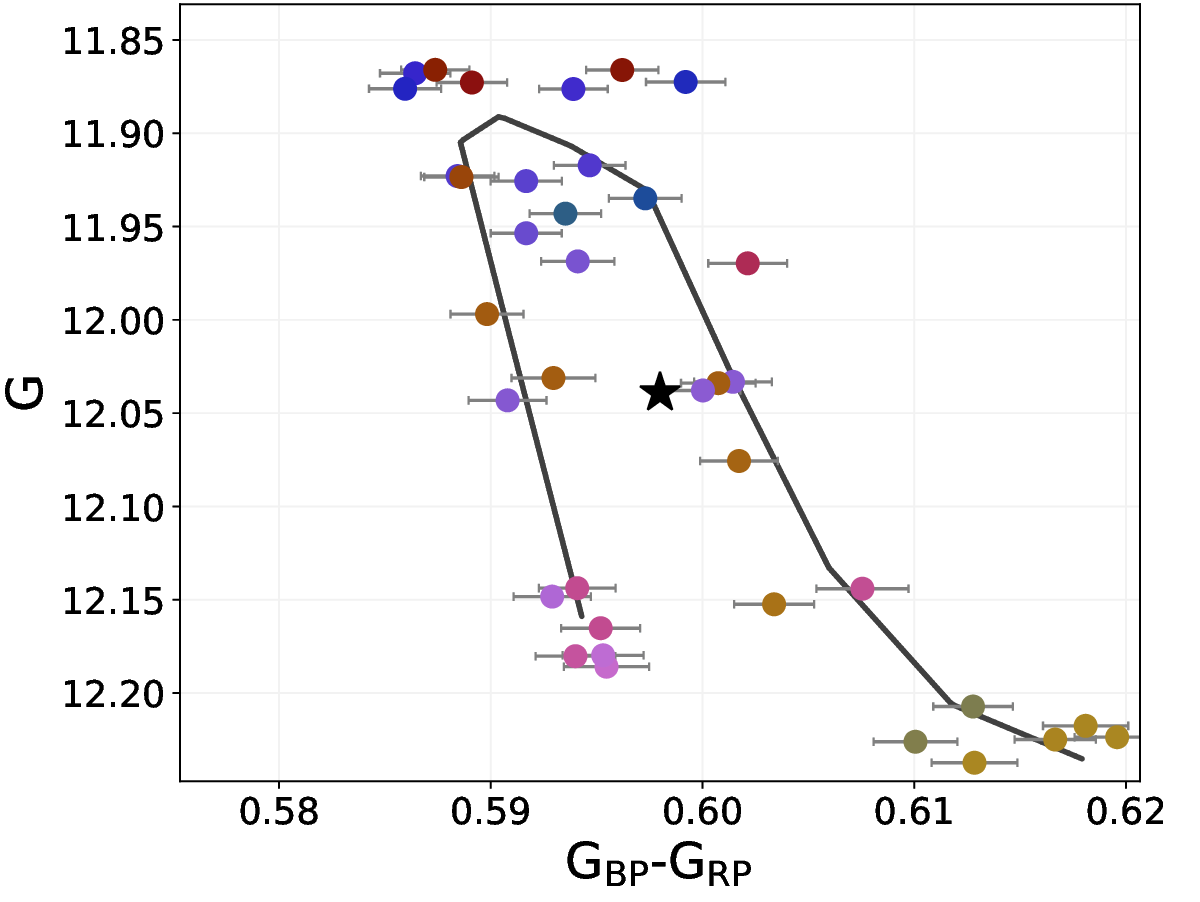} 
\includegraphics[width=0.515\columnwidth]{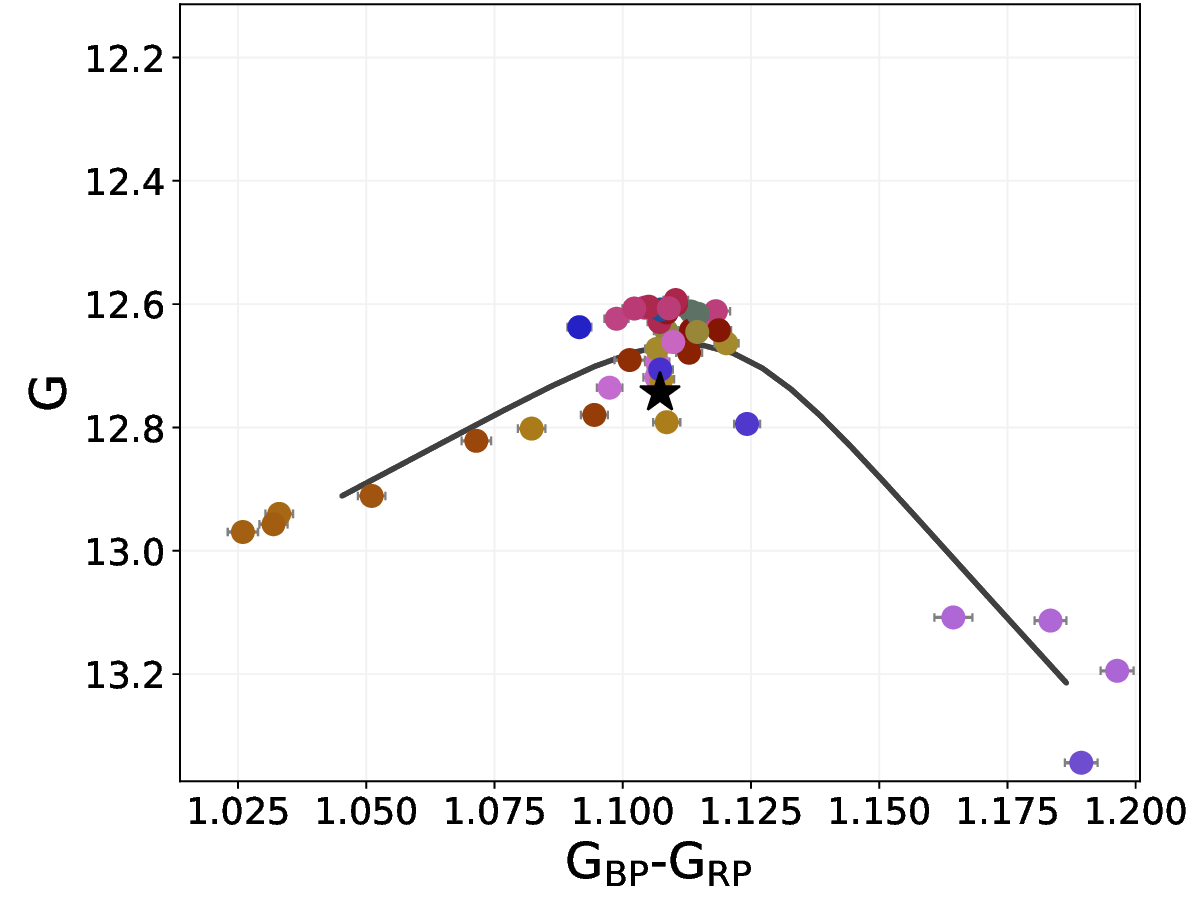}  

\caption{\gaia\ photometry and CMD trajectories for four example binaries, including a reflection effect sdB$+$dM/BD binary, ellipsoidally modulated sdB$+$WD binary, contact binary, and Algol-type binary. Lines and colors have the same meaning as in Figure \ref{fig:pulsators}. Unlike radial mode pulsators, the CMD trajectories of binaries do not typically show loops. Instead, they retrace back over their paths each orbital period. The reflection effect binary grows redder as it brightens with a slightly non-linear trend; this results from irradiation more efficiently reprocessing incident UV , the light to longer wavelengths. The ellipsoidal binary does not meaningfully change in color as it grows brighter, due to the physics governing ellipsoidal modulation; its CMD trajectory is, in essence, a vertical line. The CMD trajectory of the contact binary is a consequence of the temperature difference between stars during eclipse, gravity darkening due to tidal deformation, limb darkenning, and mutual heating. Finally, the CMD motion path of the Algol-type binary is primarily driven by the temperature difference between stars during the primary and second eclipses.}
\label{fig:binaries}
\end{figure*}

\subsection{Binary Systems}
\label{sec:binaries}
Binary systems trace a family of CMD trajectories that differ wildly from those of pulsating stars, due to a variety of effects that may be present. Some examples are shown in Figure \ref{fig:binaries}. In irradiated (reflection effect) binaries, in which a hot component heats the dayside hemisphere of a cooler companion, the reprocessed light causes the system to appear both brighter and redder when the heated face is in view. Some prominent examples include short-period sdB+dM and WD+dM binaries, which typically show reflection effect strengths around 10-20\% (e.g., \citealt{Schaffenroth2023}). They have tight, tilted trajectories in the VECMD with a strong color-magnitude correlation showing they are brightest when reddest. Unlike radial pulsators, their CMD trajectories generally retrace themselves over an orbital period and do not show loops. Because the effect is smooth and periodic, their correlations are generally quite clean. The shape of the reflection effect is such that the light curve spends slightly more time at lower fluxes than higher fluxes. This can also be seen in the CMD trajectory plot as an asymmetric distribution of points along the main trend. When irradiated binaries display eclipses, the CMD trajectories show additional vertical features that drop down from the ends of the reflection effect trajectory. Their vertical nature arises from the eclipses not generating significant color variations. Ellipsoidal systems like sdB+WD binaries, in which tidal distortion of the sdB produces two brightness maxima per orbit with little temperature change (e.g., \citealt{Pelisoli2021}), trace a nearly vertical line in the VECMD. Main sequence contact binaries, in which both stars fill their Roche lobes and share a common envelope, show yet another unique CMD trajectory. Because of their extreme proximity, their eclipses and mutual ellipsoidal effects blur into one another, generating a light curve that spends more time at higher fluxes than lower fluxes. This manifests in a non-linear CMD trajectory that contains more epochs on the higher flux end.

The value of the time-independent trajectory representation for binaries is that it disentangles mechanisms that can look similar in a single light curve. Reflection and ellipsoidal modulation both produce quasi-sinusoidal variations, but they occupy different directions in the VECMD since one is driven by a temperature change and the other is not. Within CMD phase space, a large separation between trajectory shapes appears, and classifications that would otherwise require detailed light-curve fitting become far simpler, even without knowing the orbital period.

\begin{figure*}[ht!]
\centering
\includegraphics[width=0.515\columnwidth]{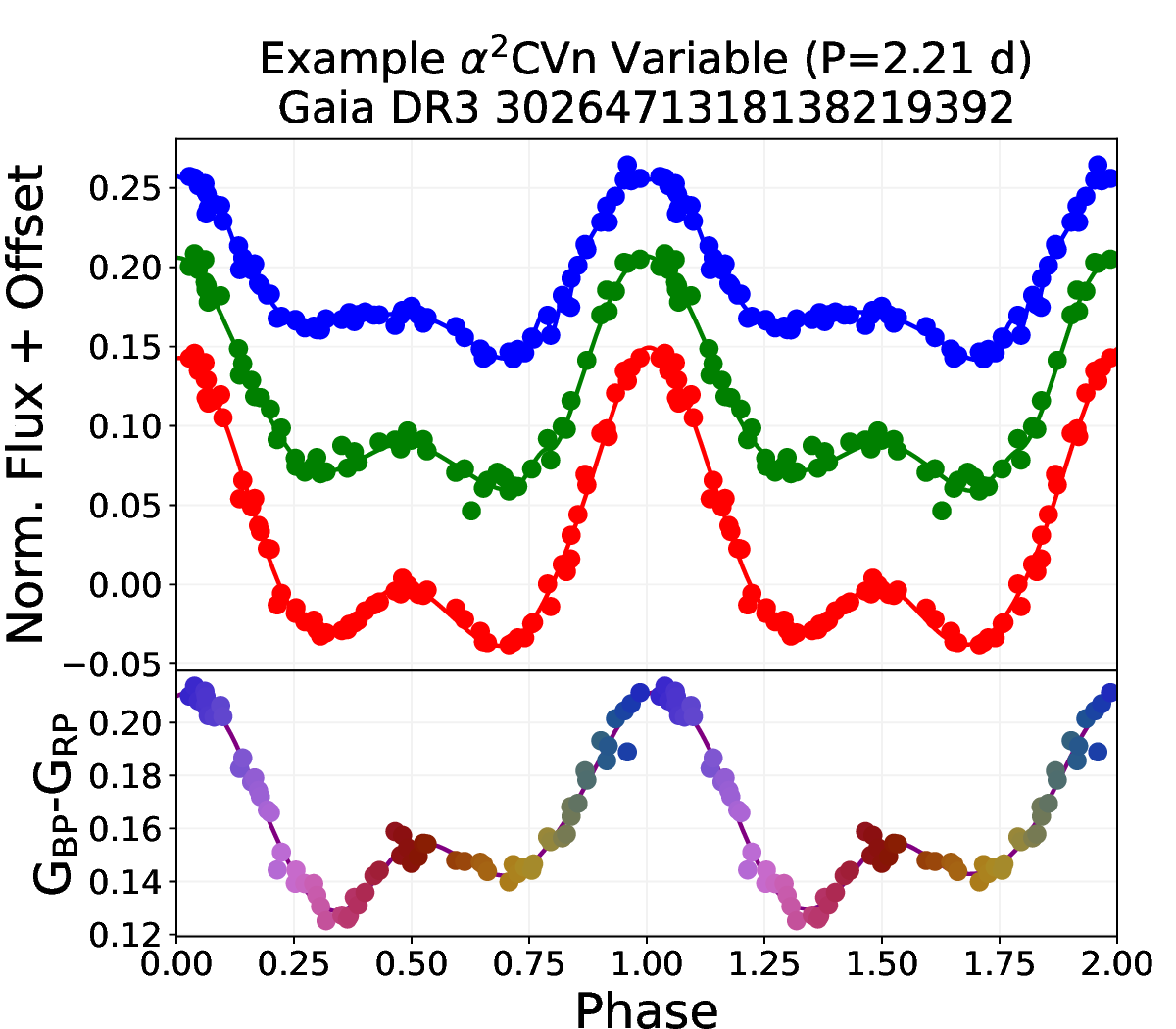} 
\includegraphics[width=0.515\columnwidth]{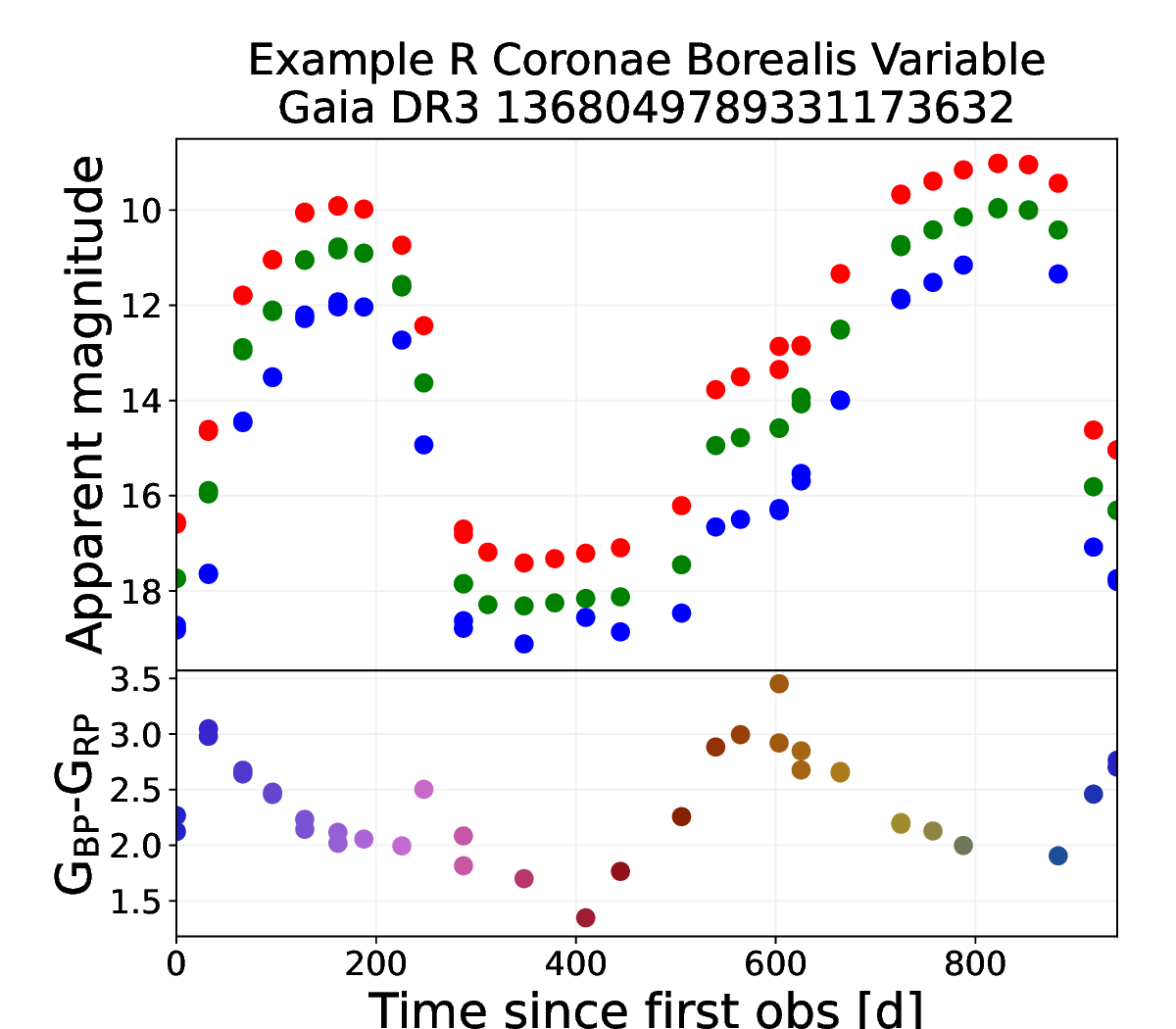}  
\includegraphics[width=0.515\columnwidth]{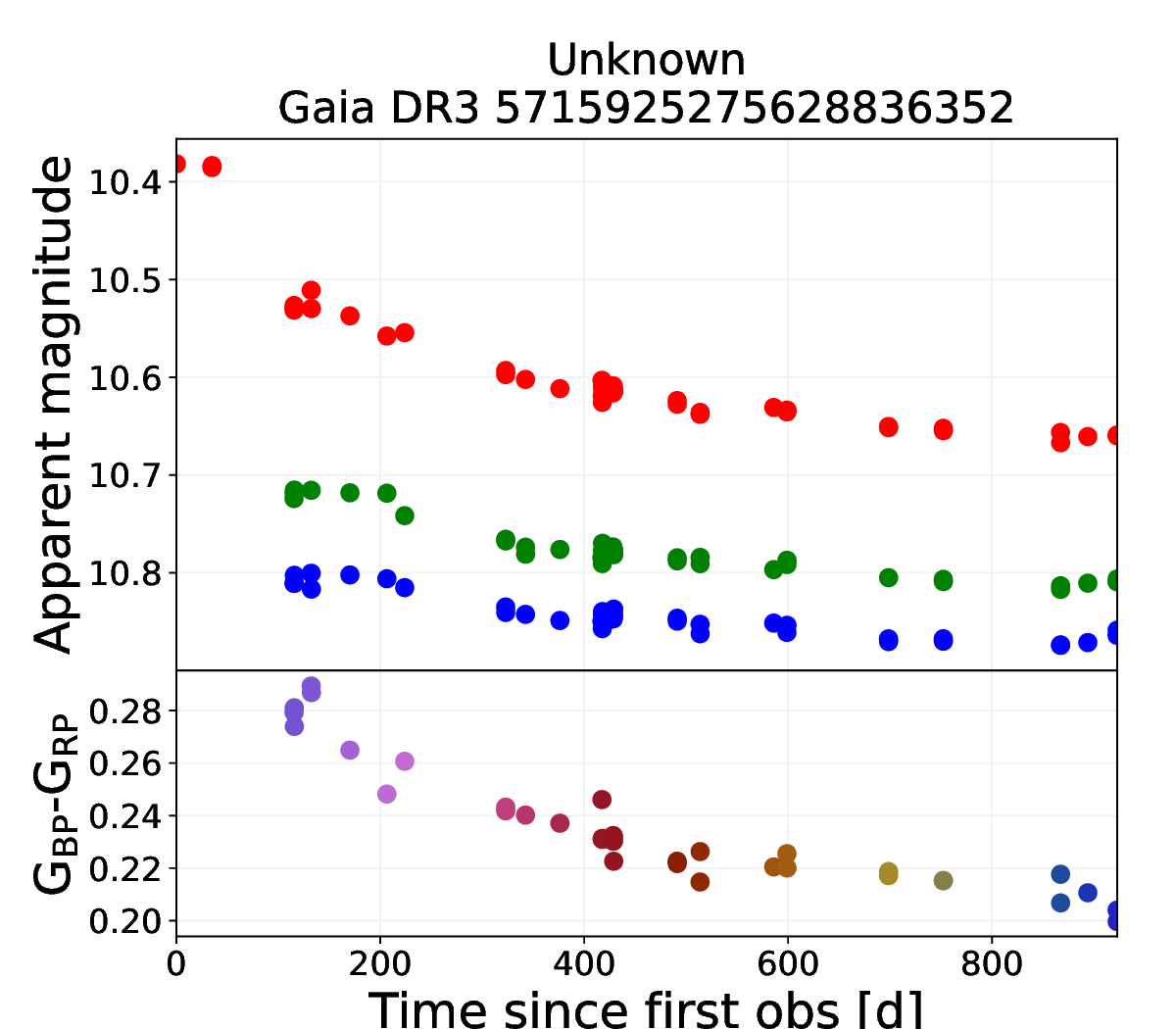} 
\includegraphics[width=0.515\columnwidth]{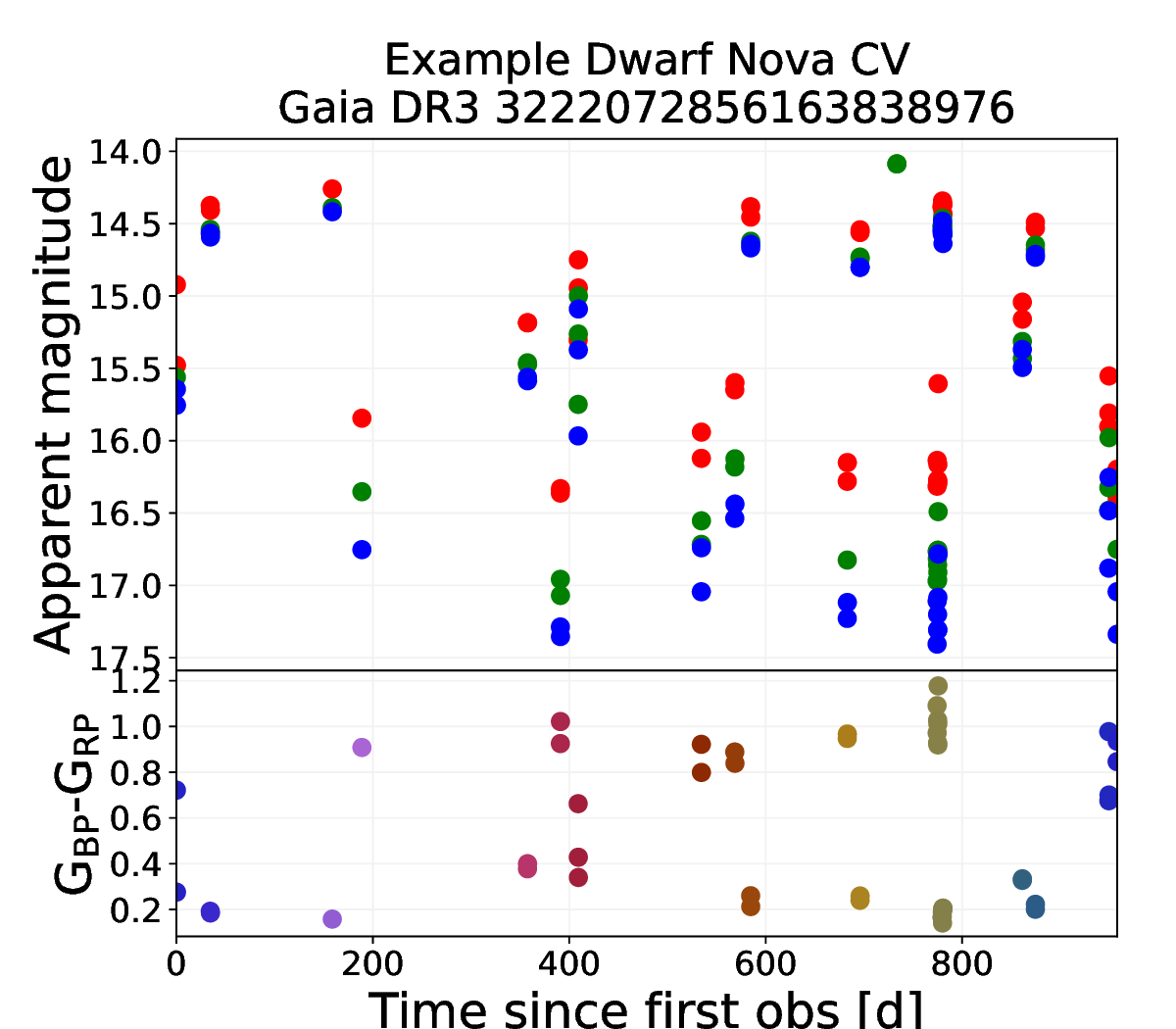}  \\

\includegraphics[width=0.515\columnwidth]{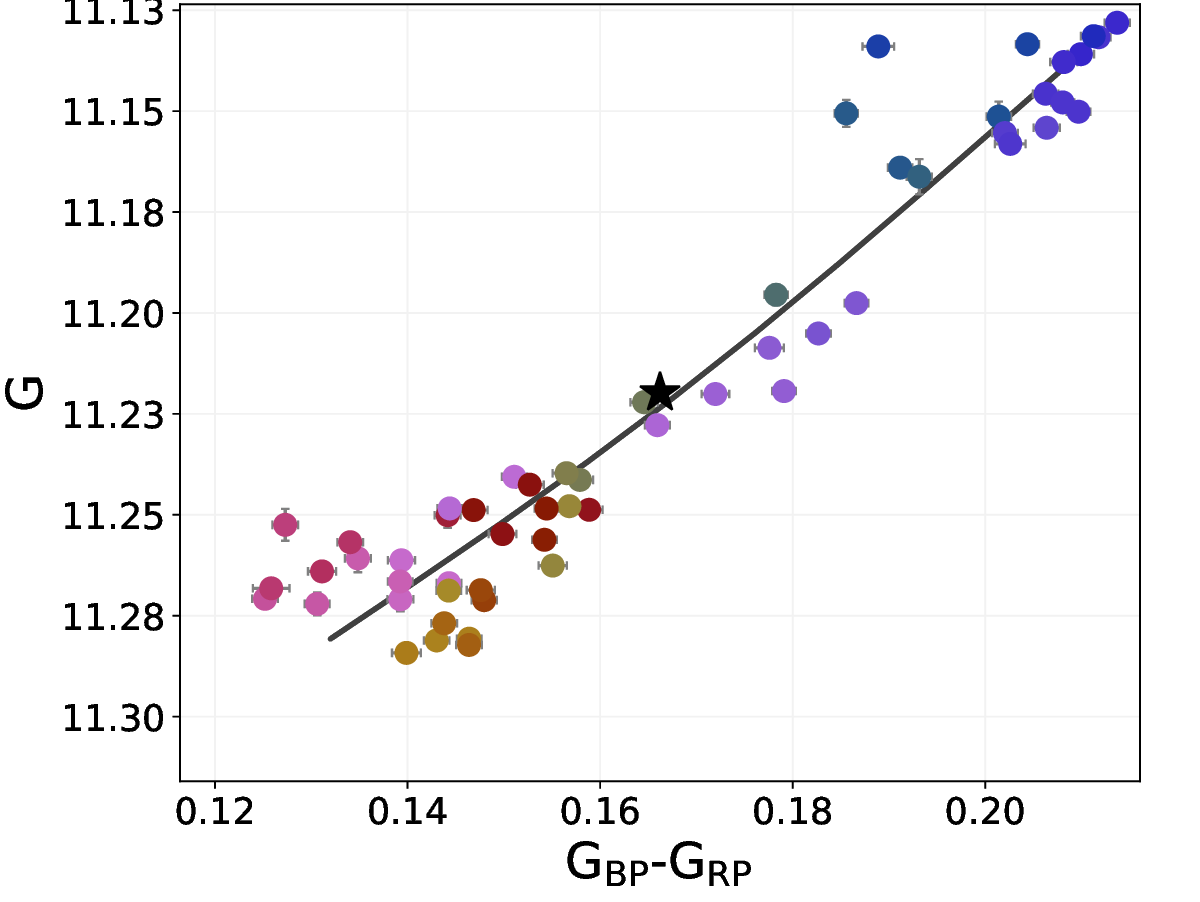} 
\includegraphics[width=0.515\columnwidth]{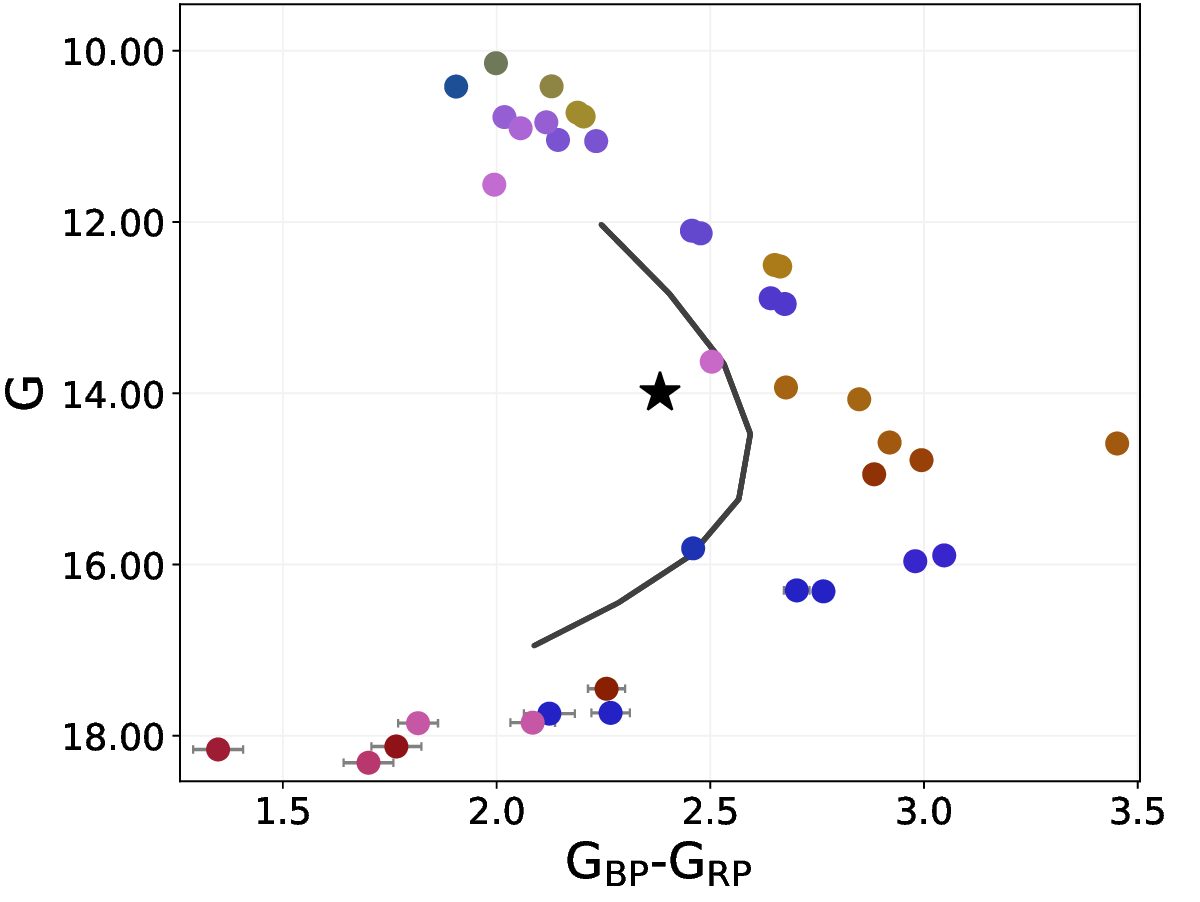} 
\includegraphics[width=0.515\columnwidth]{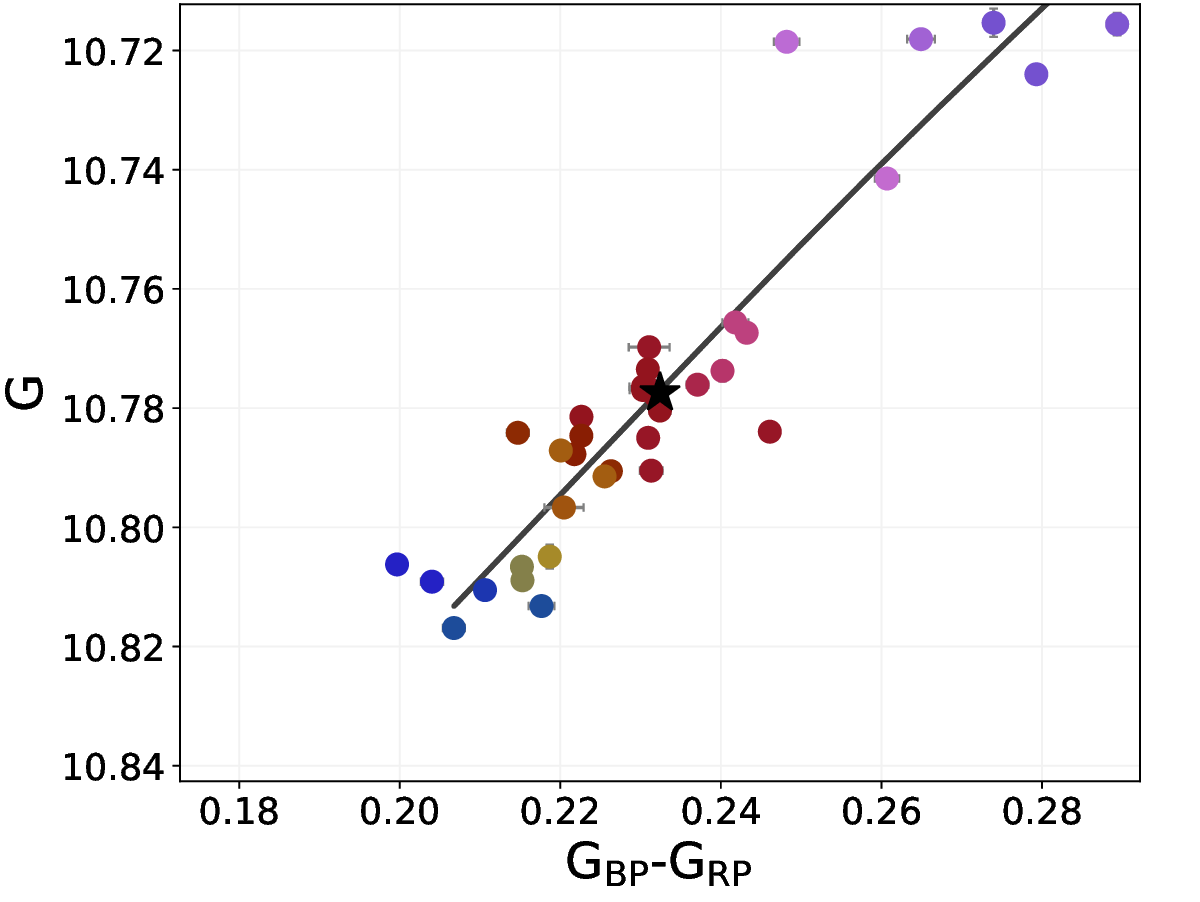} 
\includegraphics[width=0.515\columnwidth]{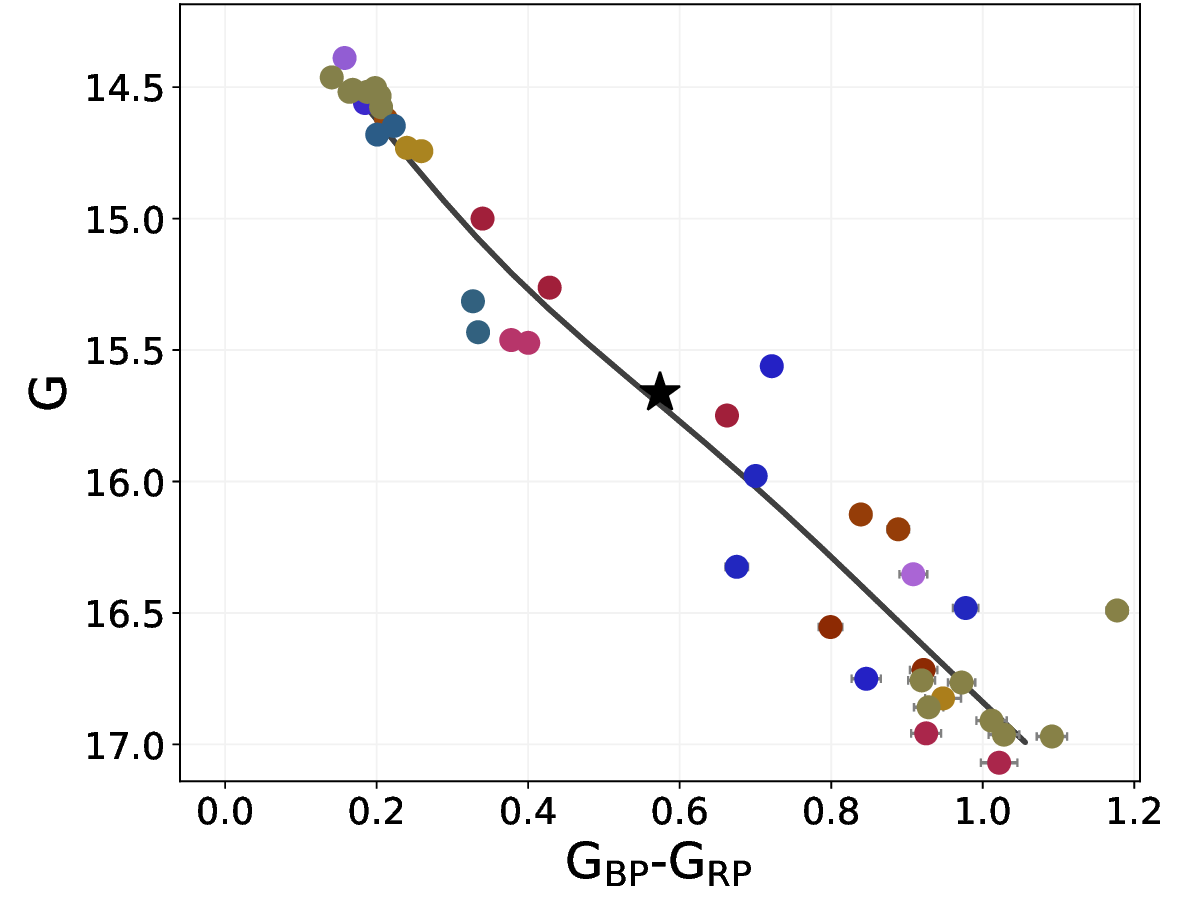}

\caption{\gaia\ photometry and CMD trajectories for example rotating stars and aperiodic variables, including a $\alpha^2$CVn variable, R Coronae Borealis star, an object of unknown type, and a dwarf nova CV (from left to right). Lines and colors have the same meaning as in Figure \ref{fig:pulsators}. For the aperiodic systems, the light and color curves are shown in relative time units, not in phase. Moreover, splines fits are not included for the aperiodic systems.}
\label{fig:aperiodic}
\end{figure*}

\subsection{Rotators \& Aperiodic Variables}
\label{sec:aperiodic}

Finally, there are many systems that exhibit rotational modulation or strong, aperiodic changes over the lifetime of the \gaia\ mission that are visible as CMD trajectories. While some aperiodic trajectories are due to secular, evolutionary effects (e.g., \citet{Conroy2018}), others are due to eruptive, accreting, or dust-obscured systems. Some example rotational and aperiodic CMD trajectories are shown in Figure \ref{fig:aperiodic}.


The CMD trajectory of the example $\alpha^2$CVn star in Figure \ref{fig:aperiodic} shows that it is brightest when reddest, which is consistent with rotational modulation of chemically peculiar abundance spots.
Dust-obscured eruptive systems can trace out trajectories that are generally aligned with a reddening vector. R Coronae Borealis (RCB) stars, which are hydrogen-deficient carbon supergiants that fade by up to several magnitudes when carbon dust condenses along the line of sight, move down and to the red as they enter a decline. The CMD track is often not, however, a simple back-and-forth along the reddening vector. Near deep minimum the color can hook back toward the blue as scattered light from the circumstellar shell and chromospheric emission begin to dominate the direct, heavily extincted photospheric light. The same blueward hook at minimum is the defining photometric signature of UX Ori stars and related young dipper systems, in which orbiting disk material occults a Herbig Ae/Be or T Tauri photosphere. In both classes the VECMD recovers the extinction-then-scattering behavior as a characteristic hooked trajectory that a single reddening-aligned line segment cannot capture. Some cataclysmic variables, like dwarf novae, show large changes in both brightness and color during eruptions. Like in the example system shown in Figure \ref{fig:aperiodic}, these systems become bluer as they brighten during outbursts, due to the accretion disk heating from a few thousand to tens of thousands of Kelvin. They show some of the most impressive CMD trajectories across the entire CMD.



\section{Future Directions}
\label{sec:future}

The \dcmd\ presented here is a modernized representation of the CMD, not yet a measurement pipeline. Its greatest value lies in what it enables. We highlight four different directions for our future work. Each will be developed in subsequent papers in this series.

\begin{figure}[t!]
\centering
\includegraphics[width=0.99\columnwidth]{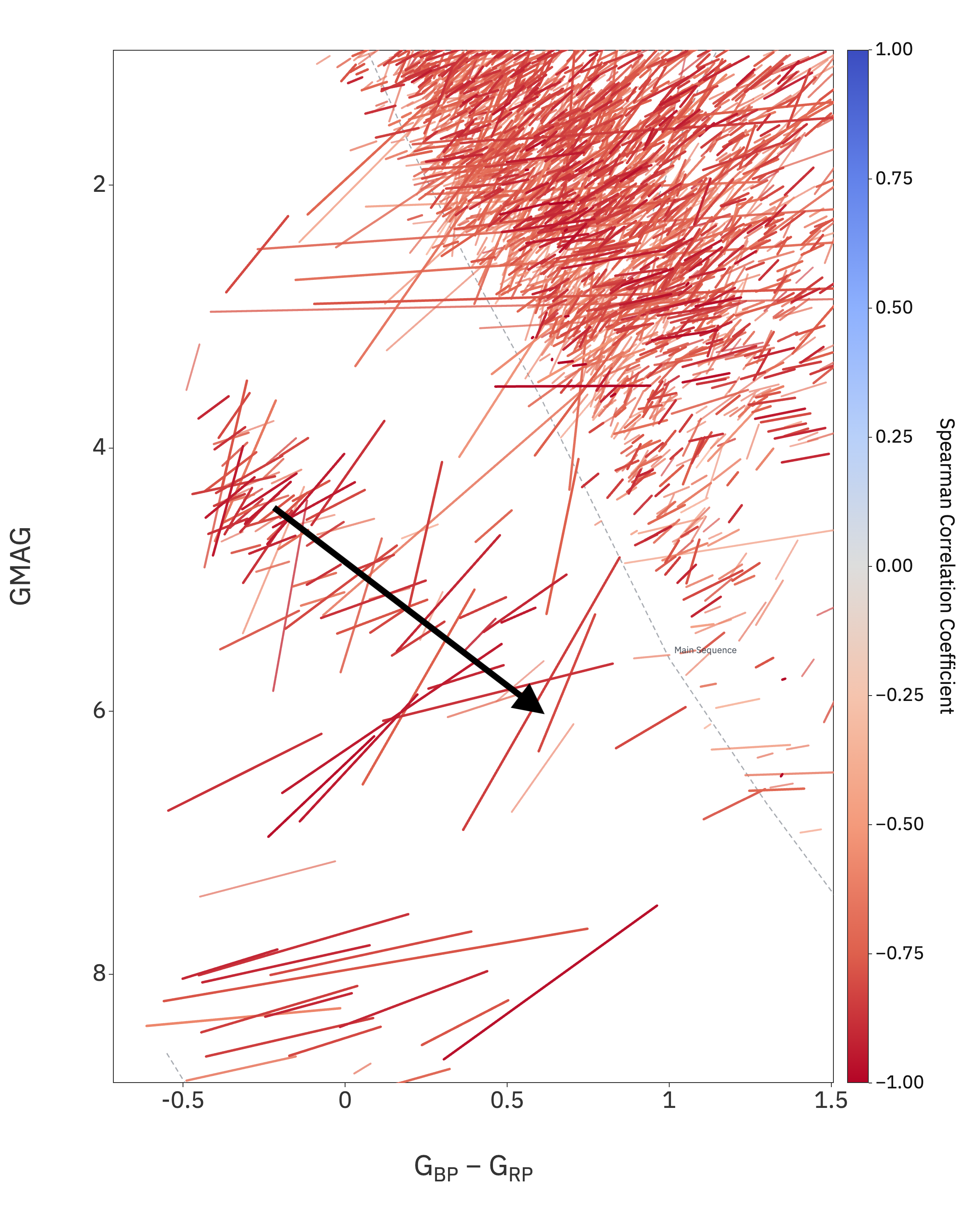}
\caption{Region of the VECMD near the extended horizontal branch, where hot subdwarf B stars reside clump. Blue line segments are turned off in the plot for better visualization. The slanted sequence of red line segments is primarily made up of reflection-effect-dominated sdB$+$dM/BD binaries and follows the reddening vector (black line), which is shown for $A_V$=2. While these reflection effect binaries are typically found via photometric searches centered on the blue end of the reddening vector (the unreddened sdB clump position) we can easily find high-reddened sdB$+$dM/BD binaries all the way down to the main sequence (gray dotted line) by looking for red line segments with similar lengths and slopes that fall along the reddening vector. These systems would otherwise be lost sitting amongst millions of main sequence stars in a traditional CMD. }
\label{fig:extinction}
\end{figure}

\begin{figure}[t!]
\centering
\includegraphics[width=0.99\columnwidth]{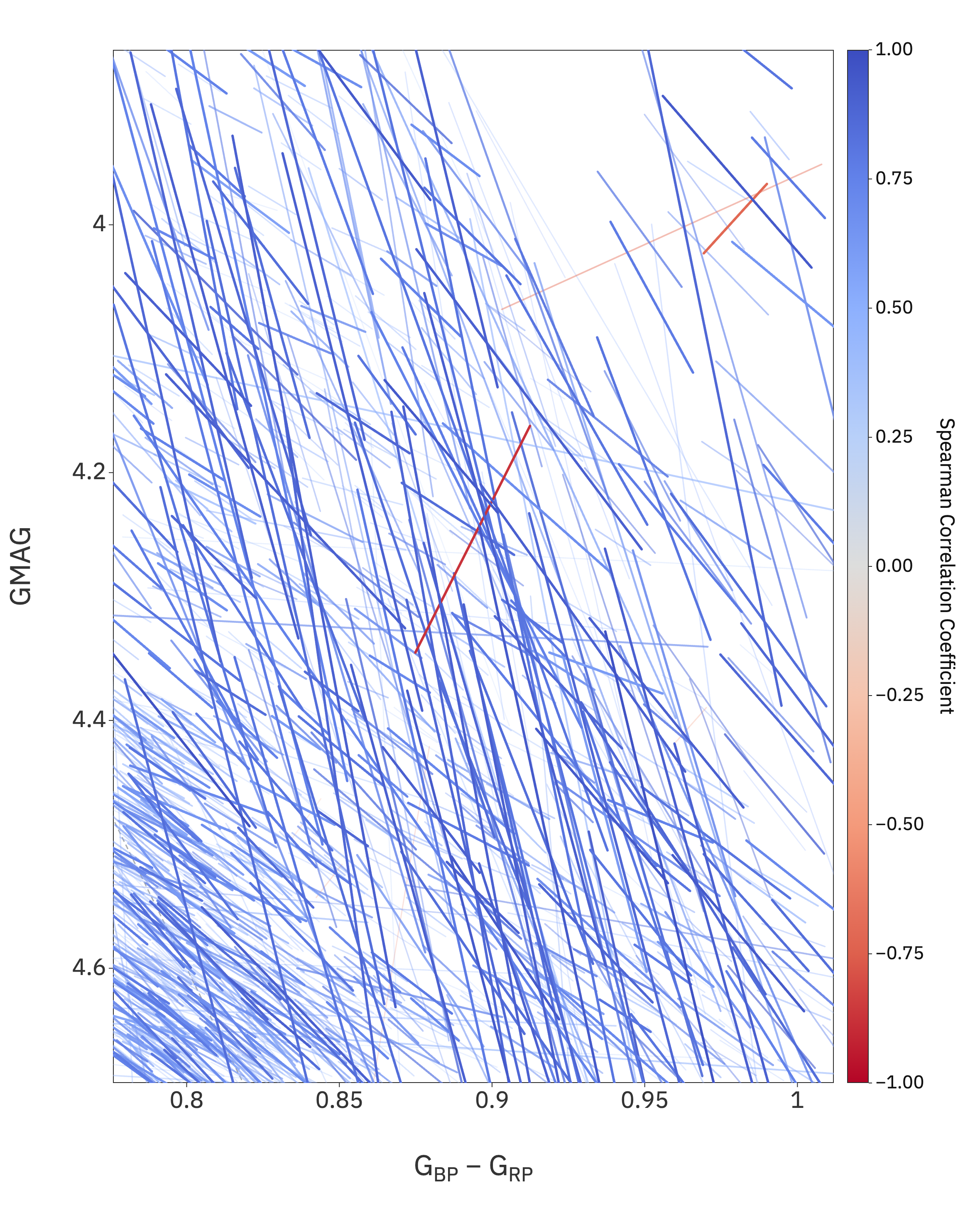}
\caption{Region of the VECMD population by early-G to early-K main sequence stars, near-equal-mass main sequence binaries, and some subgiants evolving off the turn-off. The VECMD makes it easy to spot objects `going against the grain.' One red line segment stands out amongst a sea of blue line segments, and in this case it indicates a W UMa overcontact binary (Gaia DR3 5001230863456552064; V*CR Scl).}
\label{fig:againstthegrain}
\end{figure}

\subsection{Variable Discovery \& Classification through Visual Inspection of the VECMD}

First, visual inspection of the \vecmd, even with a simple line-segment representation of the CMD trajectories, makes discovering new variable stars quick, easy, and fun. Since constant or nearly-constant stars are suppressed in the diagram, variable stars visually stick out. We have already been using \texttt{IVECMD} to manually investigate the VECMD, click on variable stars within CMD areas of interest to us, investigate the raw \gaia\ photometry, and select objects of interest for additional follow-up. We note that the CMD trajectory shapes of variable stars do not significantly change with increasing extinction, nor do they change with decreasing parallax precision (outside of a degrading S/N). This means the VECMD can be used to identify variables reddened beyond the CMD boundaries they normally reside in, or new variables that might not have survived the parallax precision cuts of catalog surveys. Hot subdwarfs serve as a good example of this. The widely used \citet{Culpan2022} catalog of hot subdwarf stars employed both color cuts and parallax precision cuts to ensure they maintained the purest sample of hot subdwarf stars as possible. Understandably, as with any sample cuts, this leads to many bona fide hot subdwarf stars being left out of the catalog, primarily due to reddening or poor parallax constraints. \texttt{IVECMD} can be used to quickly pick out and classify such systems if they are variables. \citet{Kirby2026}, for example, present the discovery of over two dozen compact post common envelope hot subdwarf binaries found using the VECMD and CMD trajectory shapes. Related to this study, we plot in Figure \ref{fig:extinction} the VECMD region surrounding hot subdwarf stars. One can see the VECMD is dominated by red line segments primarily originating from variable eclipsing and non-eclipsing sdB$+$dM/BD reflection effect binaries. Following the direction of the reddening vector, these line segments of similar slopes and lengths extend down and to the right in the CMD, even overstepping the red edge color cut used by \citet{Culpan2022} for their hot subdwarf catalog. Thus, the VECMD helps us efficiently catch and classify variables in parts of the CMD they do not belong.

Figure \ref{fig:againstthegrain} shows another example of the power of visual inspection of the VECMD. In a given local region of the VECMD, most line segments follow the same trend, having similar slopes and lengths. Objects that deviate from said trend stick out like a sore thumb, making it easy to discover rare or anomalous objects.

\subsection{Variable Discovery \& Classification through Machine Learning on CMD Trajectories}

Second, detailed CMD trajectories are ideally suited to machine learning. They are compact, low-dimensional and physically interpretable objects (a curve in a two-dimensional plane), rather than high-dimensional light curves or hand-assembled feature vectors. This makes them amenable to both supervised classification, in which trajectory shape is used to assign a variability class, and unsupervised discovery, in which clustering on trajectory shapes reveals new or anomalous morphologies. We are currently developing methods to fit and parameterize the curves that reduce each trajectory to a small set of shape descriptors (e.g. loop area, orientation, curvature, bump phase) suitable for machine-learning inputs. These methods range from simple linear regression, through traveling-salesman orderings of the epochs, to principal-curve techniques, and more. As shown in Figures \ref{fig:pulsators}, \ref{fig:binaries}, and \ref{fig:aperiodic}, we have already experimented with fitting elastic principal curves and closed elastic principal curves to the CMD trajectories with some success. The ElPiGraph fits to different types of variables require different numbers of nodes to fit correctly. More nodes increases the ability to capture fine details in the CMD trajectories (e.g., Hertzsprung bump), while too many nodes can cause overfitting, overbranching, and is more susceptible to outliers. Because the CMD trajectory shapes are period-agnostic, a classifier trained on them can operate even on stars for which no period can be measured. Since many of the CMD trajectory shapes features encode the physics behind the stars' variability (e.g., the Hertzsprung) bump, future machine learning algorithms may very well extract stellar properties directly from \gaia\ MEP.

\subsection{\texttt{MESA Custom Colors} Modeling}

Third, the CMD trajectories, which encode a significant amount of physics, naturally connect to stellar modeling. Codes such as \mesa\ \citep{Paxton2011} and \gyre\ \citep{Townsend2013} calculate the structure, evolution, and pulsations of stars. However, their native outputs are {\em theoretical} quantities (e.g., temperatures, luminosities, radii) rather than {\em observed} magnitudes and colors. The recently released \texttt{MESA Custom Colors} module permits synthetic CMD trajectories to be created for stars using magnitudes and colors in any filter combination \citep{Miller2026}. This ability turns the CMD trajectory into a powerful testbed for pulsation physics, in which features like the predicted loop shape, distribution of points along the loop, and Hertzsprung bump phase all become direct observables. 
\mesa's \texttt{RSP} module \citep{Paxton2019} solves the equations of nonlinear radial pulsation with a time-dependent treatment of convection, self-consistently following the variations in \teff, $L$, and $R$ through each hydrodynamic timestep \citep{Smolec2008}. When coupled with \texttt{MESA Custom Colors}, synthetic photometry is produced at the cadence of the pulsation integration, so that the resulting magnitudes require no temporal interpolation and remain exactly aligned with the underlying stellar state. \citet{Miller2026} present example calculations for RR Lyrae pulsations that illustrate \mesa's ability (when coupled with \texttt{GYRE}, \texttt{RSP}, and other modules) to produce the pulsation-parameter-dependent, Lissajous-like patterns we see in \gaia\ CMD trajectories. While \texttt{MESA Custom Colors} cannot currently handle binary light curve features like eclipses, ellipsoidal modulation, and the reflection effect, future, expanded versions of the module may open up binaries to CMD trajectory analyses in the near future.

\subsection{Extension to \gaia\ DR4 MEP}

Fourth, the VECMD and analysis of CMD trajectories scales with relative ease. The $\sim$11.7 million DR3 sources with released MEP represent a small and non-random slice of the \gaia\ catalog. \gaia\ DR4 is scheduled for release on 2 December 2026 and will publish MEP for a significantly larger sample of stars covering a baseline twice as long as DR3. The VECMD of DR4 will therefore populate the entire CMD with trajectories at a density well beyond DR3, transforming the atlas presented here into a statistical map of the entire variable sky. The future work presented in the previous subsections, once applied to \gaia\ DR4, should lead to the discovery, classification, and analysis of countless new variable stars across the CMD, perhaps some of unknown types.

\section{Summary}
\label{sec:summary}

We have introduced a \dcmd\ in which each star is represented not by a point but by a simple line segment representing its approximate change in magnitude and color due to intrinsic variability. Its construction is made possible by the near simultaneous, three-band, multi-epoch photometry released for $\sim$11.7 million sources in \gaia\ DR3, which samples a contemporaneous point on the CMD at every observation epoch. Replacing each star's point with a simple slope-and-amplitude line segment produces a diagram in which the relatively `quiet' main sequence recedes while the loci of variable stars dominate. Encoding the sense and strength of their magnitude-color correlations with the proper color and opacity mappings makes visual inspection of the diagram quick and easy. 
To assist with this inspection, we also built and presented an interactive tool called \texttt{IVECMD} that allows users to explore the VECMD and the \gaia\ multi-epoch photometry underlying each line segment. This application makes it fun and easy to find and classify new variable stars based upon their CMD trajectories.

Quick inspection of the VECMD reveals that different types of variables show different clusters of line segments. Behind each of these  segments lies a more complex CMD trajectory whose shape encodes the physics underlying each star's variability. Radial pulsators exhibit Lissajous-like loops that can even display features like the Hertzsprung bump directly. Binaries trace tilted, vertical, or hooked line segements that separate common features like irradiation, tidal distortion, and eclipses geometrically. Dust-obscured stars and other aperiodic accretors trace long tracks aligned with the outburst or reddening vector. Because these trajectories are built in the color-magnitude plane, they are agnostic to pulsation or orbital period, and they remain informative even for aperiodic stars or systems whose light curves cannot be phase-folded. The representation is compact, interpretable, and highly amenable to machine learning. We are currently working on developing methods to fit and parameterize the CMD trajectories, with the goal of utilizing machine learning algorithms to find, classify, and extract physics for as many targets as possible. Applying such methods to future large photometric data releases like \gaia\ DR4 and eventually the Argus array \citep{Argus2022} should be fruitful.

\begin{acknowledgments}
This project is supported by the College of Arts and Science’s Seed Funding Grant program at The University of North Carolina at Chapel Hill and the Wells Family. B.K. acknowledges support from the Summer Undergraduate Research Fellowship (SURF) program at UNC-CH. We would like to thank Tarek Zikry (UNC-CH) for drawing our attention to ElPiGraph and Niall Miller (Argus Array, UNC-CH) for discussions surrounding the capabilities of the new \texttt{MESA colors} module. Finally, B.B. would like to thank Ludwig G\"{o}ransson and Daniel Pemberton for their musical scores to {\em The Odyssey} and {\em Project Hail Mary}, both of which made working on this project and manuscript even more enjoyable.

This work has made use of data from the European Space Agency (ESA) mission {\it Gaia} (\url{https://www.cosmos.esa.int/gaia}), processed by the {\it Gaia} Data Processing and Analysis Consortium (DPAC, \url{https://www.cosmos.esa.int/web/gaia/dpac/consortium}). Funding for the DPAC has been provided by national institutions, in particular the institutions participating in the {\it Gaia} Multilateral Agreement.

Portions of the data-analysis and plotting code used to generate the figures in this work were developed with the assistance of generative AI tools (Anthropic’s Claude). These tools were also used in a limited capacity to help outline, proofread, and edit select passages of the manuscript text, which was otherwise written by the authors. All code and text were reviewed and validated by the authors, and all figures were produced by executing this code on real \gaia\ data. The authors take full responsibility for the content of this work.

\end{acknowledgments}

\section*{Data and Software Availability}

The \texttt{IVECMD} application, together with the code used to construct the variability-encoded CMD, is publicly available on GitHub\footnote{\url{https://github.com/KirbyBennett/IVECMD}} and  archived on Zenodo\footnote{\url{https://doi.org/10.5281/zenodo.23025193}} \citep{iVECMD_zenodo}. The macOS .dmg and Windows .exe are both available on the GitHub site and are free to download. Both apps require $\sim4$ GB of disk space to operate. The underlying \gaia\ DR3 multi-epoch photometry \citep{Eyer2023} is available through the ESA \gaia\ Archive\footnote{\url{https://gea.esac.esa.int/archive/}} and the VizieR catalog I/355/epphot.

\software{Astropy \citep{astropy2013, astropy2018, astropy2022}, SciPy \citep{scipy2020}, NumPy \citep{numpy2020}, Matplotlib \citep{matplotlib2007}.}
\appendix

\section{An Interactive GAIA VECMD Explorer}
\label{sec:ITCMD}

We introduce the Interactive Variability-Encoded CMD application (\texttt{IVECMD}) to streamline interactions with the large amount of complex data presented above. Sifting through dense data to find interesting systems is tedious and time-consuming with standard plotting conventions. By implementing a unique set of features to speed up searches, including the ability to isolate types of known variables, sliders filtering many statistics, a search feature based on Gaia DR3 IDs, and more, we create an app that allows researchers and students to learn about and explore the VECMD efficiently. Our hope is that \texttt{IVECMD} will also be used to discover new variable stars.  

\subsection{Features}

\begin{figure}
    \centering
    \includegraphics[width=0.89\linewidth]{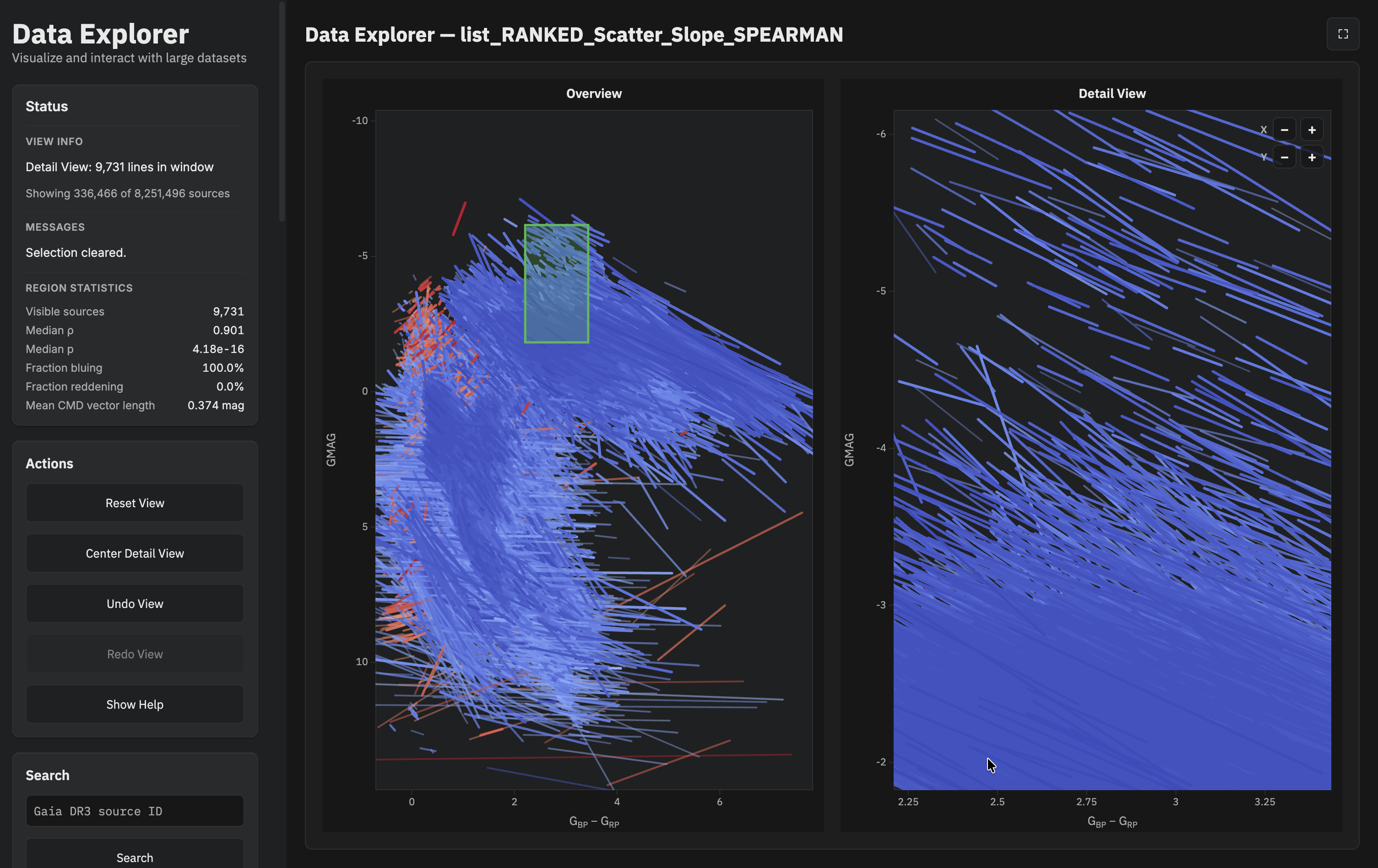}
    \caption{A full view of the app when it is launched. The interactive features appear on the left, the overview plot of the VECMD is in the middle, and the detailed view plot is on the right. The detail view plot is where a user can interact with the data and click on lines to learn more about a specific system.}
    \label{fig:fullvecmd}
\end{figure}

When users open the app, they see two VECMD plots: an overview and a detailed view as seen in Figure \ref{fig:fullvecmd}. The overview plot presents the full VECMD and shows the region the app is zoomed into via a moveable green box.  The detailed view magnifies the zoomed-in region from the overview plot and allows the user to directly interact with the dataset by clicking on the plot. In the top right of both plots, a reddening vector corresponding to an extinction of $\mathrm{A_V=1\ mag}$ and direction taken from \citet{WangandChen2019} is shown. Users can drag the reddening vector in the overview plot to any location they like. Additionally, users have the ability to turn the green zoom box as well as the reddenning vectors off. Hovering a cursor over the detailed view panel presents a gold circle indicating which system your cursor latches onto. The app looks within a snap radius and will favor more significant lines that your cursor is close to. The nearness calculation is based on the center of the line or the average of the $\mathrm{M_G}$ and $\mathrm{G_{BP}} - \mathrm{G_{RP}}$ values from Gaia DR3 MEP measurements computed in Section \ref{sec:analysis}. Hovering over a line also shows a tooltip with the system's ID, opacity, spearman p value, y-intercept, and measured slope.

Additionally, there are many features that allow for easy data extraction from the images. First, there is an export CSV of visible, selected, or filtered sources. This function allows for the quick accessibility of all data and even gives the option to include the statistics column that the app uses. Users can then perform their own analysis on select system using these CSV files. There is also a button that allows users to download a PNG, JPEG, or PDF of both the overview and detailed view window. Figures \ref{fig:extinction} and \ref{fig:againstthegrain} show example outputs from \texttt{IVECMD} that allow for quick figure making with features such as the reddening vector, main sequence line, and instability-strip overlay.

When users click on a system, the app queries the VizieR catalog ``I/355/epphot" for the Gaia DR3 MEP associated with the system's ID, as well as parallax from the catalog ``I/355/gaiadr3" to convert apparent G magnitude to absolute $\mathrm{M_G}$. The app also includes a built-in fallback to query the ESA Gaia archive for the necessary MEP if the VizieR TAP service returns an error. Users are then presented with a window displaying individual epochs of data as seen in the bottom panel of Figure \ref{fig:mep_example}. The app checks to see if there are any outliers detected by our outlier removal algorithm discussed in Section \ref{sec:mecmd}. After each removal process, each point is given a label depending on where in the process it was removed (after the 4-$\sigma$ cut, after the standard deviation-ratio slope-stability clip, after both, etc.). Points that were flagged as outliers for the 4-$\sigma$ cut in G and at least one of $\mathrm{G_{BP}}$ or $\mathrm{G_{RP}}$ in the same direction are given the designation as a candidate eclipse or flare. Candidate eclipse and flare points are kept when conducting fits and analysis on systems. Different shapes and colors highlight each outlier depending on where in the process the point was cut, as well as the labels ``used, flare/eclipse candidate, std-slope clipped, sigma-clipped, clipped by both". The gold star represents the mean of the G and $\mathrm{G_{BP}} - \mathrm{G_{RP}}$ points after the sigma-clipping algorithm has taken place. Links to the SIMBAD page and the VizieR Gaia catalog page, ``I/355/gaiadr3", are available at the top for more information on the target. The bottom of the new window shows detailed statistics pertinent to that target: the spearman correlation coefficient, spearman p-value, RA, Dec, galactic latitude $b$, parallax, distance, number of epochs and how many epochs were used, the count in each category, the low-SNR count, slope $m$ and the data source. The other buttons in the top bar produce the full light and color curve plots as well as phase-folded light and color curves, allow users to download a png of the plot, and a button to reset the zoom for the window.

\begin{figure}
    \centering
    \includegraphics[width=0.89\linewidth]{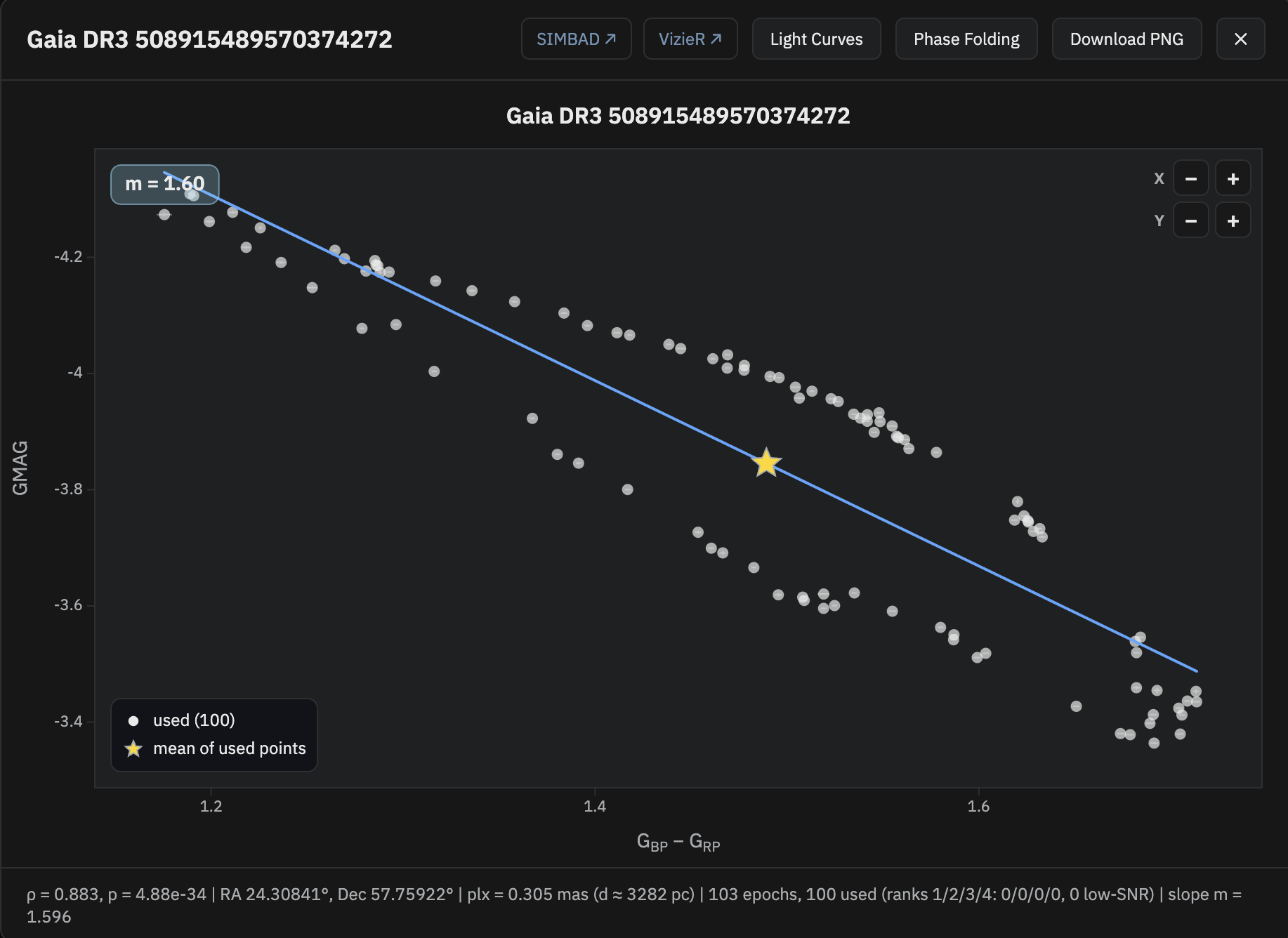}
    \caption{An example of the window that appears when a line is clicked on. At the top, we show the system’s Gaia ID, links to the SIMBAD and Gaia DR3 VizieR page for that object, and buttons to view the full light curve, phase-fold the object, and download an image of the trajectory in the CMD. In the bottom left of this figure, the legend shows the rank each point was assigned during both clipping algorithms, so you can see which points were clipped and why. Additionally, at the bottom of the page, we see more statistics for the system shown.}
    \label{fig:epochs}
\end{figure}

When the ``Light Curves" button is pressed, a new window appears showing the target's stage-1 sigma-clipped Gaia DR3 MEP data for each filter. This plot only omits points that have been cut through the standard 4-$\sigma$ clipping procedure. In the bottom panel, we also present the color data. We compute the color data ourselves by finding matched epochs of $\mathrm{G_{BP}}$ and $\mathrm{G_{RP}}$ points and computing $\mathrm{G_{BP}} -\mathrm{G_{RP}}$. The ``Phase Fold'' button phase-folds each filter on the highest peak in a generalized Lomb-Scargle Periodogram with methods provided by \citet{Zechmeister2009}. The periodogram has a frequency grid spanning $f_{\mathrm{min}} = 1/T\ \mathrm{day}^{-1}$, where $T$ is the total baseline, to $f_{\mathrm{max}} = 24\ \mathrm{day}^{-1}$ with the number of frequencies calculated by $N = 30\times T\times \left(f_{\mathrm{max}} - f_{\mathrm{min}}\right)$. For a standard \gaia\ MEP dataset, this usually comes out to about 700,000 points. We choose to oversample by 30 to allow for smooth peaks throughout the periodogram. The bottom panel shows the 3 periodograms above multiplied, and phase-folds the $\mathrm{G_{BP}} - \mathrm{G_{RP}}$ color on the highest point in the new periodogram. A button to compute a high-resolution periodogram is available and will search the frequency grid spanning $f_{\mathrm{min}} = 0\ \mathrm{day}^{-1}$ to $f_{\mathrm{max}} = 288\ \mathrm{day}^{-1}$ with 50 times  oversampling ($N = 50\times T\times \left(f_{\mathrm{max}} - f_{\mathrm{min}}\right)$). This higher resolution and wider frequency range permits a periodogram sampling of around 14,000,000 points. There are also buttons to multiply the found period by a factor of 2 (or divide the period by a factor of 2) for ellipsoidal, eclipsing, contact binary, or other systems for which the first harmonic might be stronger than the fundamental frequency. Finally, you can right-click to re-fold the periodogram at a different peak in the periodogram.

\begin{figure}
    \plottwo{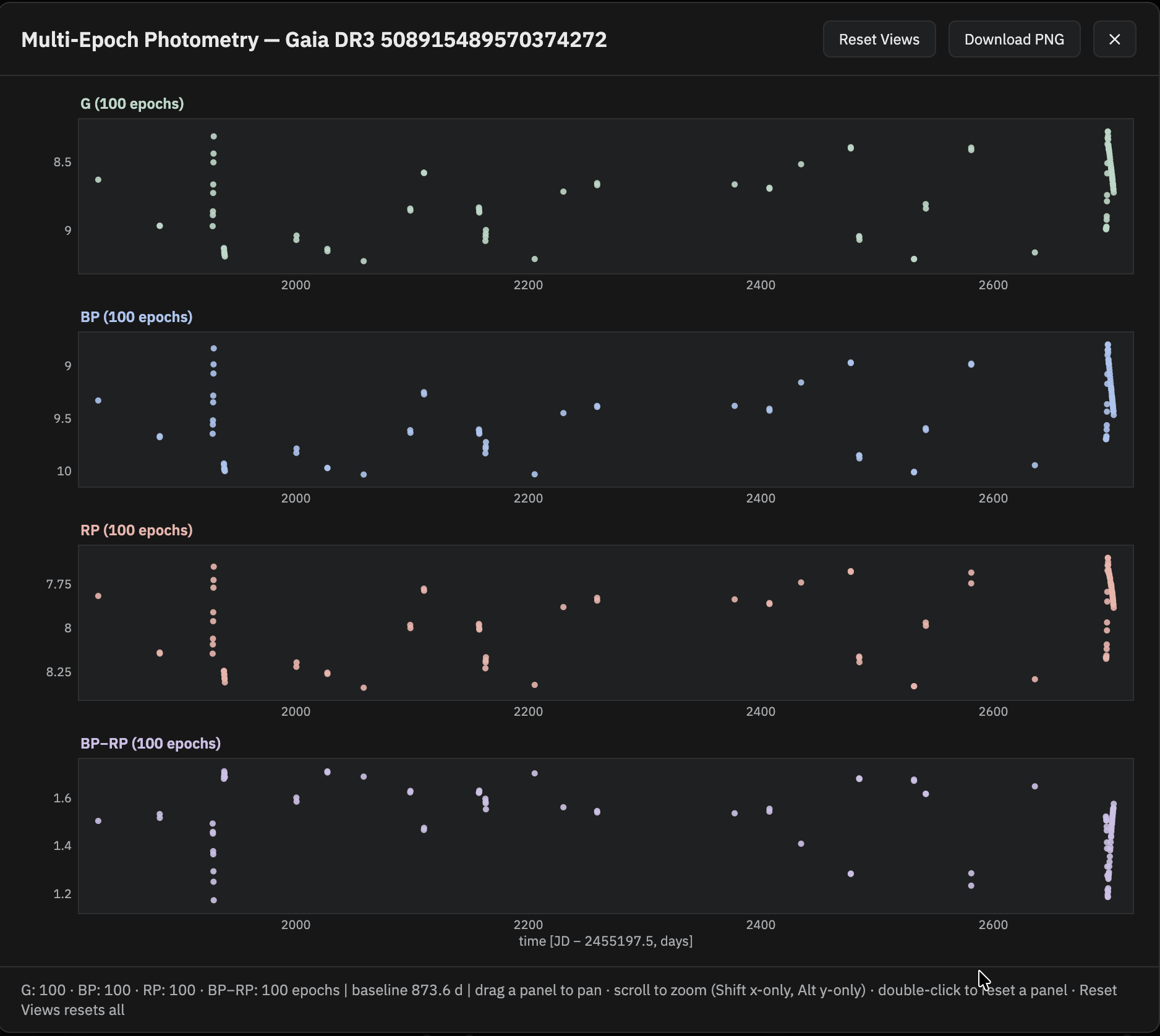}{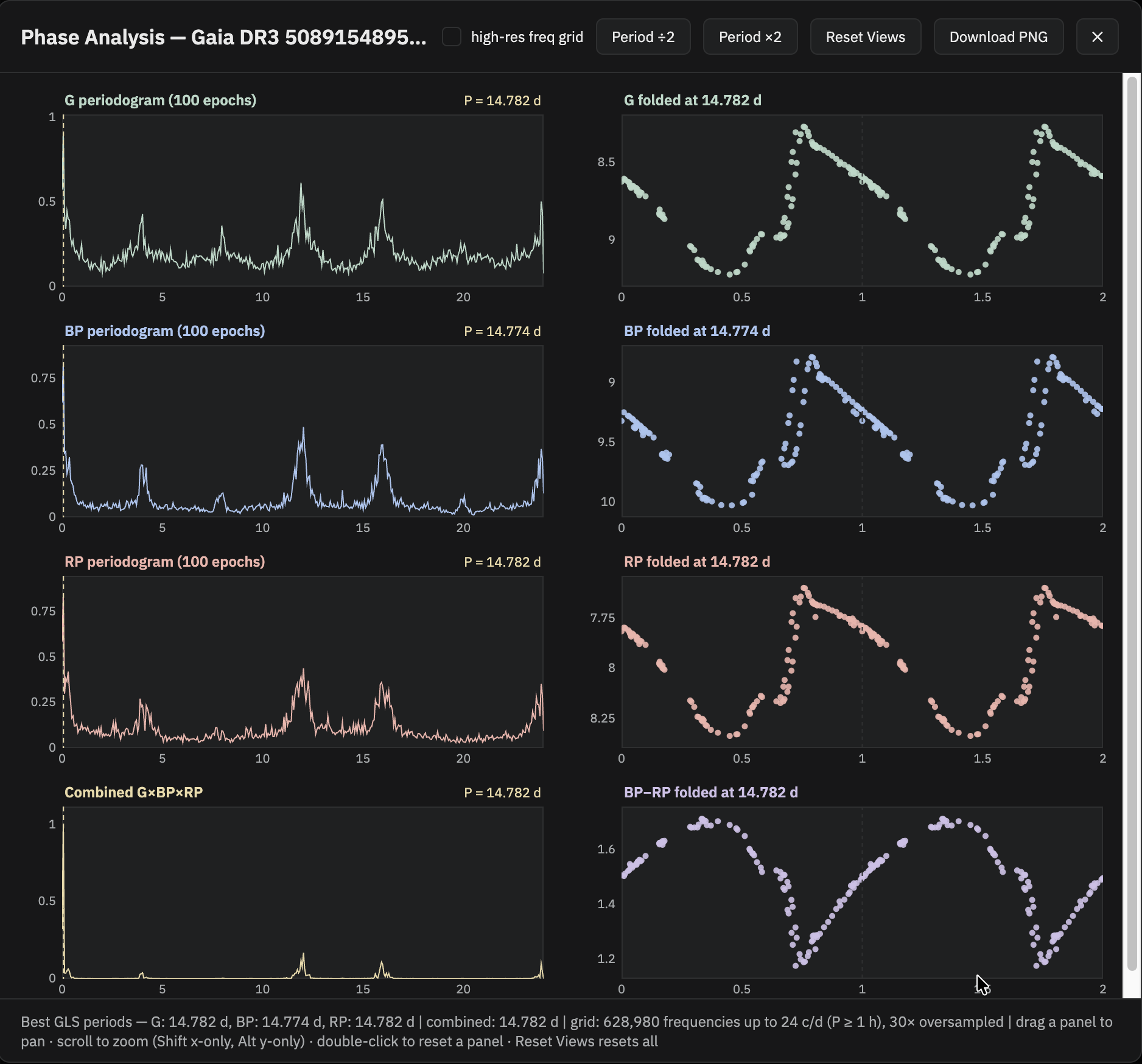}
    \caption{Left: The full sigma-clipped Gaia MEP for the above system. We show this for each band, along with the matched $\bprp$ values showing the overall color curve. Right: The phase-folded MEP for each filter and the color curve are shown next to their corresponding periodograms. For the bottom panel, the three periodograms for each filter were multiplied together to create a combined periodogram. At the top, options let you use the high-resolution frequency grid to sample each periodogram, sample at half or double the current period, and save the image.}
    \label{fig:combined}
\end{figure}

Pulsator lists (mainly focused on radial mode pulsators) have been provided to give users a jumping-off point for exploring different well-known system types, as well as some in-depth subtypes. These lists were provided from the Gaia RR Lyrae and Cepheid catalogs as well as the International Variable Star Index to get a wide breadth of different pulsator types \citep{Gaia2023Cepheid, Gaia2023RRLyrae, Watson2006}. In total, we used the classes, Cepheids (Type I/II/Anomalous), RR Lyrae (RRab/c/d), $\delta$ Scuti (SX Phe), Mira, Semiregular (SRA–SRS), BLAPs, and sdBV. The ``only'' button restricts the view to one list, and several lists can be shown at once in different colors. Additionally, users can upload their own lists of Gaia IDs to highlight lines on the plot that match the lists they upload.
    
The app provides numerous sliders for instantly filtering the dataset as a whole and isolate features a user might want to search for. First, we allow users to filter by line direction showing only blue or red lines.
We also included a slope filter that allows users to search for specific fitted slope ranges. A significance cut based on the p-value of the Spearman coefficient allows for users to analyze systems with high spearman correlation. Additionally, an opacity slider is present isolating the strongest lines on the diagram, as opacity relates to the p-value through a logarithm. A minimum parallax-over-error filter allows users to parse photometry that may be significant but lack a well-defined parallax. The distance cut lets users search for systems near or far, enabling volume-limited searches in Gaia MEP. Finally, a galactic latitude filter lets users parse different levels of the galactic disk and halo more easily and find systems that were once hidden. These filters can be saved and revisited anytime using the ``Saved Settings" box.

Our testing of this tool has already yielded many new variable star discoveries. While exploring the extreme horizontal branch (EHB) of the CMD with the ``Red" line filter activated, we serendipitously discovered the longest-period HW Vir-type (sdB$+$dM) system currently known, Gaia DR3 4176942419339829504. The system sits at the lower end of the EHB and has a period of $0.5734$ days, or 13.76 hours. Its apparent G magnitude is $15.99$ with a $\mathrm{G_{BP}}-\mathrm{G_{RP}} = 0.60$. Additionally, using the reddening vector as a guide, we found an HW Vir-type system with a galactic latitude of $\mathrm{b} = +3.75^{\circ}$. Gaia DR3 2202073408192045056 has an apparent G magnitude of $17.05$, a $\mathrm{G_{BP}} - \mathrm{G_{RP}} = 0.96$, and a period of $P = 0.3111$ days. Originally classified as an RRc type star by \citet{Sesar2017}, when we phase-fold the system, we clearly see a shallow, short eclipse indicative of a consistently recurring primary eclipse. The ease of finding these systems show how many important discoveries are accessible using the \texttt{IVECMD} app.

\subsection{Improvements for the Future}

With the preparation of our machine learning algorithm based on Gaia DR3 MEP, we look to add initial classifications to every system in our database to allow users to search for candidate systems of their choosing. Additionally, given the wide variety of CMD trajectory shapes that exist across the diagram, we plan to implement a shape filter, along with points and lines, in the near future to show the majority of shapes on the VECMD. This will show where trajectories dominate in the VECMD, as well as isolated searches for different paths. We also plan to add searching for Gaia DR3 XP Spectra for each target to allow initial, brief spectral analysis of individual systems.

The \texttt{IVECMD} app is a powerful tool for discovering and even classifying stars through their CMD motions, as well as their phase-folded light curves. Using various filters, users can discover variable stars with unprecedented efficiency. With the advent of Gaia MEP DR4 on December 2\textsuperscript{nd}, 2026, \texttt{IVECMD} provides a way to parse Gaia MEP data quickly and view all the information available for millions of targets. The app will become even more powerful with Gaia DR4 MEP, and, with our plans for the future, classification and searching will become even simpler.

\section{Hertzsprung Progression}
\label{appendix:hertzsprung}
\begin{figure*}[ht!]
\centering
\includegraphics[width=0.24\columnwidth]{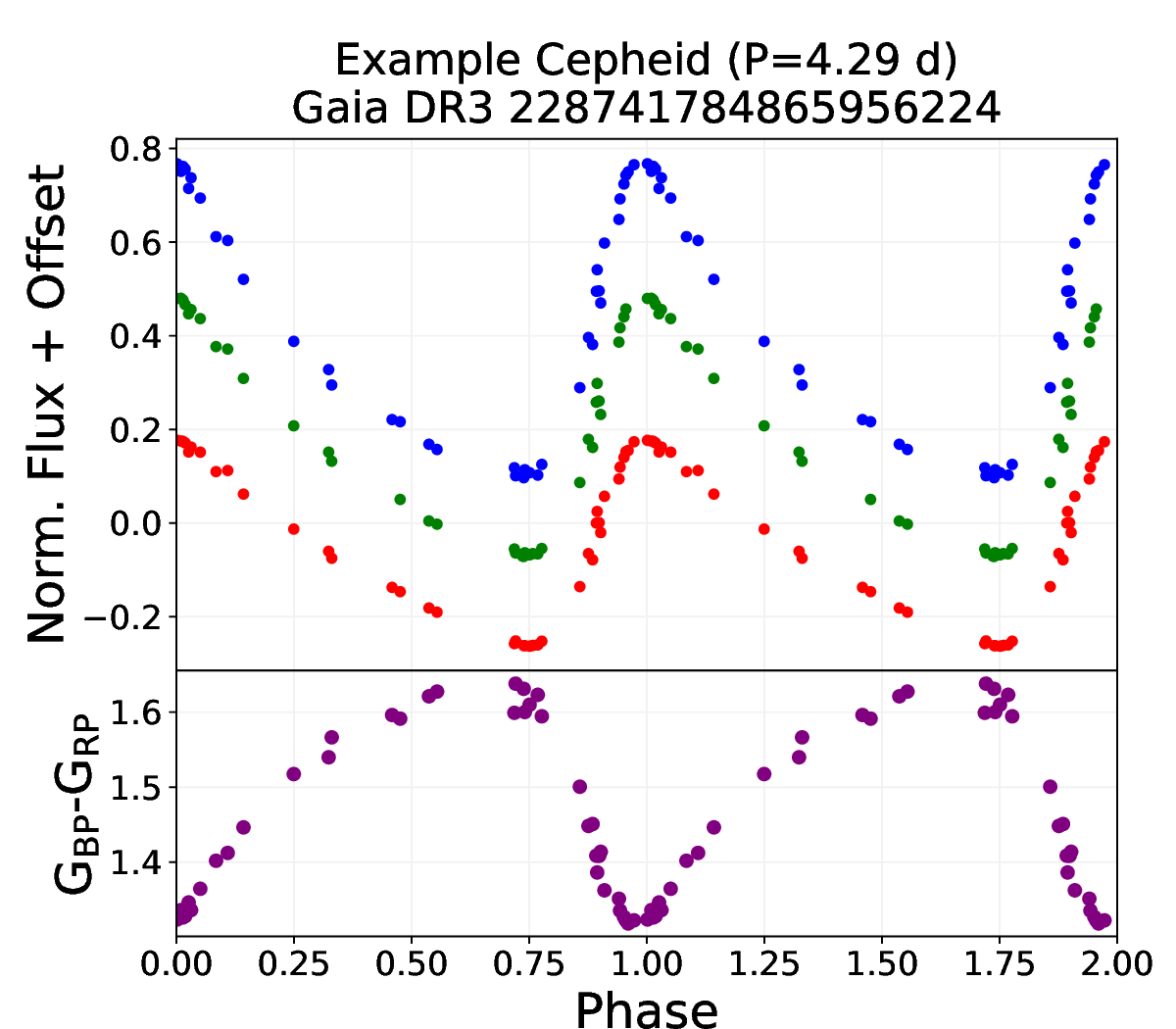} 
\includegraphics[width=0.24\columnwidth]{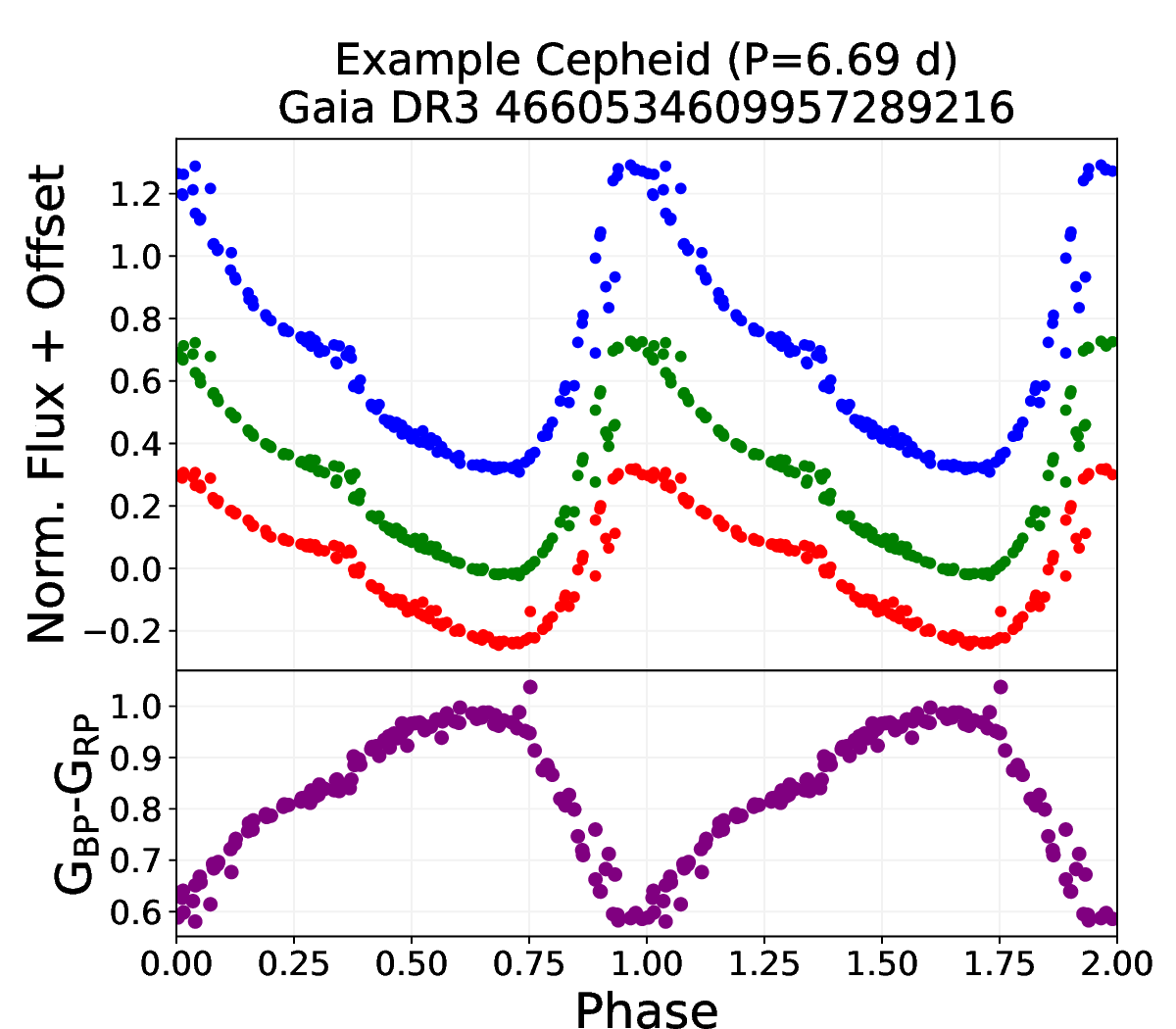} 
\includegraphics[width=0.24\columnwidth]{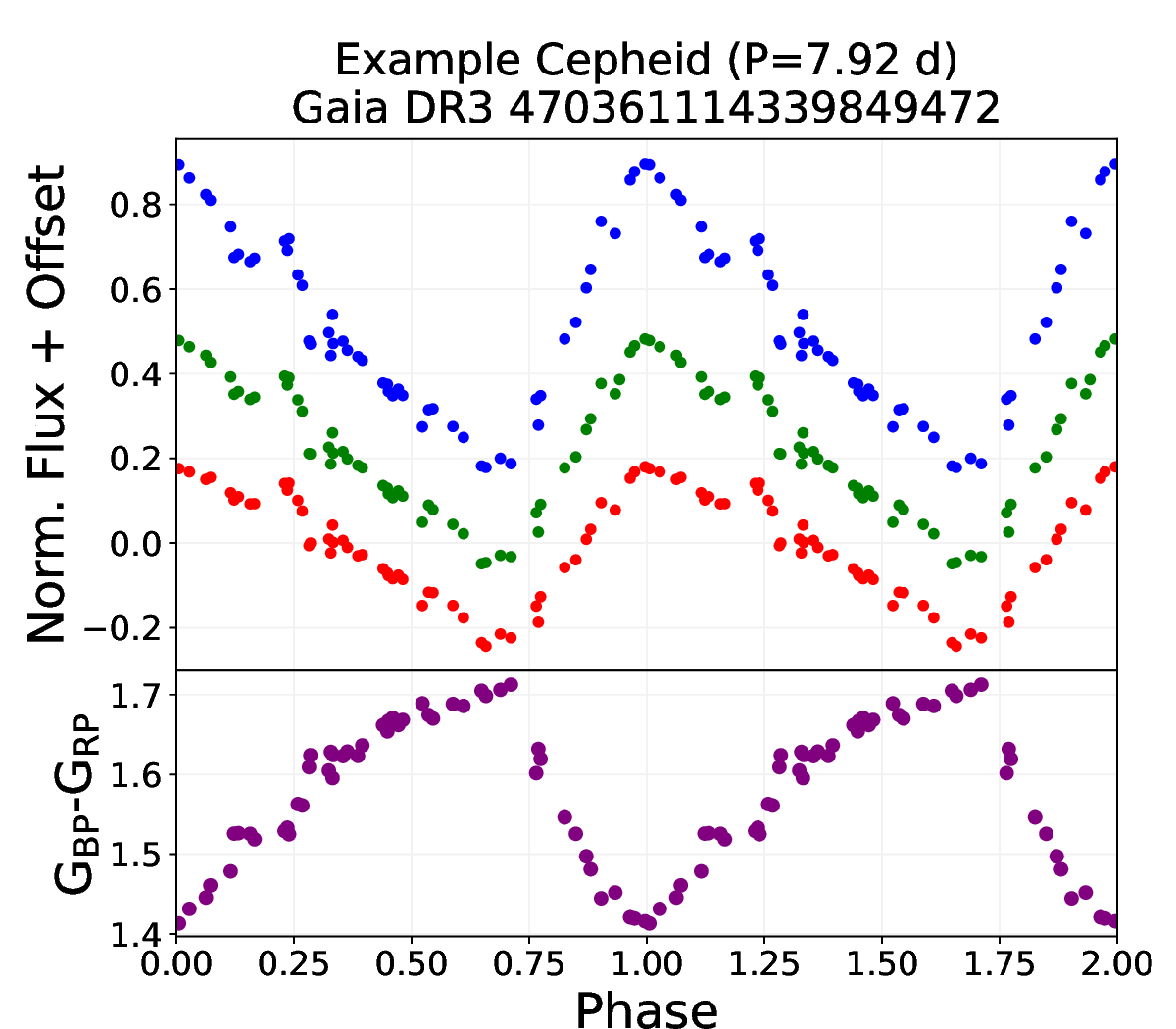} \\
\includegraphics[width=0.24\columnwidth]{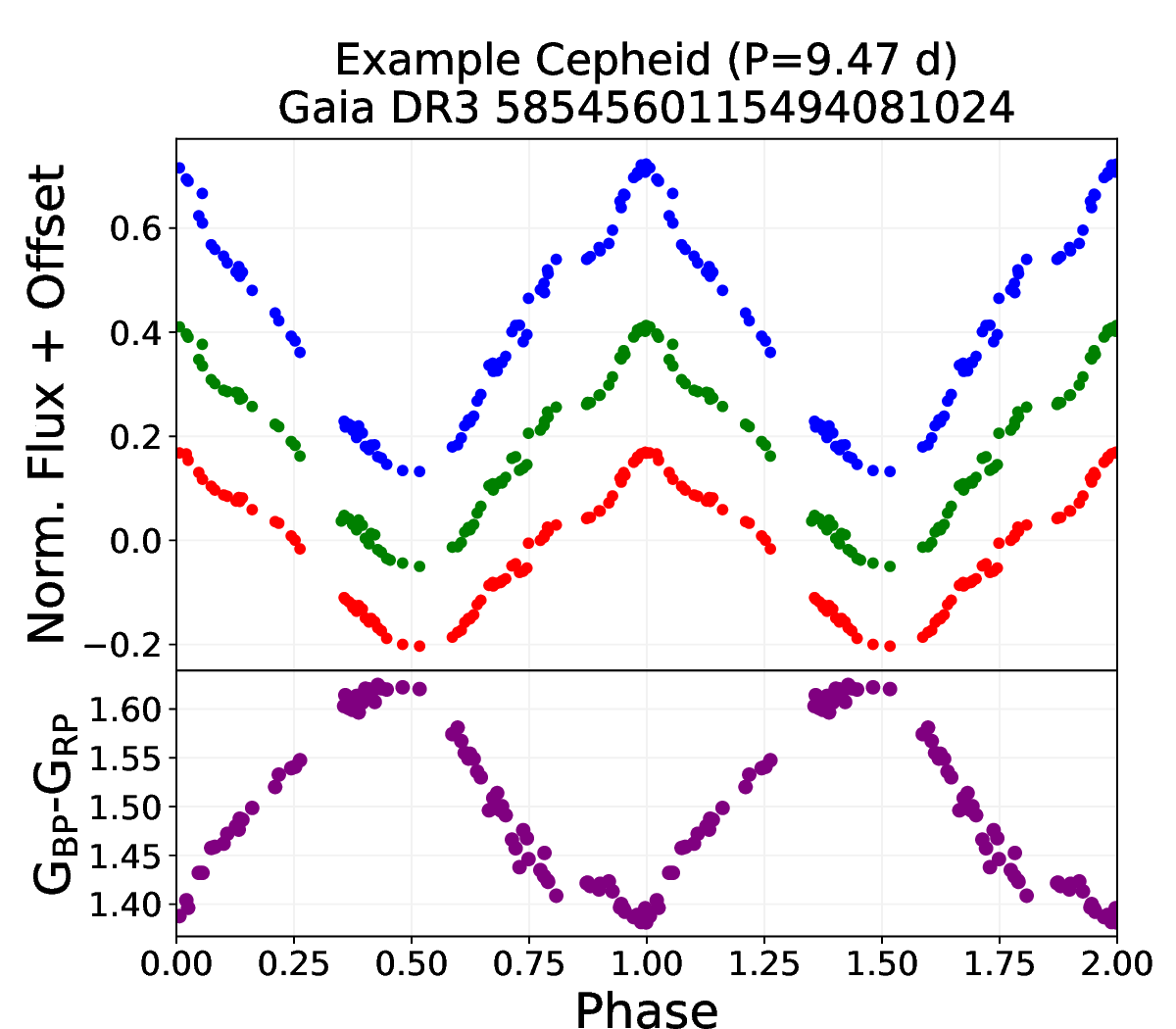} 
\includegraphics[width=0.24\columnwidth]{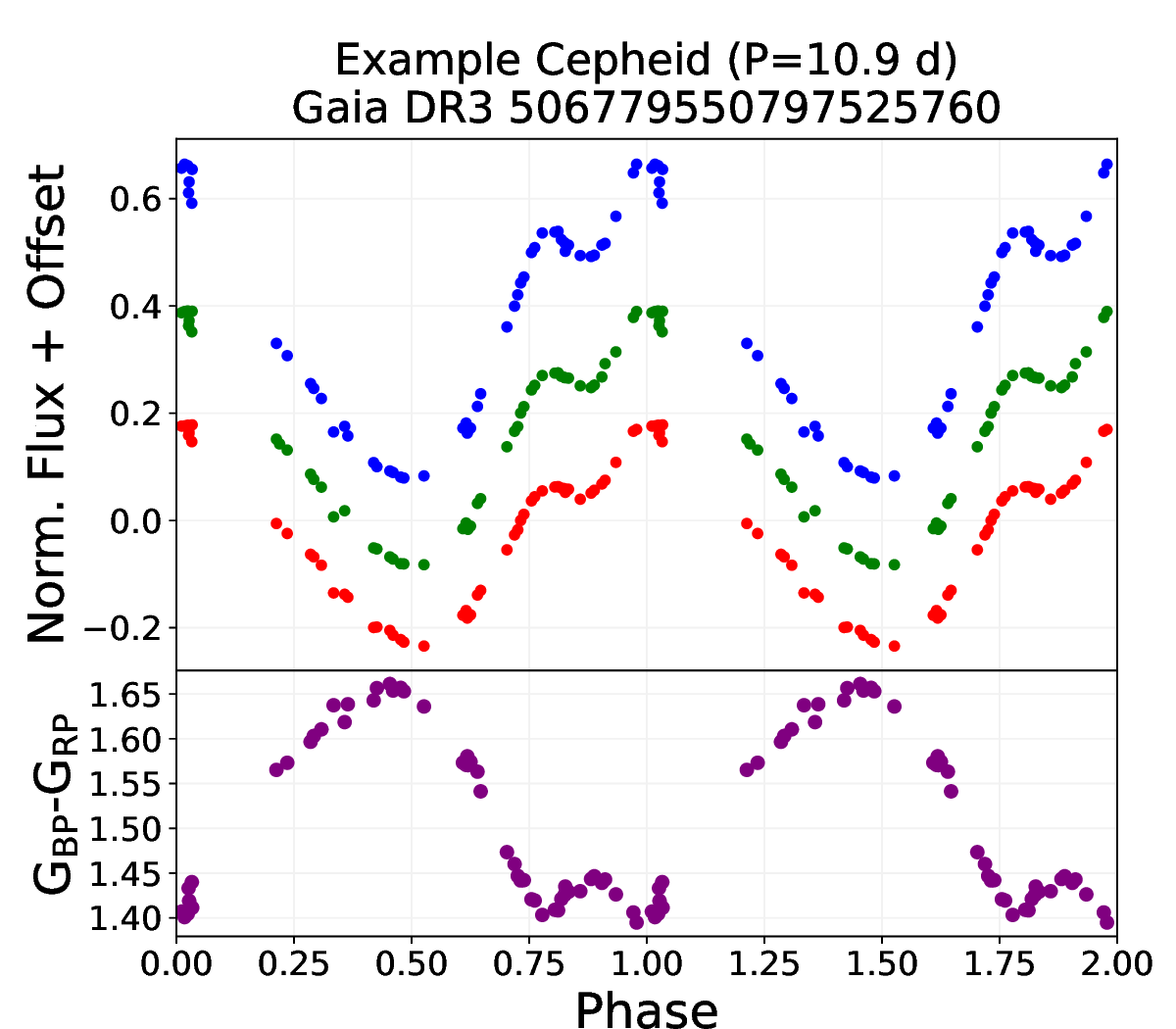} 
\includegraphics[width=0.24\columnwidth]{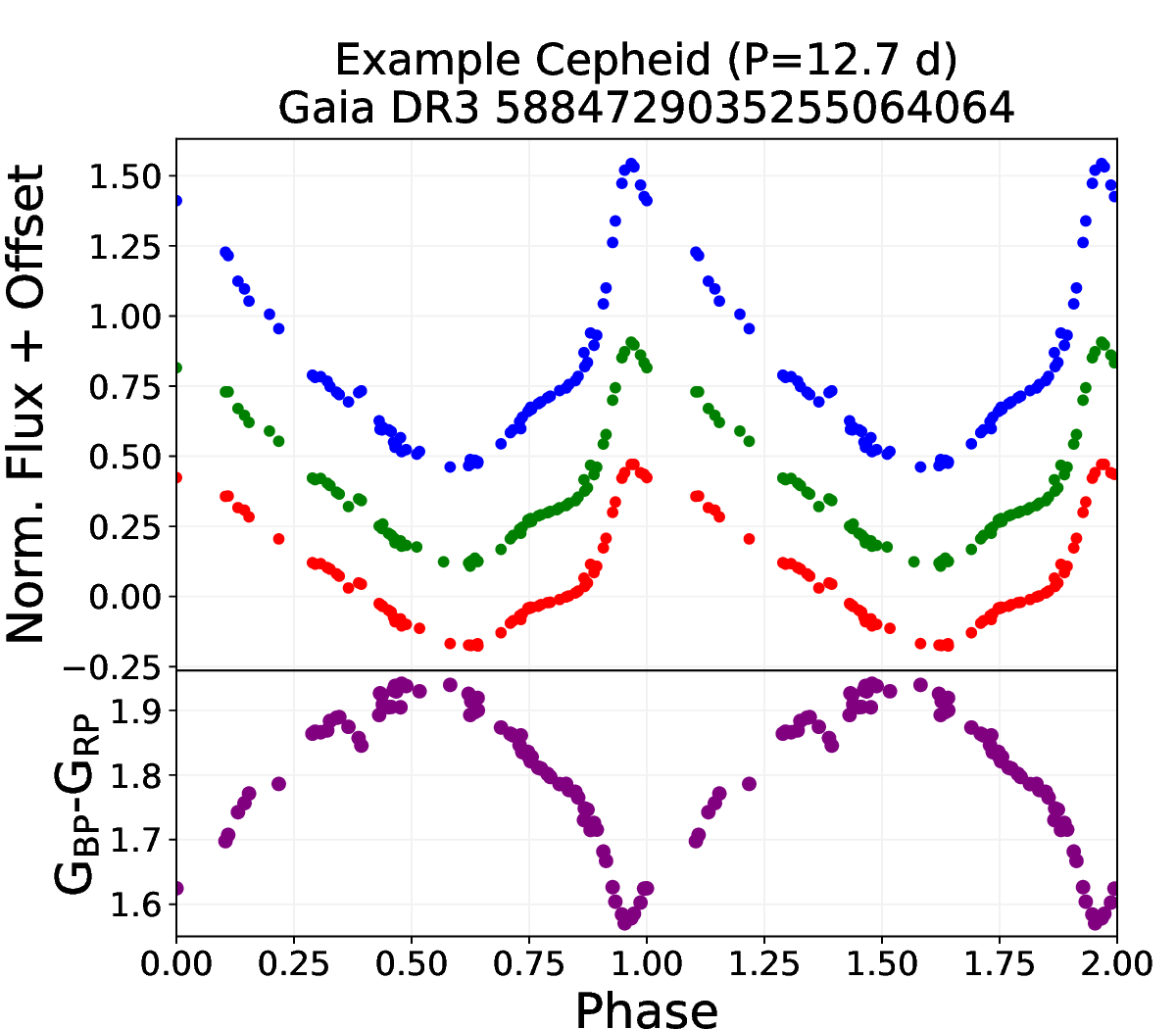} 
\includegraphics[width=0.24\columnwidth]{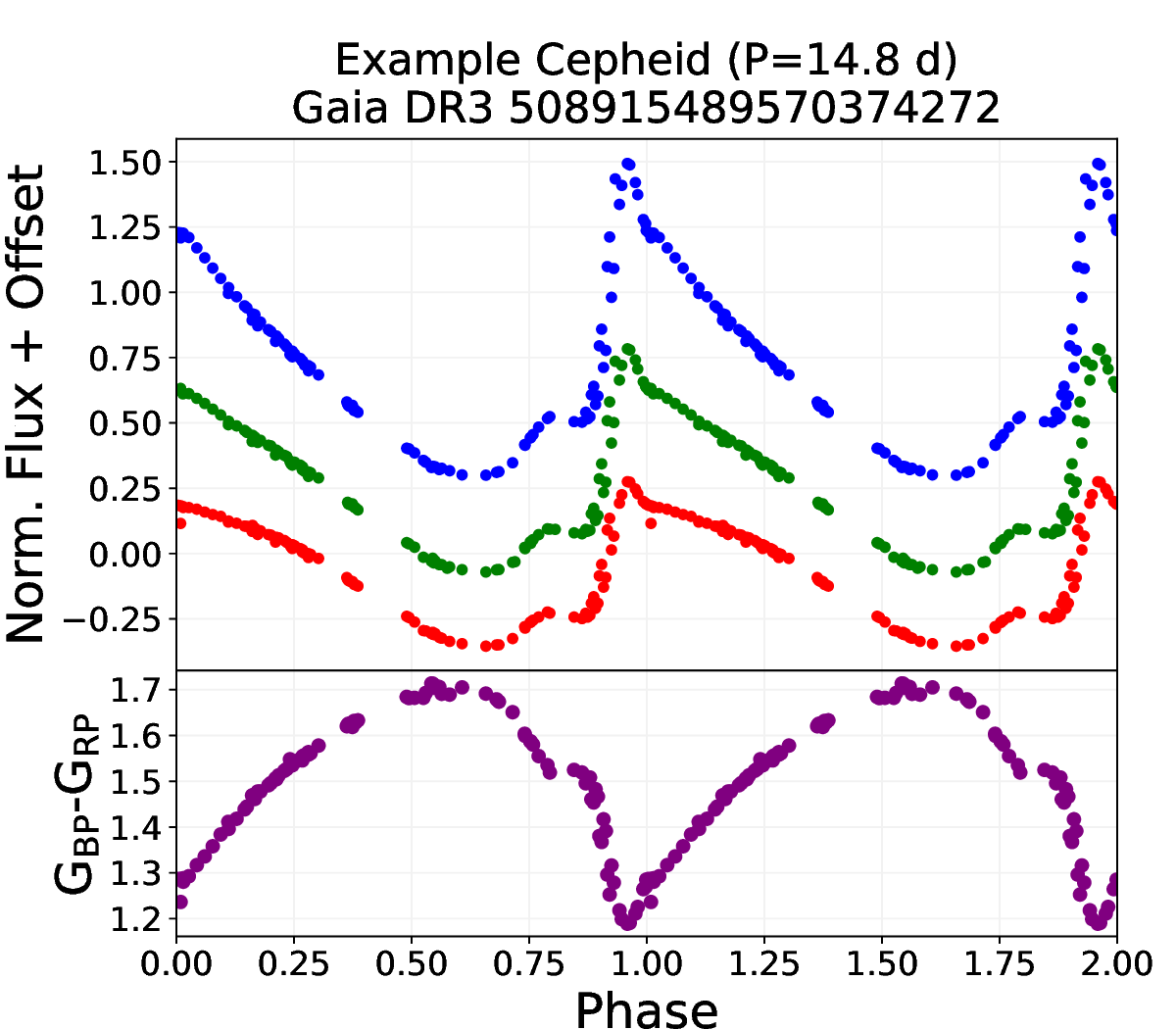} 
\caption{\gaia\ photometry for a series of Classical Cepheids across the Hertzsprung progression, with periods ranging from 4 d to 15 d (from left to right, top to bottom). The well-known correlation between the light curve shapes and oscillation periods can be seen. Here we show that a similar correlation also exists between the CMD trajectory shapes and oscillation periods. Since the creation of CMD trajectory shapes is period agnostic, one could theoretically use the shape to determine period (and, it follows, luminosity).}
\label{fig:hertzsprung1}
\end{figure*}

\begin{figure*}[ht!]
\centering
\includegraphics[width=0.24\columnwidth]{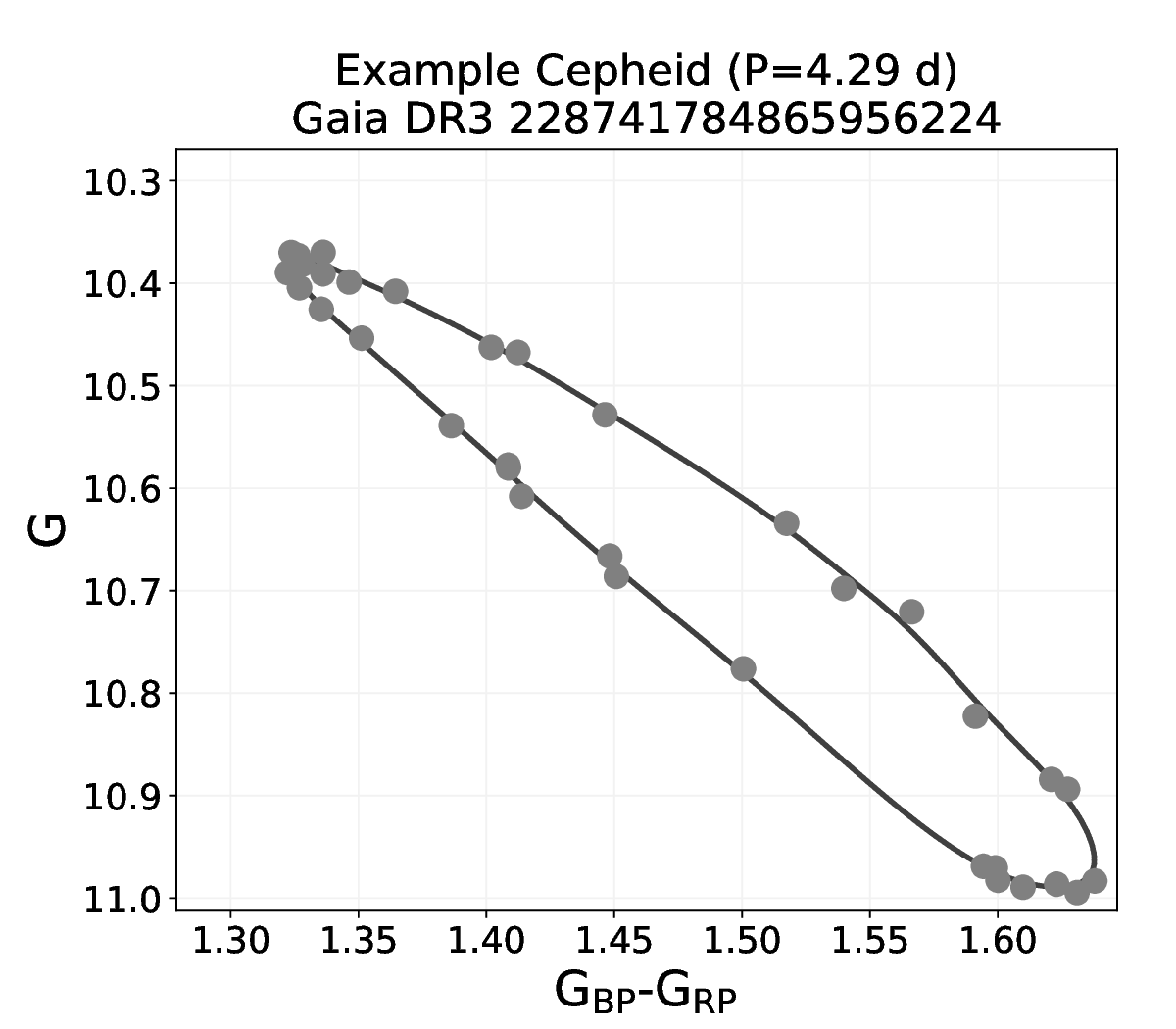} 
\includegraphics[width=0.24\columnwidth]{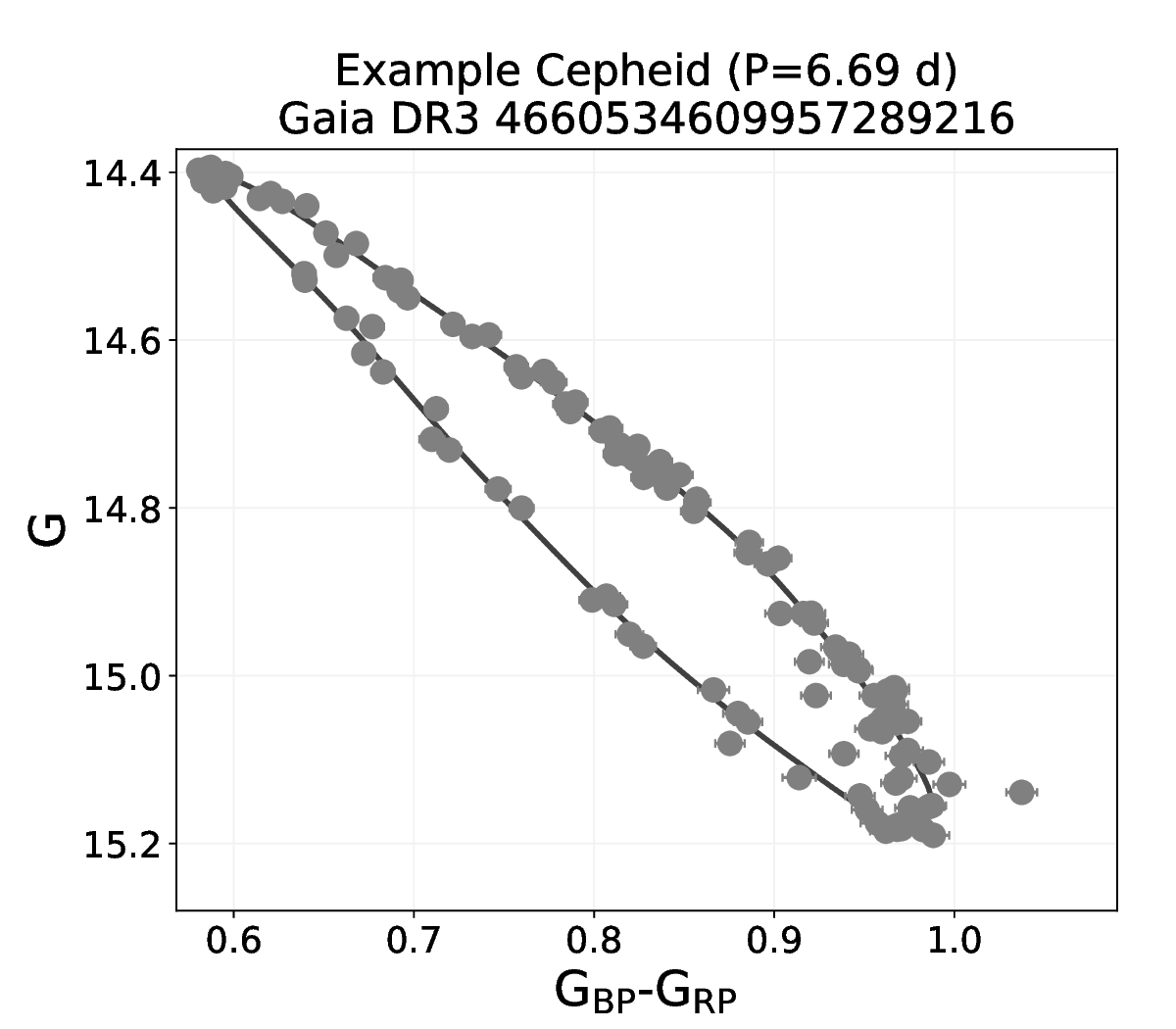} 
\includegraphics[width=0.24\columnwidth]{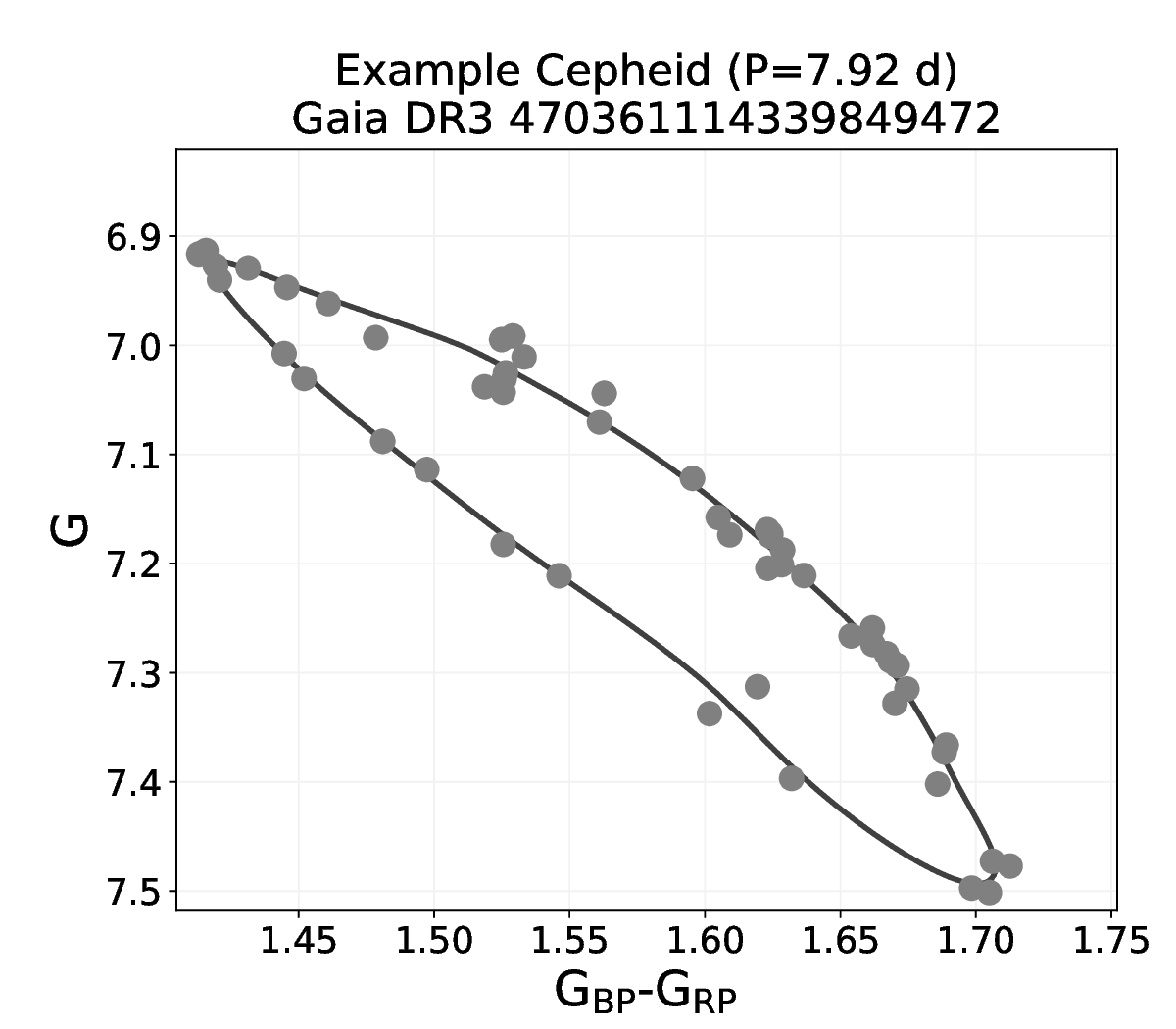}\\ 
\includegraphics[width=0.24\columnwidth]{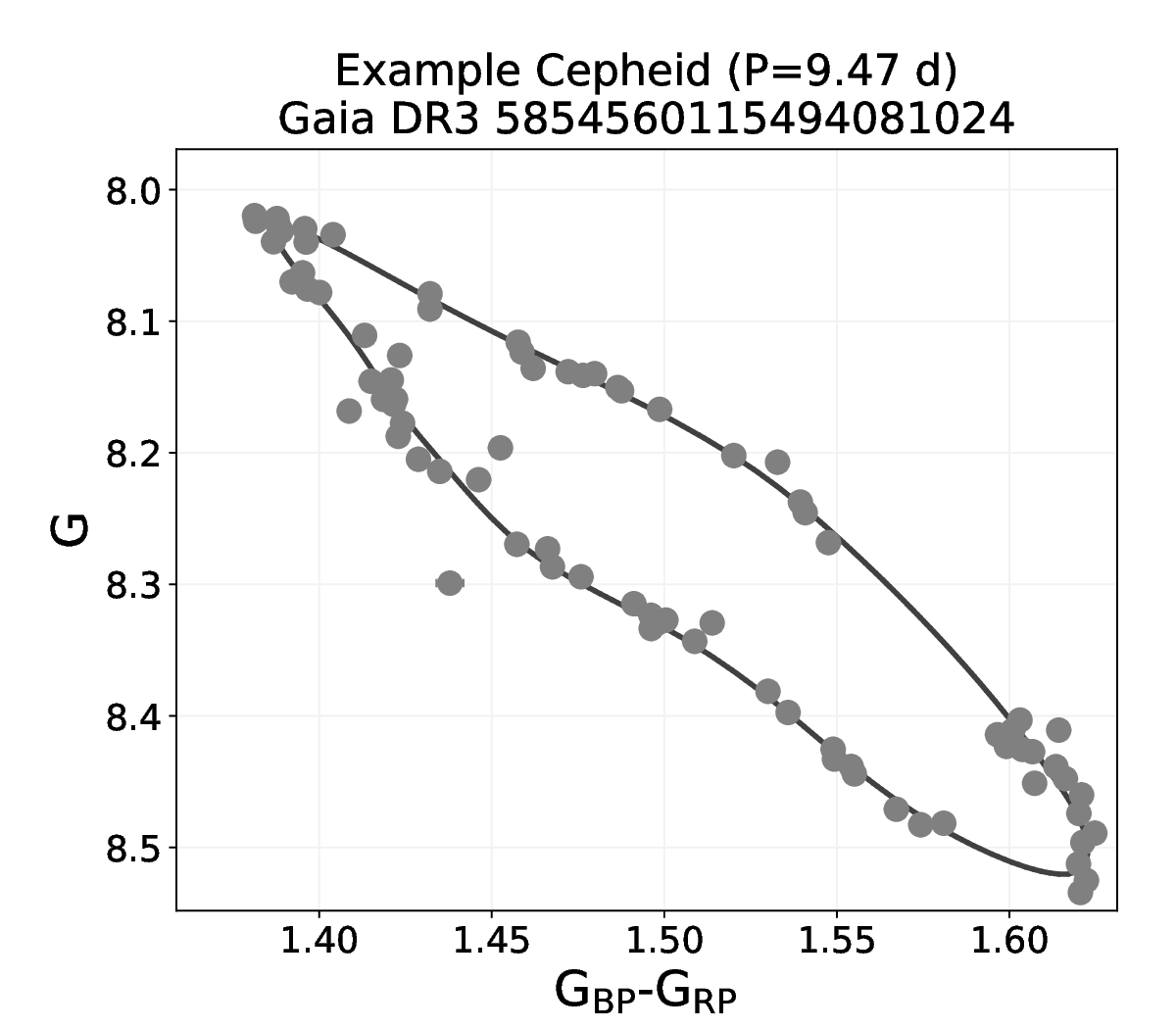} 
\includegraphics[width=0.24\columnwidth]{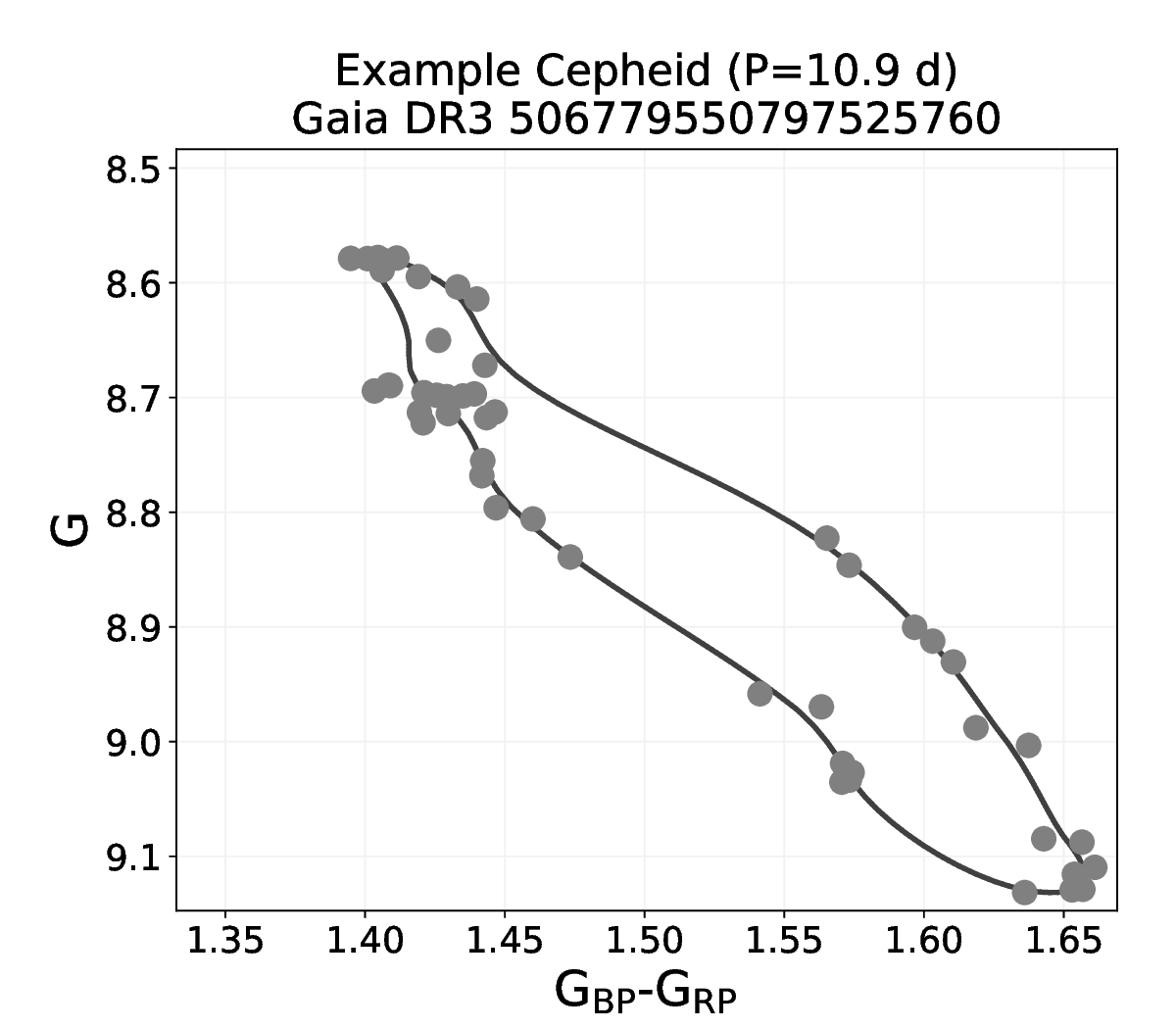} 
\includegraphics[width=0.24\columnwidth]{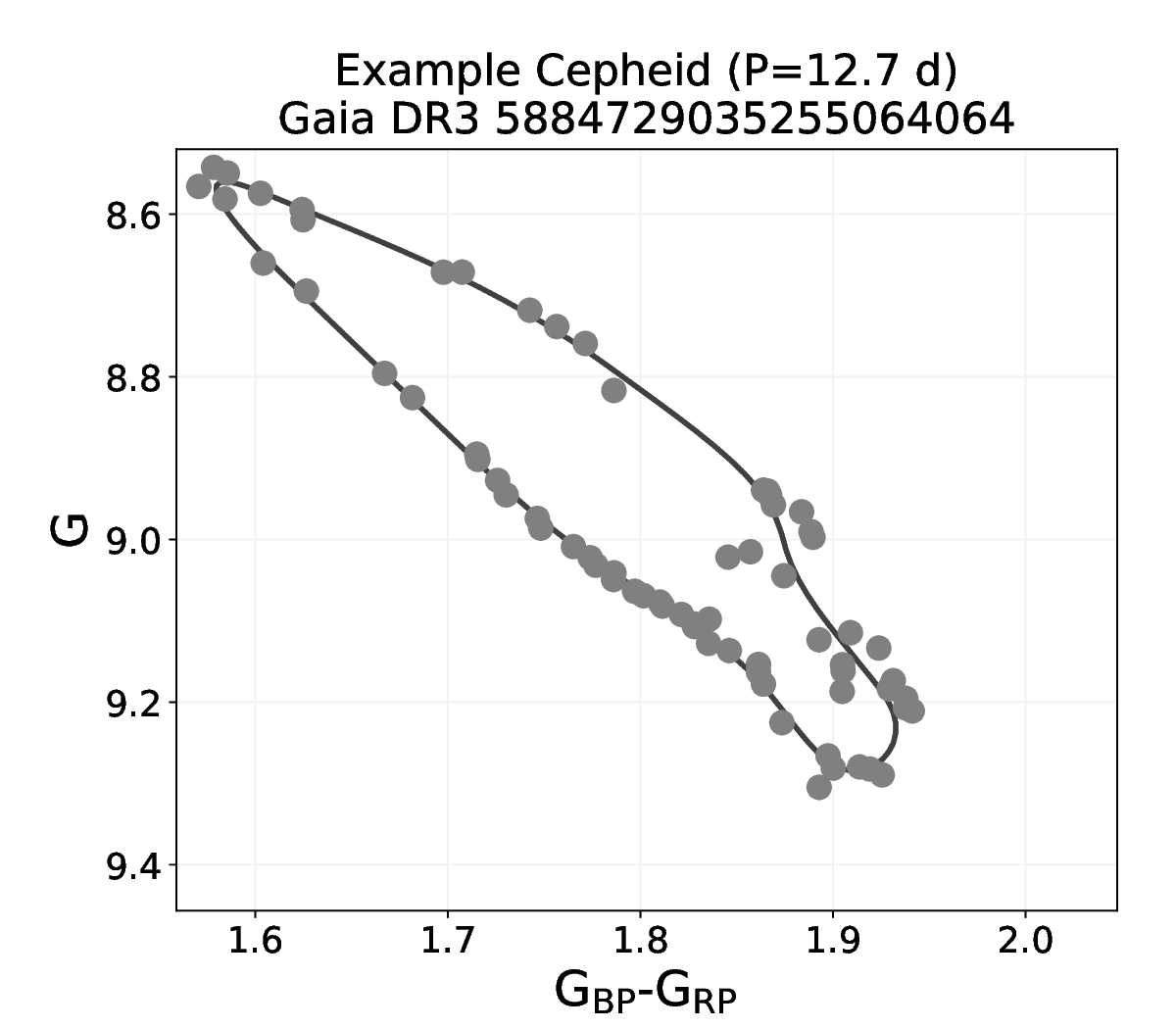} 
\includegraphics[width=0.24\columnwidth]{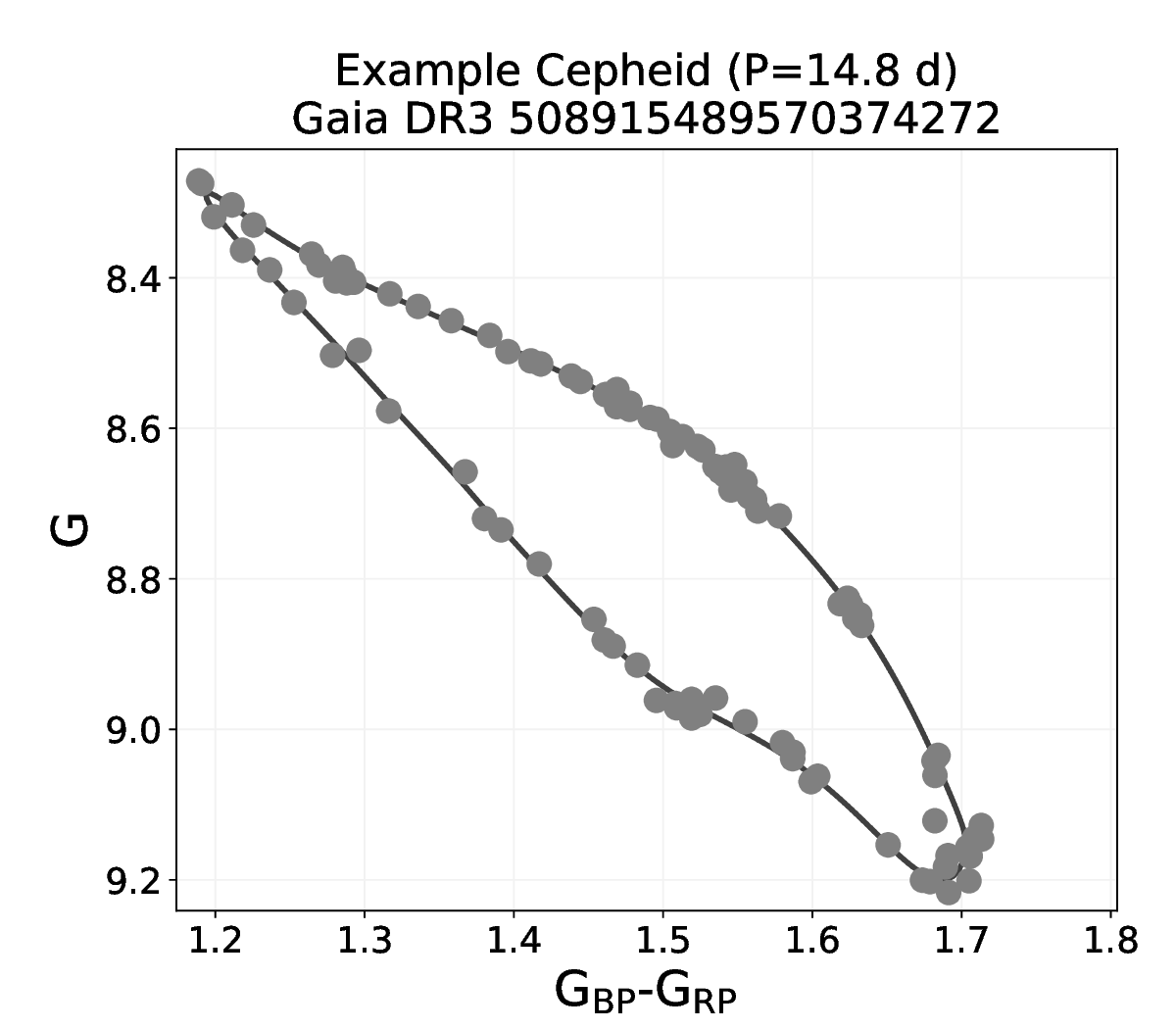} 
\caption{\gaia\ CMD trajectories with ElPiGraph fits for the series of Classical Cepheids shown above across the Hertzsprung progression, with periods ranging from 4 d to 15 d (from left to right, top to bottom). In addition to the well-known correlation between the light curve shapes and oscillation periods, there also exists a correlation between the CMD trajectory shapes and oscillation periods. Since the creation of CMD trajectory shapes is period agnostic, one could theoretically use the shape to determine period (and, it follows, luminosity).}
\label{fig:hertzsprung2}
\end{figure*}

Figure \ref{fig:hertzsprung1} and \ref{fig:hertzsprung2} illustrate how the light curve shapes and CMD trajectories of Classical Cepheids vary with pulsational period across the Hertzsprung progression.

\bibliography{main}
\bibliographystyle{aasjournal}

\end{document}